\documentclass[rmp,twocolumn,showpacs,floatfix]{revtex4}
\usepackage{amssymb}
\usepackage{natbib}
\usepackage{amsmath}
\usepackage{amsfonts}
\usepackage{graphicx}
\usepackage{mathrsfs}
\usepackage{dcolumn}
\usepackage{bm}
\usepackage{mathbbol}
\usepackage{color}
\usepackage{graphics}
\usepackage{epsfig}

\def\Tr{\mbox{Tr}}

\begin{document}
\title{{\it Colloquium:}  Quantum Fluctuation Relations:
Foundations and Applications}

\author{Michele Campisi, Peter H\"anggi, and Peter Talkner}
\affiliation{Institute of Physics, University of Augsburg,
  Universit\"atsstr. 1, D-86135 Augsburg, Germany}
\date{\today}

\begin{abstract}
Two fundamental ingredients play a decisive role in the foundation of fluctuation relations: the principle of microreversibility and the fact
that thermal equilibrium is described by the Gibbs canonical
ensemble. Building on these two pillars we guide the reader through
a self-contained exposition of the theory and applications of
quantum fluctuation relations. These are exact results that
constitute the fulcrum of the recent development of nonequilibrium
thermodynamics beyond the linear response regime. The material is
organized in a way that emphasizes the historical connection between
quantum fluctuation relations and  (non)-linear response theory. We
also attempt to clarify a number of fundamental issues which were
not completely settled in the prior literature. The main focus is on
(i) work fluctuation relations for transiently driven closed or open
quantum systems, and (ii) on fluctuation relations for heat and
matter exchange in quantum transport settings. Recently performed
and proposed experimental applications are presented and discussed.
\end{abstract}

\maketitle

\tableofcontents

\section{Introduction}
\label{sec:hist}

\begin{figure}[t]
\includegraphics[width=8.5cm]{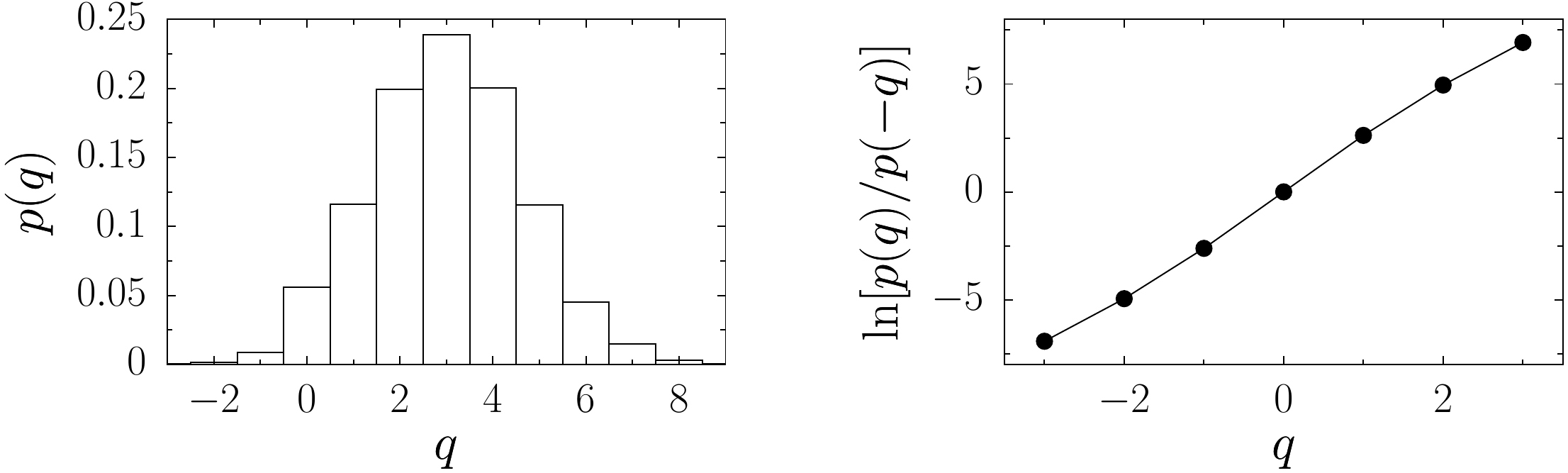}
\caption{Example of statistics obeying the fluctuation relation,
Eq. (\ref{eq:FT-general}).
Left panel: Probability distribution $p(q)$ of number $q$ of
electrons, transported through a nano-junction subject to an
electrical potential difference. Right panel: the linear behavior
of $\ln[p(q)/p(-q)]$ evidences that $p(q)$ obeys the fluctuation
relation, Eq. (\ref{eq:FT-general}).  In this example forward and backward protocols coincide yielding
$p_B=p_F\equiv p$, and consequently $b=0$ in Eq. (\ref{eq:FT-general}). Data courtesy of \textcite{Utsumi10PRB81}.}
\label{fig:histogram}
\end{figure}

This Colloquium focuses on fluctuation relations and in
particular on their quantum versions. These relations constitute a
research topic that recently has attracted a great deal of
attention. At the microscopic level, matter is in a permanent state
of agitation; consequently many physical quantities of interest
continuously undergo random fluctuations. The purpose of statistical
mechanics is the characterization of the statistical properties of
those fluctuating quantities from the known laws of classical and
quantum physics that govern the dynamics of the constituents of
matter. A paradigmatic example is the Maxwell distribution of
velocities in a rarefied gas at equilibrium, which follows from the
sole assumptions that the micro-dynamics are Hamiltonian, and that
the very many system constituents interact via negligible, short
range forces \cite{Khinchin49Book}. Besides the fluctuation of
velocity (or energy) at equilibrium, one might be interested in the
properties of other fluctuating quantities, e.g. heat and work,
characterizing non-equilibrium transformations. Imposed by the
reversibility of microscopic dynamical laws, the fluctuation
relations put severe restrictions on the form that the probability
density function (pdf) of the considered non-equilibrium fluctuating
quantities may assume. Fluctuation relations are typically expressed
in the form
\begin{equation}
 p_F(x)=p_B(-x)\exp[a(x-b)],
\label{eq:FT-general}
\end{equation}
where $p_F(x)$ is the probability density function (pdf) of the
fluctuating quantity $x$ during a nonequilibrium thermodynamic
transformation -- referred to for simplicity as the forward ($F$)
transformation --, and $p_B(x)$ is the pdf of $x$ during the
reversed (backward, $B$) transformation. The precise meaning of
these expressions will be amply clarified below. The real-valued
constants $a,b$, contain information about the \emph{equilibrium}
starting points of the $B$ and $F$ transformations. Figure
\ref{fig:histogram} depicts a probability distribution satisfying
the fluctuation relation, as measured in a recent experiment of
electron transport through a nano-junction \cite{Utsumi10PRB81}. We
shall analyze this experiment in detail in Sec. \ref{sec:exp}.

As often happens in science, the historical development of theories
is quite tortuous. Fluctuation relations are no exception in this
respect. Without any intention of providing a thorough and complete
historical account, we will mention below a few milestones that, in
our view, mark crucial steps in the historical development of
quantum fluctuation relations. The beginning of the story might be
traced back to the early years of the last century, with the work of
\textcite{Sutherland02PHILMAG3,Sutherland05PHILMAG9} and
\textcite{Einstein05AP17,Einstein06AP19a,Einstein06AP19b}
 first, and of \textcite{Johnson28PR32} and \textcite{Nyquist28PR32}
later, when it was found that the linear response of a system in
thermal equilibrium as it is driven out of equilibrium by an
external force, is determined by the fluctuation properties of the
system in the initial equilibrium state. Specifically,
\textcite{Sutherland02PHILMAG3,Sutherland05PHILMAG9} and
\textcite{Einstein05AP17,Einstein06AP19a,Einstein06AP19b} found a
relation between the mobility of a Brownian particle (encoding
information about its response to an externally applied force) and
its diffusion constant (encoding information about its equilibrium
fluctuations). \textcite{Johnson28PR32} and
\textcite{Nyquist28PR32}\footnote{Nyquist already discusses in his
Eq. (8) a precursor of the quantum fluctuation-dissipation theorem
as developed later by \textcite{Callen51PR83}. He only missed the
correct form by omitting in his result the zero-point energy
contribution, see also in \textcite{Hanggi05CHAOS2}.} discovered the
corresponding relation between the resistance of a circuit and the
spontaneous current fluctuations occurring in absence of an applied
electric potential.

The next prominent step was taken by
\textcite{Callen51PR83} who derived the previous results within a
general quantum mechanical setting. The starting point of their
analysis is a quantum mechanical system described by a Hamiltonian
$\mathcal{H}_0$. Initially this system stays in a thermal
equilibrium state at the inverse temperature $\beta
\equiv{(k_BT)^{-1}}$, wherein $k_B$ is the Boltzmann constant. This
state is described by a density matrix of canonical form; i.e., it
is given by a Gibbs state
\begin{equation}
 \varrho_0=e^{-\beta \mathcal{H}_0}/\mathcal{Z}_0\, ,
\label{eq:varrho_0}
\end{equation}
where $\mathcal{Z}_0=\Tr \, e^{-\beta \mathcal{H}_0}$
denotes the partition function of the unperturbed system
and $\Tr$ denotes trace over its Hilbert space.
At later times $t>0$ the system is perturbed by the action of an
external, in general time-dependent force $\lambda_t$ that
couples to an observable $\mathcal{Q}$ of the system. The
dynamics of the system then is governed by the modified,
time-dependent Hamiltonian
\begin{equation}
   \mathcal{H}(\lambda_t) =\mathcal H_0 - \lambda_t \mathcal Q.
\label{eq:H-quantum}
\end{equation}
The approach of \textcite{Callen51PR83} was further systematized by
\textcite{Green52JCP20, Green54JCP22} and in particular by
\textcite{Kubo57aJPSJ12} who proved that the linear response is
determined by a response function
$\phi_{\mathcal{B}\mathcal{Q}}(t)$, which gives the deviation
$\langle \Delta B(t) \rangle$ of the expectation value of an
observable $\mathcal B$ to its unperturbed value as
\begin{equation}
 \langle \Delta B(t) \rangle=\int_{-\infty}^{t}
\phi_{\mathcal{B}\mathcal{Q}}(t-s)\lambda_s\mathrm{d}s .
\label{eq:Green-Kubo}
\end{equation}
\textcite{Kubo57aJPSJ12} showed that the response function can be
expressed in terms of the commutator of the observables $\mathcal Q$
and  $\mathcal{B}^H(t)$, as
$\phi_{\mathcal{B}\mathcal{Q}}(s)=\langle
[\mathcal{Q},\mathcal{B}^H(s)]\rangle/i\hbar$ (the superscript $H$
denoting the Heisenberg picture with respect to the unperturbed
dynamics.)  Moreover, Kubo  derived the general relation
\begin{equation}
\langle \mathcal{Q} \mathcal{B}^H(t) \rangle = \langle
\mathcal{B}^H(t-i\hbar \beta) \mathcal{Q} \rangle
\label{eq:KMS}
\end{equation}
between differently ordered thermal correlation functions and
deduced from it the celebrated quantum fluctuation-dissipation
theorem \cite{Callen51PR83}, reading:
\begin{equation}
\hat{\Psi}_{\mathcal{B}\mathcal{Q}}(\omega)= (\hbar/2i) \,
\coth(\beta \hbar \omega/2)\,
\hat{\phi}_{\mathcal{B}\mathcal{Q}}(\omega) ,
\label{eq:FDT}
\end{equation}
where
$\hat{\Psi}_{\mathcal{B}\mathcal{Q}}(\omega)=\int_{-\infty}^{\infty}
e^{i\omega s}\Psi_{\mathcal{B}\mathcal{Q}}(s)\mathrm{d}s$, denotes
the Fourier transform of the symmetrized, stationary equilibrium
correlation function $\Psi_{\mathcal{B}\mathcal{Q}}(s) = \langle
\mathcal{Q} \mathcal{B}^H(s) + \mathcal{B}^H(s)\mathcal{Q}
\rangle/2$, and
$\hat{\phi}_{\mathcal{B}\mathcal{Q}}(\omega)=\int_{-\infty}^{\infty}
e^{i\omega s}\phi_{\mathcal{B}\mathcal{Q}}(s)\mathrm{d}s$ the
Fourier transform of the response function
$\phi_{\mathcal{B}\mathcal{Q}}(s)$. Note that the
fluctuation-dissipation theorem is valid also for many-particle
systems independent of the respective particle statistics. Besides
offering a unified and rigorous picture of the
fluctuation-dissipation theorem, the theory of Kubo also included
other important advancements in the field of non-equilibrium
thermodynamics. Specifically, we note the celebrated Onsager-Casimir
reciprocity relations (\citealt{Onsager31PR37,Onsager31PR38};
\citealt{Casimir45RMP17}). These relations state that, as a
consequence of \emph{microreversibility}, the matrix of transport
coefficients that connects applied forces to so-called fluxes in a
system close to equilibrium  consists of a symmetric and an
anti-symmetric block. The symmetric block couples forces and fluxes
that have same parity under time-reversal and the antisymmetric
block couples forces and fluxes that have different parity.

Most importantly, the analysis of \textcite{Kubo57aJPSJ12} opened
the possibility for a systematic advancement of response theory,
allowing in particular to investigate the existence of higher order
fluctuation-dissipation relations, beyond linear regime. This task
was soon undertaken by \textcite{Bernard59RMP31}, who pointed out a
hierarchy of irreversible thermodynamic relationships. These higher
order fluctuation dissipation relations were investigated in detail
by Stratonovich for Markovian system, and later by
\textcite{Efremov69ZETF55} for non-Markovian systems, see \cite[Ch.
I]{Stratonovich92Book} and references therein.

Even for arbitrary systems far from equilibrium the linear
response to an applied force can likewise be related to
tailored two-point correlation functions of
corresponding stationary nonequilibrium fluctuations of the
underlying {\it unperturbed}, stationary nonequilibrium system
\cite{Hanggi78HPA51, Hanggi82PREP88}. These authors coined the expression ``fluctuation theorems'' for these relations.
As in the near thermal equilibrium case, also in this case higher
order nonlinear response can be linked to corresponding higher order
correlation functions of those nonequilibrium fluctuations
\cite{Hanggi78HPA51,Prost09PRL103}.

At the same time, in the late seventies of the last century
\textcite{Bochkov77SPJETP45} provided a single compact
\emph{classical} expression that contains fluctuation relations of
all orders for systems that are at thermal equilibrium when unperturbed. This expression, see Eq.
(\ref{eq:BK-gen-functional-identity}) below, can be seen as a fully
nonlinear, exact and universal fluctuation relation.
This \citeauthor{Bochkov77SPJETP45} formula, Eq.
(\ref{eq:BK-gen-functional-identity}) below, soon turned out useful in
addressing the problem of connecting the deterministic and the
stochastic descriptions of \emph{nonlinear} dissipative systems
\cite{Bochkov78RQE21,Hanggi82PRA25}.

As it often happens in physics,
the most elegant, compact and universal relations, are consequences
of general physical symmetries. In the case of
\textcite{Bochkov77SPJETP45} the fluctuation relation follows from
the time reversal invariance of the equations of microscopic motion,
combined with the assumption that the system initially resides in
thermal equilibrium described by the classical analogue of the Gibbs
state, Eq. (\ref{eq:varrho_0}).
\textcite{Bochkov77SPJETP45,Bochkov79SPJETP49,Bochkov81aPHYSA106,
Bochkov81bPHYSA106} proved Eq. (\ref{eq:BK-gen-functional-identity})
below for classical systems. Their derivation will be reviewed in
the next section. The quantum version, see Eq.
(\ref{eq:q-J-gen-functional-identity}), was not reported until very
recently \cite{Andrieux08PRL100}. In Sec.
\ref{subsec:work-not-observable} we shall discuss the fundamental
obstacles that prevented
\textcite{Bochkov77SPJETP45,Bochkov79SPJETP49,Bochkov81aPHYSA106,
Bochkov81bPHYSA106} and \textcite{Stratonovich94Book} who also
studied this very quantum problem.

A new wave of activity in fluctuation relations was initiated by the
works of \textcite{Evans93PRL71} and \textcite{Gallavotti95PRL74}
 on the statistics of the entropy
produced in non-equilibrium steady states, and of
\textcite{Jarzynski97PRL78} on the statistics of work performed by a
transient, time-dependent perturbation. Since then, the field has
generated grand interest and flourished considerably. The existing
reviews on this topic mostly cover classical fluctuation relations
\cite{Rondoni07NL20, Jarzynski08EPJB64, Marconi08PREP111, Jarzynski11ARCMP2, Seifert08EPJB64}, while the
comprehensive review by \textcite{Esposito09RMP81} provides a solid,
though in parts technical account of the state of the art
of quantum fluctuation theorems. With this work we want to
present a widely accessible introduction to quantum fluctuation
relations,  covering as well the most recent decisive advancements.
Particularly, our  emphasis will be on (i) their connection to the
linear and non-linear response theory, Sec. \ref{sec:nonlinear},
(ii) the clarification of fundamental issues that relate to the
notion of  ``work'', Sec. \ref{sec:fund}, (iii) the derivation of
quantum fluctuation relations for both, closed and open quantum
systems, Sec. \ref{sec:QFT} and \ref{sec:XFT}, and also (iv) their
impact  for experimental applications and validation, Sec.
\ref{sec:exp}.

\section{Nonlinear response theory and classical fluctuation relations}
\label{sec:nonlinear}

\subsection{Microreversibility of non autonomous classical
systems}
Two ingredients are at the heart of fluctuation relations. The first
one concerns the initial condition of the system under study. This
is supposed to be in thermal equilibrium described by a canonical
distribution of the form of Eq. (\ref{eq:varrho_0}). It hence is of
{\it statistical} nature. Its use and properties are discussed in
many textbooks on statistical mechanics. The other ingredient,
concerning the \emph{dynamics} of the system is the principle of
microreversibility. This point needs some clarification since
microreversibility is customarily understood as a property of
autonomous (i.e., non-driven) systems described by a
time-independent Hamiltonian \cite[Vol. 2, Ch. XV]{Messiah62Book}.
On the contrary, here we are concerned with non-autonomous systems,
governed by explicitly time-dependent Hamiltonians. In the following
we will analyze  this principle for classical systems in a way that
at first glance may appear rather formal but will prove
indispensable later on. The analogous discussion of the quantum case
will be given next in Sec. \ref{subsec:q-microrev}.

We deal here
with a classical system characterized by a Hamiltonian that consists
of an unperturbed part $H_0(\mathbf{z})$ and a perturbation
$-\lambda_t Q(\mathbf{z})$ due to an external force $\lambda_t$ that
couples to the conjugate coordinate $Q(\mathbf{z})$. Then the total
system Hamiltonian becomes\footnote{The generalization to the case
of several forces coupling via different conjugate coordinates is
straightforward.}
\begin{equation}
 H(\mathbf{z},\lambda_t)=H_0(\mathbf{z})-\lambda_t
Q(\mathbf{z}),
\label{eq:H}
\end{equation}
where $\mathbf z=(\mathbf q,\mathbf p)$ denotes a point in the phase
space of the considered system. In the following we assume that the
force acts within a temporal interval set by a starting
time $0$ and a final time $\tau$. The instantaneous force values $\lambda_t$ are
specified by a function $\lambda$, which we will refer to as the
\emph{force protocol}. In the sequel, it will turn out necessary to
clearly distinguish between the function $\lambda$ and the value
$\lambda_t$ that it takes at a particular instant of time $t$.

\begin{figure}[t]
\includegraphics[width=8.5cm]{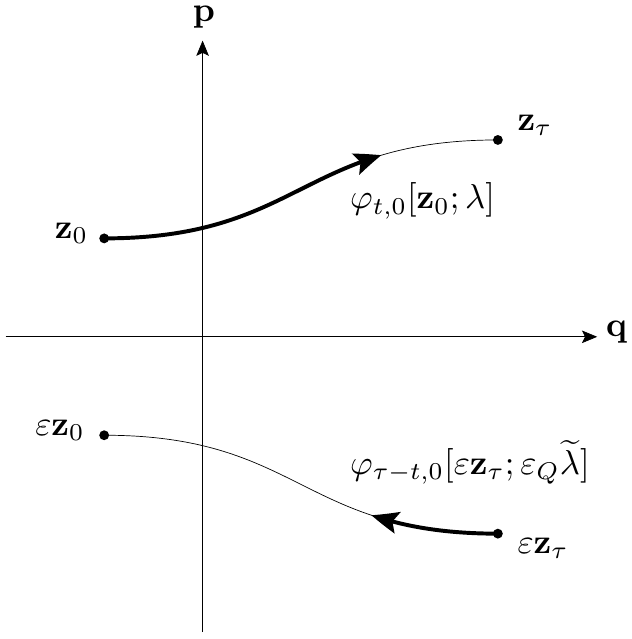}
\caption{Microreversibility for non-autonomous classical
(Hamiltonian) systems. The initial condition $\mathbf z_0$ evolves,
under the protocol $\lambda$, from  time $t=0$ until time $t$ to
$\mathbf{z}_t=\varphi_{t,0}[\mathbf{z}_0;\lambda]$ and until time
$t=\tau$ to $\mathbf{z}_\tau$. The time-reversed final condition
$\varepsilon \mathbf{z}_\tau$ evolves, under the protocol
$\varepsilon_Q \widetilde \lambda$ from  time $t=0$  until $\tau-t$
to $\varphi_{\tau-t,0}[\varepsilon
\mathbf{z}_\tau;\varepsilon_Q\widetilde \lambda]=\varepsilon
\varphi_{t,0}[\mathbf{z}_0;\lambda]$, Eq. (\ref{eq:microreversibility}),
and until time $t=\tau$ to the time-reversed initial condition
$\varepsilon \mathbf z_0$.} \label{fig:microreversibility}
\end{figure}

For these systems the principle of microreversibility holds in the
following sense. The solution of Hamilton's equations of motion
assigns to each initial point in phase space
$\mathbf{z}_0=(\mathbf{q}_{0},\mathbf{p}_{0})$ a point
$\mathbf{z}_t$ at the later time $t\in[0,\tau]$, which is specified
by the values of the force in the order of their appearance within
the considered time span. Hence, the position
\begin{equation}
\mathbf{z}_{t}=\varphi_{t,0}[\mathbf{z}_0;\lambda]
\label{eq:varphi}
\end{equation}
at time $t$ is determined by the flow $\varphi_{t,0}
[\mathbf{z}_0;\lambda]$ which is a function of the initial point
$\mathbf{z}_0$ and a functional of the force protocol
$\lambda$.\footnote{Due to causality
$\varphi_{t,0}[\mathbf{z}_0,\lambda]$ may of course only depend on
the part of the protocol including times from $t=0$ up to time $t$.}
In a computer simulation one can invert the direction of time and
let run the trajectory backwards without problem. Although, as
experience evidences, it is impossible to actively revert the
direction of time in any experiment, there is yet a way to run a
time reversed trajectory in real time. For the sake of simplicity we
assume that the Hamiltonian $H_0$ is time reversal invariant, i.e.
that it remains unchanged if the signs of momenta are reverted.
Moreover we restrict ourselves to conjugate coordinates
$Q(\mathbf{z})$ that transform under time reversal with a definite
parity $\varepsilon_Q=\pm 1$.
\textcite[Sec.1.2.3]{Stratonovich94Book} showed that then the flow
under the backward protocol $\tilde \lambda$, with
\begin{equation}
\widetilde \lambda_t= \lambda_{\tau-t},
\label{eq:lambda-tilde}
\end{equation}
is related to the flow under the forward protocol $\lambda$, via
the relation
\begin{equation}
\varphi_{t,0}[\mathbf{z}_0;\lambda]=\varepsilon\varphi_{\tau-t,0}
[\varepsilon \mathbf{z}_\tau;\varepsilon_Q\widetilde \lambda],
\label{eq:microreversibility}
\end{equation}
where $\varepsilon$ maps any phase space point $\mathbf{z}$ on its
time reversed image $\varepsilon \mathbf{z}=\varepsilon
(\mathbf{q},\mathbf{p})=(\mathbf{q},-\mathbf{p})$. Equation
(\ref{eq:microreversibility}) expresses the principle of
microreversibility in driven systems. Its meaning is illustrated in
Fig. \ref{fig:microreversibility}. Particularly, it states that in
order to trace back a trajectory, one has to reverse the sign of the
velocity, as well as the temporal succession of the force values
$\lambda$, and the sign of force, $\widetilde \lambda$, if $\varepsilon_Q=-1$.

\subsection{\label{subsec:BK}Bochkov-Kuzovlev approach}
We consider a phase-space function $B(\mathbf{z})$ which has a
definite parity under time reversal  $\varepsilon_B=\pm 1$, i.e.
$B(\varepsilon\mathbf{z})=\varepsilon_BB(\mathbf{z})$. Let
$B_t=B(\varphi_{t,0}[\mathbf{z}_0;\lambda])
$
denote its temporal evolution. Depending on the initial condition
$\mathbf z_0$ different trajectories $B_t$ are realized. Under
the above stated assumption that at time $t=0$ the system is
prepared in a Gibbs equilibrium, the  initial conditions  are
randomly sampled from the distribution
\begin{equation}
 \rho_0(\mathbf{z}_0)={e^{-\beta H_0(\mathbf{z_0})}}/{Z_0},
\label{eq:rho_0}
\end{equation}
with $Z_0=\int \mathrm{d}\mathbf z_0 e^{-\beta
H_0(\mathbf{z_0})}$. Consequently the trajectory $B_t$ becomes a
random quantity. Next we introduce the quantity:
\begin{equation}
W_0[\mathbf{z}_0;\lambda] = \int_{0}^{\tau}\mathrm{d}t \lambda_t
\dot Q_t,
\label{eq:ex-work-integral}
\end{equation}
where $\dot Q_t$ is the time derivative of
$Q_t=Q(\varphi_{t,0}[\mathbf{z}_0;\lambda])$. From Hamilton's
equations it follows that \cite{Jarzynski07CRPHYS8}:
\begin{equation}
W_0[\mathbf{z}_0;\lambda]=H_0(\varphi_{\tau,0}[\mathbf{z}
_0;\lambda])-H_0(\mathbf{z}_0).
\label{eq:ex-work}
\end{equation}
Therefore, we interpret $W_0$ as the work injected in the system
described by $H_0$ during the action of the force
protocol.\footnote{ Following \cite{Jarzynski07CRPHYS8} we refer to
$W_0$  as the \emph{exclusive} work, to distinguish from the
\emph{inclusive} work $W=H(\mathbf{z}_\tau,\lambda_\tau)-H(\mathbf{z}_0,\lambda_0)$,
Eq. (\ref{eq:in-work}), which
accounts also for the coupling between the external source and the
system. We will come back later to these two definitions of work in
Sec. \ref{subsec:Inc-Exc}.} The central finding of
\textcite{Bochkov77SPJETP45} is a formal relation between the
generating functional for multi-time correlation functions of the
phase space functions $B_t$ and $Q_t$ and the generating functional
for the time-reversed multi-time auto-correlation functions of
$B_t$, reading
\begin{align}
\langle e^{\int_{0}^{\tau}\mathrm{d}t u_t B_t} e^{-\beta
W_0}\rangle_{\lambda}=&\langle e^{\int_{0}^{\tau}\mathrm{d}t
\widetilde u_t\varepsilon_B B_t}\rangle_{\varepsilon_Q\widetilde
\lambda},
\label{eq:BK-gen-functional-identity}
\end{align}
where $u_\tau$ is an arbitrary test-function, $\widetilde u_t=
u_{\tau-t}$ is its temporal reverse, and the average denoted by
$\langle \cdot \rangle$ is taken with respect to the Gibbs
distribution $\rho_0$ of Eq. (\ref{eq:rho_0}). On the left hand side,
the time evolutions of $B_t$ and $Q_t$ are governed by the full
Hamiltonian (\ref{eq:H}) in presence of the forward protocol as
indicated by the subscript $\lambda$  while on the right hand side
the dynamics is determined by the time-reversed protocol indicated
by the subscript $\varepsilon_Q \widetilde \lambda$. The derivation
of Eq. (\ref{eq:BK-gen-functional-identity}), which is based on the
microreversibility principle, Eq. (\ref{eq:microreversibility}), is
given in Appendix \ref{app:BK}. The importance of Eq.
(\ref{eq:BK-gen-functional-identity}) lies in the fact that it
contains  the Onsager reciprocity relations and fluctuation
relations of all orders within a single compact formula
\cite{Bochkov77SPJETP45}. These relations may be obtained by means
of functional derivatives of both sides of the Eq.
(\ref{eq:BK-gen-functional-identity}), of various orders, with
respect to the force-field $\lambda$ and the test-field $u$ at
vanishing fields $\lambda=u=0$. The classical limit of the
\textcite{Callen51PR83} fluctuation-dissipation theorem, Eq. (\ref{eq:FDT}), for
instance, is obtained by differentiation with respect to $u$,
followed by a differentiation with respect to $\lambda$
\cite{Bochkov77SPJETP45}, both at vanishing fields $u$ and
$\lambda$.

Another remarkable identity is achieved from Eq.
(\ref{eq:BK-gen-functional-identity}) by putting $u=0$, but leaving
the force $\lambda$ finite. This yields the Bochkov-Kuzovlev
equality, reading
\begin{equation}
 \langle e^{-\beta W_0} \rangle_\lambda =1 .
\label{eq:BK-identity}
\end{equation}
In other words, for any system that initially stays in thermal
equilibrium at a temperature $T=1/(k_B\beta)$ the work, Eq.
(\ref{eq:ex-work-integral}), done on the system by an external
force
is a random quantity with an ``exponential expectation value''
$\langle e^{-\beta W_0} \rangle_\lambda$ that is independent of
any detail of the system and the force acting on it. This of
course does not hold for the individual moments of work.
Since the exponential function is concave, a direct consequence
of Eq. (\ref{eq:BK-identity}) is
\begin{equation}
 \langle W_0 \rangle_\lambda \geq 0 .
\label{eq:W0>0}
\end{equation}
That is, \emph{on average}, a driven Hamiltonian system may only
absorb energy if it is perturbed out of thermal equilibrium. This
does not exclude the existence of
energy releasing events which, in fact, must happen with certainty in
order that Eq. (\ref{eq:BK-identity}) holds if the average work is
larger than zero. Equation (\ref{eq:W0>0}) may be regarded as a
microscopic manifestation of the Second Law of thermodynamics. For
this reason \textcite[Sec. 1.2.4]{Stratonovich94Book} referred to it
as the H-theorem. We recapitulate that only two ingredients --
initial Gibbsian equilibrium and microreversibility of the dynamics
-- have led to Eq. (\ref{eq:BK-gen-functional-identity}). In conclusion,
this relation  not only contains linear and nonlinear response theory,
but also the second law of thermodynamics.

The complete information about the statistics is contained in the
 work probability density function (pdf) $p_0[W_0;\lambda]$.
The only random element entering the work, Eq. (\ref{eq:ex-work}), is
the initial phase point $\mathbf{z}_0$ which is distributed
according to Eq. (\ref{eq:rho_0}). Therefore  $p_0[W_0;\lambda]$
may formally be expressed as
\begin{equation}
p_0[W_0;\lambda]=\int \mathrm{d}\mathbf{z}_0
\rho_0(\mathbf{z}_0)\delta[W_0-H_0(\mathbf{z}_\tau)+H_0(\mathbf{z
}_0)] ,
\end{equation}
where $\delta$ denotes Dirac's delta function. The functional
dependence of $p_0[W_0;\lambda]$  on $\lambda$ is contained in
the term
$\mathbf{z}_\tau=\varphi_{\tau,0}[\mathbf{z}_0;\lambda]$.
Using the microreversibility principle, Eq.
(\ref{eq:microreversibility}), one obtains  the following
fluctuation relation
\begin{equation}
 \frac{p_0[W_0;\lambda]}{p_0[-W_0;\varepsilon_Q \widetilde
\lambda]}=e^{\beta W_0} ,
\label{eq:BK-w-fluc-theo}
\end{equation}
in a way analogous to the derivation of Eq.
(\ref{eq:BK-gen-functional-identity}).
We shall refer to this relation as the Bochkov-Kuzovlev work
fluctuation relation, although it was not explicitly given by
Bochkov and Kuzovlev, but was only recently obtained  by
\textcite{Horowitz07JSM07}. This equation has a profound physical
meaning. Consider a positive work $W_0>0$. Then Eq.
(\ref{eq:BK-w-fluc-theo}) says that the probability that this
work is injected into the system is larger by the factor
$e^{\beta W_0}$ than the probability that the same work is
absorbed under the reversed forcing: energy consuming processes
are exponentially more probable than energy releasing processes.
Thus, Eq. (\ref{eq:BK-w-fluc-theo}) expresses the second law of
thermodynamics at a very detailed level which quantifies the
relative frequency of energy releasing processes. By multiplying
both sides of Eq. (\ref{eq:BK-w-fluc-theo}) by
$p_0[-W_0;\varepsilon_Q \widetilde \lambda]e^{-\beta W_0}$ and
integrating over $W_0$, one recovers the Bochkov-Kuzovlev
identity, Eq. (\ref{eq:BK-identity}).

\subsection{Jarzynski approach}
An alternative definition of work is based on the comparison of the
{\it total} Hamiltonians at the end and the beginning of a force
protocol, leading to the notion of ``inclusive work'' in contrast to
the ``exclusive work'' defined in Eq. (\ref{eq:ex-work}). The latter
equals the energy difference referring to the unperturbed
Hamiltonian $H_0$.
Accordingly, the inclusive work is the difference of the total
Hamiltonians at the final time $t=\tau$ and the initial time $t=0$:
\begin{equation}
 W[\mathbf{z}_0;\lambda]=H(\mathbf{z}_\tau,
\lambda_\tau)-H(\mathbf{z}_0, \lambda_0).
\label{eq:in-work}
\end{equation}
In terms of the force $\lambda_t$ and the conjugate coordinate
$Q_t$, the inclusive work is expressed as\footnote{For a further
discussion of inclusive and exclusive work we refer to Sect.
\ref{subsec:Inc-Exc}.}:
\begin{align}
 W[\mathbf{z}_0;\lambda]&=\int_0^\tau dt \dot \lambda_t
\frac{\partial H(\mathbf{z}_t,\lambda_t)}{\partial \lambda_t}
\label{eq:in-work-integral}\\
&=- \int_{0}^{\tau} \mathrm{d}t \dot \lambda_t Q_t \nonumber\\
&= W_0 [\mathbf{z}_0;\lambda]-\lambda_tQ_t|_{0}^{\tau}.\nonumber
\end{align}
For the sake of simplicity we confine ourselves to the case of an
even conjugate coordinate $Q$. In the corresponding way, as
described in Appendix \ref{app:BK}, we obtain the following relation
between generating functionals of forward and backward processes in
analogy to Eq. (\ref{eq:BK-gen-functional-identity}), reading
\begin{align}
\langle e^{\int_{0}^{\tau}\mathrm{d}t u_t B_t} e^{-\beta
W}\rangle_{\lambda}=\frac{Z(\lambda_\tau)}{Z(\lambda_0)}&\langle
e^{\int_{0}^{t}\mathrm{d}t \widetilde u_t\varepsilon_B
B_t}\rangle_{\widetilde
\lambda}.
\label{eq:J-gen-functional-identity}
\end{align}
While on the left hand side the time evolution is controlled by
the forward protocol $\lambda$ and the average is performed with
respect to the initial
thermal distribution $\rho^\beta(\mathbf{z},\lambda_0)$, on the
right hand side the time evolution is governed by the reversed
protocol $\widetilde{\lambda}$ and averaged over the reference
equilibrium state
$\rho^\beta(\mathbf{z},\widetilde{\lambda}_0)=\rho^\beta(\mathbf{
z},\lambda_\tau)$. Here
\begin{equation}
\rho^\beta(\mathbf{z},\lambda_t)={e^{-\beta H(\mathbf{z},
\lambda_t)}}/{Z(\lambda_t)}
\label{eq:rho-t_f}
\end{equation}
formally describes thermal equilibrium of a system with the
Hamiltonian $H(\mathbf{z}, \lambda_t)$ at the inverse temperature
$\beta$. The partition function $Z(\lambda_t)$ is defined
accordingly as $ Z(\lambda_t) = \int \mathrm{d}\mathbf{z} \,
e^{-\beta H(\mathbf{z},\lambda_t)}$. Note that in general the
reference state $\rho^\beta (\mathbf{z},\lambda_t)$ is different
from the actual phase space distribution reached under the action of
the protocol $\lambda$ at time $t$, i.e.,
$\rho(\mathbf{z},t)=\rho^\beta(\varphi_{t,0}^{-1}[\mathbf{z}
;\lambda],\lambda_0)$, where
$\varphi_{t,0}^{-1}[\mathbf{z};\lambda]$ denotes the point in phase
space that evolves to $\mathbf{z}$ in the time $0$ to $t$ under the
action of $\lambda$.

Setting $u \equiv 0$ we obtain
\begin{equation}
 \langle e^{-\beta W} \rangle_\lambda = e^{-\beta \Delta F} ,
\label{eq:J-identity}
\end{equation}
where
\begin{equation}
\Delta F= F(\lambda_\tau)-F(\lambda_0)= -\beta \ln \frac{Z(\lambda_\tau)}{Z(\lambda_0)} .
\label{eq:F}
\end{equation}
is the free energy difference between the reference state
$\rho^\beta(\mathbf z,\lambda_t)$ and the initial equilibrium
state $\rho^\beta(\mathbf z \lambda_0)$. As a consequence of Eq.
(\ref{eq:J-identity}) we have
\begin{equation}
 \langle W \rangle_\lambda \geq \Delta F ,
\label{eq:W>deltaF}
\end{equation}
which is yet another expression of the second law of thermodynamics.
Equation (\ref{eq:J-identity}) was first put forward by
\textcite{Jarzynski97PRL78}, and is commonly referred to in the
literature as the ``Jarzynski equality''.

In close analogy to the Bochkov-Kuzovlev approach the pdf of the
inclusive work can be formally expressed as
 \begin{align}
p[W;\lambda]=\int \mathrm{d}\mathbf{z}_0 \rho(\mathbf{z}_0,
\lambda_0)\delta[W-H(\mathbf{z}_\tau,\lambda_\tau)+H(\mathbf{z}_0
,\lambda_0)] .
\label{eq:Jpw}
\end{align}
Its Fourier transform defines the characteristic function of work:
 \begin{align}
G[u;\lambda]&=\int \mathrm{d}W e^{iuW}p[W;\lambda]\nonumber\\
&=\int \mathrm{d}\mathbf{z}_0
e^{iu[H(\mathbf{z}_\tau,\lambda_\tau)-H(\mathbf{z}_0
,\lambda_0)]}  e^{-\beta H(\mathbf{z}_0,\lambda_0)}/Z(\lambda_0) \nonumber\\
=\int \mathrm{d}\mathbf{z}_0 &
\exp\left[iu\int_0^\tau \mathrm{d}t \dot \lambda_t\frac{\partial H(\mathbf{z}_t,\lambda_t)}{\partial \lambda_t}\right]
\frac{e^{-\beta H(\mathbf{z}_0,\lambda_0)}}{Z(\lambda_0)}.
\label{eq:Jgu}
\end{align}

Using the microreversibility principle, Eq.
(\ref{eq:microreversibility}), we obtain in a way similar to
Eq. (\ref{eq:BK-w-fluc-theo}) the (inclusive) work fluctuation
relation:
\begin{equation}
 \frac{p[W;\lambda]}{p[-W; \widetilde \lambda]}=e^{\beta
(W-\Delta F)} ,
\label{eq:J-w-fluc-theo}
\end{equation}
where the probability $p[-W; \widetilde \lambda]$ refers to the
backward process which for the inclusive work has to be determined
with reference to the initial thermal state $\rho^\beta(\mathbf{z},\lambda_\tau)$. First
put forward by \textcite{Crooks99PRE60},  Eq.
(\ref{eq:J-w-fluc-theo}) is commonly referred to in literature as
the ``Crooks fluctuation theorem''. The Jarzynski equality, Eq.
(\ref{eq:J-identity}), is obtained by multiplying both sides of Eq.
(\ref{eq:J-w-fluc-theo}) by $p[-W; \widetilde \lambda]e^{-\beta W}$
and integrating over $W$. Equations
(\ref{eq:J-gen-functional-identity}, \ref{eq:J-identity},
\ref{eq:J-w-fluc-theo}) continue to hold also when $Q$ is odd
under time reversal, with
the provision that $\widetilde \lambda$ is replaced by $-\widetilde
\lambda$.

We here point out the salient fact that, within the inclusive
approach, a connection is established between the {\it
nonequilibrium} work $W$ and the  difference of  free energies
$\Delta F$, of the corresponding {\it equilibrium states}
$\rho^\beta(\mathbf{z}, \lambda_\tau)$ and
$\rho^\beta(\mathbf{z},\lambda_0)$. Most remarkably, Eq.
(\ref{eq:W>deltaF}) says that the average (inclusive) work is
always larger than or equal to the free energy difference, no matter
the form of the protocol $\lambda$;
even more surprising is the content of Eq. (\ref{eq:J-identity})
saying that
 the equilibrium free
energy difference may be inferred by measurements of nonequilibrium
work in many realizations of the forcing experiment
\cite{Jarzynski97PRL78}. This is similar in spirit to the
fluctuation-dissipation theorem, also connecting an equilibrium
property (the fluctuations), to a non-equilibrium one (the linear
response), with the major difference that Eq. (\ref{eq:J-identity})
is an exact result, whereas the fluctuation-dissipation theorem
holds only to first order in the perturbation. Note that as a
consequence of Eq. (\ref{eq:J-w-fluc-theo}) the forward and backward
pdf's of exclusive work take on the same value at $W=\Delta F$. This
property has been used in experiments \cite{Douarche05EPL70,
Collin05NAT437, Liphardt02SCIENCE296} in order to determine free
energy differences from nonequilibrium measurements of work.
Equations (\ref{eq:J-identity}, \ref{eq:J-w-fluc-theo}) have further
been employed to develop efficient numerical methods for the
estimation of free energies \cite{Jarzynski02PRE65, Hahn09PRE79,
Hahn10PRE81, Vaikuntanathan08PRL100, Minh08PRL100}.

Both the Crooks fluctuation theorem, Eq.
(\ref{eq:J-w-fluc-theo}), and the Jarzynski equality, Eq.
(\ref{eq:J-identity}), continue to hold for any time dependent
Hamiltonian $H(\mathbf z,\lambda_t)$ without restriction to
Hamiltonians of the form in Eq. (\ref{eq:H}). Indeed no
restriction of the form in Eq. (\ref{eq:H}) was imposed in the
seminal paper by \textcite{Jarzynski97PRL78}. In the original works
of \textcite{Jarzynski97PRL78} and \textcite{Crooks99PRE60}, Eqs.
(\ref{eq:J-identity}) and (\ref{eq:J-w-fluc-theo}) were obtained
directly, without passing through the more general formula in Eq.
(\ref{eq:J-gen-functional-identity}). Notably, neither these seminal
papers, nor the subsequent literature, refer to such
general functional identities as Eq.
(\ref{eq:J-gen-functional-identity}). We introduced them here to
emphasize the connection between the recent results, Eqs.
(\ref{eq:J-identity}) and (\ref{eq:J-w-fluc-theo}), with the older
results of \textcite{Bochkov77SPJETP45}, Eqs.
(\ref{eq:BK-identity}, \ref{eq:BK-w-fluc-theo}). The latter ones were
practically ignored, or sometimes misinterpreted as special instances
of the former ones for the case of cyclic protocols ($\Delta F = 0$), by those working in the
field of non-equilibrium work fluctuations. Only recently
\textcite{Jarzynski07CRPHYS8} pointed out the differences
and analogies between the inclusive and exclusive approaches.

\section{Fundamental issues}
\label{sec:fund}

\subsection{Inclusive, exclusive and dissipated work}
\label{subsec:Inc-Exc}
As we evidenced in the previous section, the studies of
\textcite{Bochkov77SPJETP45} and \textcite{Jarzynski97PRL78} are
based on   different definitions of work, Eqs. (\ref{eq:ex-work},
 \ref{eq:in-work}), reflecting two different viewpoints
\cite{Jarzynski07CRPHYS8}. From the ``exclusive'' viewpoint of
\textcite{Bochkov77SPJETP45} the change in the energy $H_0$ of the
unforced system is considered, thus the forcing term $(-\lambda_tQ)$
of the total Hamiltonian is not included in the computation of work.
From the ``inclusive'' point of view the definition of work, Eq.
(\ref{eq:in-work}), is based on the change of the total energy $H$
including the forcing term $(-\lambda_tQ)$. In experiments and
practical applications of fluctuation relations, special care must be
paid in properly identifying the measured work with either the
inclusive ($W$) or exclusive ($W_0$) work, bearing in mind that
$\lambda$ represents the prescribed parameter progression and $Q$ is the
measured conjugate coordinate.

The experiment of
\textcite{Douarche05EPL70} is very well suited to illustrate this
point. In that experiment a prescribed torque $M_t$ was applied to a
torsion pendulum whose angular displacement $\theta_t$ was
continuously monitored. The Hamiltonian of the system is
\begin{align}
 H(\mathbf{y},p_\theta,\theta,M_t)=&H_B(\mathbf{y})+H_{SB}
(\mathbf{y},p_\theta,\theta) \nonumber \\
&+\frac{p_\theta^2}{2I}+\frac{I\omega^2\theta^2}{2}- M_t\theta ,
\label{eq:H-torsion}
\end{align}
where $p_\theta$ is the canonical momentum conjugate to $\theta$,
 $H_B(\mathbf y)$ is the Hamiltonian of the thermal bath to which
the pendulum is coupled via the Hamiltonian $H_{SB}$, and $\mathbf
y$ is a point in the bath phase space. Using the definitions of
inclusive and exclusive work, Eqs. (\ref{eq:ex-work-integral},
\ref{eq:in-work-integral}), and noticing that $M$ plays the role of
$\lambda$ and $\theta$ that of $Q$, we find in this case $W=-\int
\theta \dot M \mathrm{d}t$ and $W_0=\int M \dot \theta \mathrm{d}t$.

Note that the work $W=-\int \theta \dot M \mathrm{d}t$, obtained by
monitoring the pendulum  degree of freedom only, amounts to the
energy change of the total pendulum+bath system. This is true in
general \cite{Jarzynski04JSM04}. Writing the total Hamiltonian as
\begin{equation}
 H(\mathbf x,\mathbf y, \lambda_t)= H_S(\mathbf
x,\lambda_t)+H_{BS}(\mathbf x,\mathbf y)+H_B(\mathbf y) ,
\label{eq:HS+HB+HSB}
\end{equation}
with $H_S(\mathbf x,\lambda_t)$ being the Hamiltonian of the
system of interest, one obtains
\begin{align}
\int_{0}^{\tau} \mathrm{d}t\frac{\partial H_S}{\partial
\lambda_t}\dot{\lambda}_t
= \int_{0}^{\tau} \mathrm{d}t  \frac{\partial H}{\partial t}
=\int_{0}^{\tau} \mathrm{d}t \frac{\mathrm{d} H}{\mathrm{d}t}=W ,
\label{eq:Wm=W}
\end{align}
 because $H_{BS}$ and $H_B$ do not depend on time, and as a
consequence of Hamilton's equations of motion
$\mathrm{d}H/\mathrm{d}t=\partial H/\partial{t}$.

Introducing the notation
$
 W_{\text{diss}}=W-\Delta F
$, for the dissipated work, one deduces that the Jarzynski equality
can be re-expressed in a way that looks exactly as the
Bochkov-Kuzovlev identity, namely:
\begin{equation}
 \langle e^{-\beta W_{\text{diss}}}\rangle_\lambda =1 .
\end{equation}
This might let one believe that the dissipated work coincides with
$W_0$. This, however, would be incorrect. As discussed in
\cite{Jarzynski07CRPHYS8} and explicitly demonstrated by
\textcite{Campisi10arXiv} $W_0$ and $W_{\text{diss}}$ constitute
distinct stochastic quantities with different pdf's. The inclusive,
exclusive and dissipated work coincide only in the case of cyclic
forcing $\lambda_\tau=\lambda_0$ \cite{Campisi10arXiv}.

\subsection{The problem  of gauge freedom}
\label{subsec:gauge}
We point out that the inclusive work $W$, and free energy difference
$\Delta F$, as defined in Eqs. (\ref{eq:in-work}, \ref{eq:F}), are
-- to use the expression coined by \textcite{CohenTannoudji77Book}
-- not ``true physical quantities.'' That is to say they are not
invariant under gauge transformations that lead to a time-dependent
shift of the energy reference point. To elucidate this, consider a
mechanical system whose dynamics are governed by the Hamiltonian
$H(\mathbf z,\lambda_t)$. The new Hamiltonian
\begin{equation}
 H'(\mathbf{z},\lambda_t)=H(\mathbf{z},\lambda_t)+g(\lambda_t),
\label{eq:gauge}
\end{equation}
where $g(\lambda_t)$ is an arbitrary function of the time dependent
force, generates the {\it same}  equations of motion as $H$.
However, the work
$W'=H'(\mathbf{z}_\tau,\lambda_{\tau})-H'(\mathbf{z}_0,\lambda_{0
})$ that one obtains from this Hamiltonian differs from the one that
follows from $H$, Eq. (\ref{eq:in-work}): $
 W'=W+g(\lambda_\tau)-g(\lambda_0).
$
Likewise we have, for the free energy difference
$
 \Delta F'= \Delta F+g(\lambda_\tau)-g(\lambda_0) .
$ Evidently the Jarzynski equality, Eq. (\ref{eq:J-identity}),
\emph{is invariant under such gauge transformations}, because the
term $g(\lambda_\tau)-g(\lambda_0)$ appearing on both sides of
the identity in the primed gauge,  would cancel; explicitly this
reads:
\begin{equation}
\langle e^{-\beta W'}\rangle_\lambda=e^{-\beta \Delta F'} \iff \langle
e^{-\beta W}\rangle_\lambda=e^{-\beta \Delta F} .
\end{equation}
Thus, there is no fundamental problem associated with the gauge
freedom.

However one must be aware that, in each particular
experiment, the very way by which the work is measured implies a
specific gauge. Consider for example the torsion pendulum
experiment of \cite{Douarche05EPL70}.
The inclusive work was computed as: $W=-\int \theta \dot M
\mathrm{d}t$. The condition that this measured work is related to
the Hamiltonian of Eq. (\ref{eq:H-torsion}) via the relation
$W=H(\mathbf{z}_\tau,\lambda_{\tau})-H(\mathbf{z}_0,\lambda_{0})$
, Eq. (\ref{eq:in-work}), is equivalent to
$ -\int_0^\tau \theta \dot M \mathrm{d}t=\int_0^\tau (\partial
H/\partial M) \dot M \mathrm{d}t$,
see Eq. (\ref{eq:Wm=W}). If this is required for all $\tau$ then the
stricter condition $\partial H/\partial M=-\theta$ is implied,
restricting the remaining gauge freedom to the choice of
a constant function $g$.  This residual freedom however is not
important as it does neither affect work nor free energy. We now
consider a different experimental setup where the support to
which the pendulum is attached is rotated in a prescribed way
according to a protocol $\alpha_t$, specifying the angular
position of the support with respect to the lab frame. The
dynamics of the pendulum are now described by the Hamiltonian
\begin{align}
 H=H_B+H_{SB}
+\frac{p_\theta^2}{2I}+\frac{I\omega^2\theta^2}{2}-
I\omega^2\alpha_t\theta+g(\alpha_t).
\label{eq:H-torsion-2}
\end{align}
If the work $W=\int N\dot \alpha \mathrm{d}t$ done by the elastic
torque $N=I\omega^2(\alpha -\theta)$ on the support is recorded then
the requirement $\partial H/\partial \alpha=N$ singles out the gauge
$g(\alpha_t)=I\omega^2 \alpha_t^2/2+const$, leaving only the freedom
to chose the unimportant constant. Note that when
$M_t=I\omega^2\alpha_t$, the pendulum obeys exactly the same
equations of motion in the two examples above, Eqs.
(\ref{eq:H-torsion}, \ref{eq:H-torsion-2}). The gauge is irrelevant
for the law of motion but is essential for the energy-Hamiltonian
connection.\footnote{See also \textcite{Kobe81AJP49}, in the context
of non-relativistic electrodynamics.}

The issue of gauge freedom was first pointed out by
\textcite{Vilar08PRL100}, who  questioned whether a connection
between work and Hamiltonian may actually exist. Since then this
topic had been highly debated,\footnote{See
\textcite{Peliti08PRL101,
Horowitz08PRL101,Vilar08aPRL101,Vilar08bPRL101,Vilar08PRL100,Peliti08JSM08,Chen08aJCP129,Chen08bJCP129,Chen09JCP130,Adib09JCP130,Crooks09JCP130,Zimanyi09JCP130}.}
but neither the gauge invariance of fluctuation relations nor the
fact that different experimental setups imply different gauges were
clearly recognized before.

\subsection{Work is not a quantum observable}
\label{subsec:work-not-observable}
Thus far we have reviewed the general approach to work fluctuation
relations for classical systems. The question then naturally arises
of how to treat the quantum case. Obviously, the Hamilton function
$H(\mathbf z,\lambda_t)$ is to be replaced by the Hamilton operator
$\mathcal H (\lambda_t)$, Eq. (\ref{eq:H-quantum}). The probability
density $\rho(\mathbf z,\lambda_t)$ is then replaced by the density
matrix $\varrho(\lambda_t)$, reading
\begin{equation}
 \varrho(\lambda_t)=e^{-\beta \mathcal
H(\lambda_t)}/\mathcal{Z}(\lambda_t),
\label{eq:varrho_t}
\end{equation}
where $ \mathcal Z(\lambda_t)= \Tr e^{-\beta \mathcal H(\lambda_t)}$
is the partition function and $\Tr$ denotes the trace over the
system Hilbert space. The free energy is obtained from the partition
function in the same way as for classical systems, i.e., $
F(\lambda_t)=-\beta^{-1}\ln \mathcal Z(\lambda_t). $ Less obvious is
the definition of work in quantum mechanics. Originally,
\textcite{Bochkov77SPJETP45} defined the exclusive quantum work, in
analogy with the classical expression, Eqs.
(\ref{eq:ex-work-integral}, \ref{eq:ex-work}), as the operator $
  \mathcal W_0= \int_{0}^{\tau}\mathrm{d}t \lambda_t \dot{\mathcal
 Q}_t^H=  \mathcal H_{0\,\tau}^H - \mathcal H_0\, ,
$
where the superscript $H$ denotes the Heisenberg picture:
\begin{equation}
 \mathcal B^H_t=U_{t,0}^{\dagger}[\lambda]\, \mathcal B\,
U_{t,0}[\lambda].
\label{eq:Heisenberg}
\end{equation}
Here $\mathcal B$ is an operator in the Schr\"{o}dinger picture
and $U_{t,0}[\lambda]$ is the unitary time evolution operator
governed by the Schr\"{o}dinger equation
\begin{equation}
 i\hbar \frac{\partial U_{t,0}[\lambda]}{\partial t}=\mathcal
H(\lambda_t)U_{t,0}[\lambda], \quad U_{0,0}[\lambda]=\mathbb 1,
\label{eq:Schroedinger}
\end{equation}
with $\mathbb 1$ denoting the identity operator. We use the
notation $U_{t,0}[\lambda]$ to emphasize that, like the classical
evolution $\varphi_{t,0}[\mathbf{z};\lambda]$ of Eq.
(\ref{eq:varphi}), the quantum evolution operator is a functional
of the protocol $\lambda$.\footnote{Due to causality
$U_{t,0}[\lambda]$ may of course only depend on the part of the
protocol including times from $0$ up to $t$.} The time derivative $\dot{\mathcal
Q}_t^H$ is determined by the Heisenberg equation. In case of a
time-independent operator $\mathcal{Q}$ it becomes
$\dot{\mathcal{Q}}^H_t = i [\mathcal H^H_t(\lambda_t),\mathcal
Q^H_t]/\hbar$.

\textcite{Bochkov77SPJETP45} were not able to provide any
quantum analog of their fluctuation relations, Eqs.
(\ref{eq:BK-gen-functional-identity}, \ref{eq:BK-identity}), with
the classical work $W_0$ replaced by the  operator $\mathcal
W_0$.

\textcite{Yukawa00JPSJ69} and \textcite{Allahverdyan05PRE71} arrived
at a similar conclusion when attempting to define an inclusive work
operator by $\mathcal W= \mathcal H^H_\tau(\lambda_\tau) - \mathcal
H(\lambda_{0}).$ According to this definition the exponentiated work
$\langle e^{-\beta \mathcal W}\rangle =\Tr \varrho_0 e^{-\beta
\mathcal W} = \langle e^{-\beta[\mathcal H^H_\tau(\lambda_\tau) - \mathcal
H(\lambda_{0})]} \rangle $ agrees with $e^{-\beta \Delta F}$ only if
$\mathcal H(\lambda_t)$ commutes at different times $[\mathcal
H(\lambda_t),\mathcal H(\lambda_\tau)]=0$ for any $t,\tau$. This
could lead to the premature conclusion that there exists no direct
quantum analog of the Bochkov-Kuzovlev and the Jarzynski identities,
Eqs. (\ref{eq:BK-identity}, \ref{eq:J-identity}),
\cite{Allahverdyan05PRE71}.

Based on the works by \textcite{Kurchan00arXiv} and
\textcite{Tasaki00arXiv}, \textcite{Talkner07PRE75} and
\textcite{Talkner07JPA40} demonstrated that this conclusion is based
on an erroneous assumption. They pointed out that work characterizes
a process, rather than a state of the system; this is also an
obvious observation from thermodynamics: unlike internal energy,
work is not a state function (its differential is not exact).
Consequently, work cannot be represented by a Hermitean operator
whose eigenvalues can be determined in a single projective
measurement. In contrast, the energy $\mathcal H(\lambda_t)$ (or
$\mathcal H_0$, when the exclusive viewpoint is adopted) must be
measured \emph{twice} first at the initial time $t=0$ and again at
the final time time $t=\tau$.

The difference of the outcomes of these two measurements then
yields  the work performed on the system in a particular
realization \cite{Talkner07PRE75}. That
is, if at time $t=0$ the eigenvalue $E_n^{\lambda_0}$ of $\mathcal
H(\lambda_{0})$ and later, at $t=\tau$, the eigenvalue
$E_m^{\lambda_\tau}$ of $\mathcal H(\lambda_\tau)$ were obtained,\footnote{For
a formal definition of these eigenvalues see Eq.
(\ref{eq:eigenvalueEq}) below.}
the measured (inclusive) work becomes:
\begin{equation}
 w= E_m^{\lambda_\tau}-E_n^{\lambda_0}.
\label{eq:q-in-work}
\end{equation}

Equation (\ref{eq:q-in-work}) represents the quantum version of the
classical inclusive work, Eq. (\ref{eq:in-work}).
In contrast to the classical case, this energy difference, which yields the work performed
in a single realization of the protocol, cannot be expressed
in the form of an integrated power, as in Eq. (\ref{eq:in-work-integral}).

The quantum version of the exclusive work, Eq. (\ref{eq:ex-work}), is
$w_0= e_m-e_n$ \cite{Campisi10arXiv}, where now $e_l$ are the
eigenvalues of $\mathcal H_0$. As we will demonstrate in the next
section, with these definitions of work straightforward quantum
analogs of the Bochkov-Kuzovlev results, Eqs.
(\ref{eq:BK-gen-functional-identity}, \ref{eq:BK-identity},
\ref{eq:BK-w-fluc-theo}) and of their inclusive viewpoint
counterparts, Eqs. (\ref{eq:J-gen-functional-identity},
\ref{eq:J-identity}, \ref{eq:J-w-fluc-theo}) can be derived.

\section{Quantum work fluctuation relations}
\label{sec:QFT}
Armed with all the proper mathematical definitions of nonequilibrium
quantum mechanical work, Eq. (\ref{eq:q-in-work}), we next
confidently embark on the study of work fluctuation relations in
quantum systems. As in the classical case, also in the quantum case
one needs to be careful in properly identifying the exclusive and
inclusive work, and must  be aware of the gauge freedom issue. In
the following we shall adopt, except when otherwise explicitly
stated, the inclusive viewpoint. The two fundamental ingredients for
the development of the theory are, in the quantum case like in the
classical case, the canonical form of equilibrium and
microreversibility.

\subsection{Microreversibility of non autonomous quantum systems}
\label{subsec:q-microrev}
The principle of microreversibility is introduced and discussed in
quantum mechanics textbooks for autonomous (i.e., non-driven)
quantum systems \cite{Messiah62Book}. As in the classical case,
however, this principle continues to hold in a more general sense
also for non-autonomous quantum systems. In this case it can be
expressed as:
\begin{equation}
U_{t,\tau}[\lambda]=\Theta^\dagger U_{\tau-t,0}[\widetilde
\lambda] \Theta  ,
\label{eq:q-microreversibility}
\end{equation}
where $\Theta$ is the quantum mechanical time reversal operator
\cite{Messiah62Book}.\footnote{Under the action of $\Theta$
coordinates transform evenly, whereas linear and angular momenta,
as well as spins change sign. In the coordinate representation,
in absence of spin degrees of freedom, the operator $\Theta$ is
the complex conjugation: $\Theta \psi=\psi^*$.}
Note that the presence of the protocol $\lambda$ and its time
reversed image $\tilde{\lambda}$ distinguishes this generalized
version from the standard form of microreversibility for
autonomous systems.
The principle of microreversibility,
Eq. (\ref{eq:q-microreversibility}), holds under the assumption
that at any time $t$ the Hamiltonian is invariant under time
reversal,\footnote{In presence of external magnetic fields the
direction of these fields has also to be inverted in the same way
as in the autonomous case.} that is:
\begin{equation}
\mathcal{H}(\lambda_t) \Theta =\Theta \mathcal{H}(\lambda_t) .
\label{eq:[Theta-H]}
\end{equation}
A derivation of Eq. (\ref{eq:q-microreversibility}) is presented in
Appendix \ref{app:q-microreversibility}. See also
\cite{Andrieux08PRL100} for an alternative derivation.

\begin{figure}
\includegraphics[trim = 20mm 25mm 10mm 10mm,
clip,scale=1]{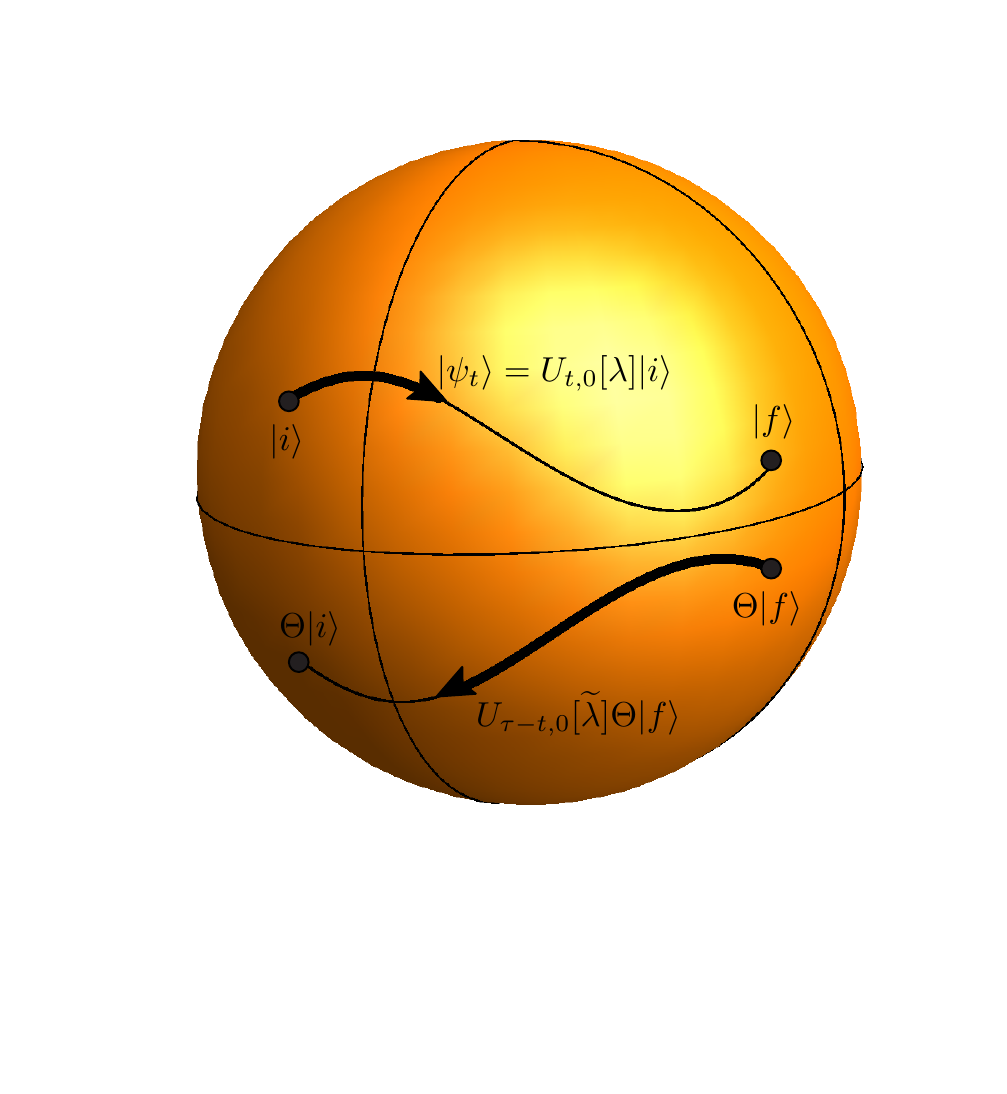} \caption{(Color online)
Microreversibility for non autonomous quantum systems. The
normalized initial condition $|i\rangle$ evolves, under the unitary
time evolution generated by $\mathcal H(\lambda_t)$, from time $t=0$
until $t$ to $|\psi_t\rangle=U_{t,0}[\lambda]|i\rangle$ and until time
$t=\tau$ to $|f\rangle$. The time-reversed final condition $\Theta
|f\rangle$ evolves, under the unitary evolution generated by
$\widetilde \lambda$ from time $t=0$ until $\tau-t$ to
$U_{\tau-t,0}[\widetilde \lambda]\Theta |f\rangle=\Theta
|\psi_t\rangle$, Eq. (\ref{eq:q-microreversibility3}), and until
time $t=\tau$ to the time-reversed initial condition $\Theta
|i\rangle$. The motion occurs onto the hypersphere of unitary radius
in the Hilbert space.} \label{fig:q-microreversibility}
\end{figure}

In order to better understand the physics behind  Eq.
(\ref{eq:q-microreversibility}) we rewrite it as: $
U_{t,0}[\lambda]= \Theta^\dagger U_{\tau-t,0}[\widetilde \lambda]
\Theta U_{\tau,0}[\lambda]  , $ where we used the concatenation rule
$U_{t,\tau}[\lambda]=U_{t,0}[\lambda]U_{0,\tau}[\lambda]$, and the
inverse $U_{0,\tau}[\lambda]=U^{-1}_{\tau,0}[\lambda]$ of the
propagator $U_{t,s}[\lambda]$. Applying it to a pure state
$|i\rangle$, and multiplying by $\Theta$ from the left, we obtain:
\begin{equation}
\Theta |\psi_t\rangle= U_{\tau-t,0}[\widetilde\lambda]
\Theta |f\rangle  ,
\label{eq:q-microreversibility3}
\end{equation}
where we introduced the notations
$|\psi_{t}\rangle=U_{t,0}[\lambda]|i\rangle$ and $|f\rangle
=U_{\tau,0}[\lambda]|i\rangle$. Equation (\ref{eq:q-microreversibility3}) says
that, under the evolution generated by the reversed protocol
$\widetilde \lambda$ the time reversed final state,
$\Theta |f\rangle$, evolves between time $0$ and $\tau-t$, to
$\Theta | \psi_t\rangle$. This is illustrated in Fig.
\ref{fig:q-microreversibility}. As for the classical case, in
order to trace a non autonomous system back to its initial state,
one needs not only to invert the momenta (applying $\Theta$), but
also to invert the temporal sequence of Hamiltonian values.

\subsection{The  work probability density function}
\label{subsec:workPDF}
We consider a system described by the Hamiltonian $\mathcal
H(\lambda_t)$ initially prepared in the canonical state
$
 \varrho(\lambda_0)={e^{-\beta \mathcal
H(\lambda_0)}}/{\mathcal Z(\lambda_0)} .
$
The instantaneous eigenvalues of $\mathcal H(\lambda_t)$ are denoted by
$E_n^{\lambda_t}$, and the corresponding instantaneous eigenstates by
$|\psi_{n,\gamma}^{\lambda_t}\rangle$:
\begin{equation}
 \mathcal
H(\lambda_t)|\psi_{n,\gamma}^{\lambda_t}\rangle=E_n^{\lambda_t}
|\psi_{n,\gamma}^{\lambda_t}\rangle .
\label{eq:eigenvalueEq}
\end{equation}
The symbol $n$ labels the quantum number specifying the energy
eigenvalues and $\gamma$ denotes all further quantum
numbers, necessary to specify an energy eigenstate in case of
$g_n$-fold degeneracy.
We emphasize that the instantaneous eigenvalue equation (\ref{eq:eigenvalueEq})
must not be confused with the Schr\"odinger equation
$i \hbar \partial_t |\psi(t)\rangle = \mathcal{H}(\lambda_t)|\psi(t)\rangle$.
The instantaneous eigenfunctions resulting from (\ref{eq:eigenvalueEq})
in particular are not solutions of the Schr\"odinger equation.

At $t=0$ the first measurement of $\mathcal H(\lambda_0)$ is
performed, with outcome $E_n^{\lambda_0}$.
This occurs with probability
\begin{equation}
 p_n^0=g_n e^{-\beta
E_n^{\lambda_0}}/\mathcal Z(\lambda_0)\, .
\end{equation}
According to the
postulates of quantum mechanics, immediately after the
measurement of the energy $E_n^{\lambda_0}$ the system is found
in the state:
\begin{equation}
 \varrho_n=\Pi_n^{\lambda_0} \varrho(\lambda_0) \Pi_n^{\lambda_0}
/p_n^0
\end{equation}
where $\Pi_n^{\lambda_0}=\sum_{\gamma}
|\psi_{n,\gamma}^{\lambda_0}\rangle \langle
\psi_{n,\gamma}^{\lambda_0} |$ is the projector onto the
eigenspace spanned by the eigenvectors belonging to the
eigenvalue $E_n^{\lambda_0}$. The system is assumed to be
thermally isolated at any time $t\geq 0$, so that its evolution
is determined by the unitary operator $U_{t,0}[\lambda]$, Eq. (\ref{eq:Schroedinger}),
 hence it evolves according to
\begin{equation}
\varrho_n(t)=U_{t,0}[\lambda]\varrho_n
U^{\dagger}_{t,0}[\lambda] .
\label{eq:rho-n}
\end{equation}
At time $\tau$ a second measurement of $\mathcal H(\lambda_\tau)$
yielding the eigenvalue $E_m^{\lambda_\tau}$ with probability
\begin{equation}
 p_{m|n}[\lambda]= \Tr \,\Pi_m^{\lambda_\tau} \varrho_n(\tau) \, .
\label{eq:p(m|n)}
\end{equation}
is performed. The pdf to observe the work $w$ is thus given by:
\begin{equation}
 p[w;\lambda]= \sum_{m,n}
\delta(w-[E_m^{\lambda_\tau}-E_n^{\lambda_0}])p_{m|n}[\lambda]p_n^0.
 \label{eq:p[w;lambda]}
\end{equation}

The work pdf has been calculated explicitly for a forced harmonic
oscillator \cite{Talkner08PRE78,Talkner09PRE79} and for a parametric oscillator
with varying frequency \cite{Deffner08PRE77, Deffner10CP375}.

\subsection{The characteristic function of work}
\label{subsec:workCharFun}
The characteristic function of work $ G[u;\lambda]$ is defined as
in the classical case as
the Fourier transform of the work probability density function
\begin{equation}
 G[u;\lambda]=\int \mathrm{d}w e^{iuw}p[w;\lambda] \;.
\end{equation}
Like $p[w;\lambda]$, $G[u;\lambda]$ contains full information
regarding the statistics of the random variable $w$.
\textcite{Talkner07PRE75} showed that the work pdf has the form of a
two-time non-stationary quantum correlation function; i.e.,
\begin{align}
G[u;\lambda]&= \langle e^{iu\mathcal H^H_\tau(\lambda_\tau)}
e^{-iu\mathcal H(\lambda_0)} \rangle  \label{eq:G[u;lambda]}\\
&=\Tr\,  e^{iu\mathcal H^H_\tau(\lambda_\tau)}
e^{-(iu+\beta)\mathcal H(\lambda_0)}/\mathcal{Z}(\lambda_0)\, , \nonumber
\label{eq:G[u;lambda]b}
\end{align}
where the average symbol stands for quantum expectation over the
initial state density matrix $\varrho(\lambda_0)$, Eq. (\ref{eq:varrho_t}), i.e., $\langle
\mathcal B \rangle=\Tr \varrho(\lambda_0) \mathcal B$, and the
superscript $H$ denotes Heisenberg picture, i.e.,
 $\mathcal{H}^H_\tau(\lambda_\tau)  = U^\dagger_{\tau,0}[\lambda] \mathcal{H}(\lambda_\tau) U_{\tau,0}[\lambda]$.

Equation (\ref{eq:G[u;lambda]}) was derived
first by \textcite{Talkner07PRE75} in the case of nondegenerate
$\mathcal H(\lambda_t)$ and later generalized by
\textcite{Talkner08PRE77} to the case of possibly degenerate
$\mathcal H(\lambda_t)$.\footnote{The derivation in
\textcite{Talkner08PRE77} is more general in that it does not assume
any special form of the initial state, thus allowing the study,
e.g., of microcanonical fluctuation relations. The formal expression,
Eq. (\ref{eq:G[u;lambda]}), remains valid for any initial state
$\varrho$, with the provision that the average is taken with respect to
$\bar \varrho=\sum_n\Pi_n^{\lambda_0} \varrho \Pi_n^{\lambda_0}$
representing the diagonal part of $\rho$ in the eigenbasis of $\mathcal{H}(\lambda_0)$.}

The product of the two exponential operators $e^{iu \mathcal{H}^H_\tau(\lambda_\tau)} e^{-iu \mathcal{H}(\lambda_0)}$ can be combined into a single exponent under the protection of the time ordering operator $\mathcal{T}$ to yield $e^{iu \mathcal{H}^H_\tau(\lambda_\tau)} e^{-iu \mathcal{H}(\lambda_0)}= \mathcal{T} e^{iu[\mathcal{H}^H_\tau(\lambda_\tau) - \mathcal{H}(\lambda_0) ]}$. In this way one may convert the characteristic function of work to a form that is analogous to the corresponding classical
expression, Eq. (\ref{eq:Jgu}),
\begin{align}
G[u;\lambda]& = \Tr\,  \mathcal{T} e^{iu\mathcal{H}^H_\tau(\lambda_\tau)- \mathcal{H}(\lambda_0)} e^{-\beta \mathcal{H}(\lambda_0)}/\mathcal{Z}_0\\
&=  \Tr\,  \mathcal{T}\exp \left [ iu \int_0^\tau \mathrm{d}t \dot{\lambda}_t \frac{\partial \mathcal{H}^H_t(\lambda_t)}{\partial \lambda_t} \right ] e^{-\beta \mathcal{H}(\lambda_0)}/\mathcal{Z}_0\nonumber
\end{align}
The second line follows from the fact that the total time-derivative of the Hamiltonian in the Heisenberg picture coincides with its partial derivative.

As a consequence of quantum microreversibility, Eq.
(\ref{eq:q-microreversibility}), the characteristic function of work
obeys the following important symmetry relation (see Appendix
\ref{app:Z0G=ZfG}):
\begin{equation}
\mathcal Z(\lambda_0)  G[u;\lambda]=\mathcal Z(\lambda_\tau)
G[-u+i\beta;\widetilde \lambda]  \;. \label{eq:Z0G=ZfG}
\end{equation}
By applying the inverse Fourier transform and using $\mathcal
Z(\lambda_t)= \Tr e^{-\beta \mathcal H(\lambda_t)}=e^{-\beta
F(\lambda_t)}$ one ends up with the quantum version of the Crooks
fluctuation theorem in Eq. (\ref{eq:J-w-fluc-theo}):
\begin{equation}
  \frac{p[w;\lambda]}{p[-w; \widetilde \lambda]}=e^{\beta
(w-\Delta F)} .
\label{eq:q-J-w-fluc-theo-lambda}
\end{equation}
This result was first accomplished by \textcite{Tasaki00arXiv} and
\textcite{Kurchan00arXiv}. Later \textcite{Talkner07JPA40} gave a
systematic derivation based on the characteristic function of work.
The quantum Jarzynski equality
\begin{equation}
 \langle e^{-\beta w}\rangle_\lambda = e^{-\beta \Delta F}
,\label{eq:q-J-identity}
\end{equation}
follows by multiplying both sides by $p[-w;\lambda]e^{-\beta w}$ and
integrating over $w$. Given the fact that the characteristic
function is determined by a two-time quantum correlation function
rather than by a single time expectation value is another clear
indication that work is not an observable but instead characterizes
a process.

As discussed in \cite{Campisi10PRL105} the Tasaki-Crooks relation,
Eq. (\ref{eq:q-J-w-fluc-theo-lambda}) and the quantum version of the
Jarzynski equality, Eq. (\ref{eq:q-J-identity}), continue to hold
even if further projective measurements of any observable
$\mathcal{A}$ are performed within the protocol duration $(0,\tau)$.
These measurements, however do alter the work pdf
\cite{Campisi11arXiv}.

\subsection{Quantum generating functional}
\label{subsec:q-GenFun}
The Jarzynski equality can also immediately been obtained from
the characteristic function by setting $u=i\beta$, in Eq.
(\ref{eq:G[u;lambda]}) \cite{Talkner07PRE75}.
In order to obtain this result it is important that the Hamiltonian
operators at initial and final times enter into the characteristic
function, Eq. (\ref{eq:G[u;lambda]}), as arguments of two
factorizing exponential functions, i.e. in the form  $e^{-\beta
\mathcal H^H(\lambda_\tau)}e^{\beta \mathcal H(\lambda_0)}$. In general,
this of course is different from a single exponential  $e^{-\beta[
\mathcal H^H(\lambda_\tau)- \mathcal H(\lambda_0)]}$. In the definitions of
generating functionals, \textcite{Bochkov77SPJETP45,Bochkov81aPHYSA106} and \cite{Stratonovich94Book} employed
yet different ordering prescriptions which though do not lead to the
Jarzynski equality. In order to maintain the structure of the
classical generating functional, Eq.
(\ref{eq:J-gen-functional-identity}), also for quantum systems the
classical exponentiated work $e^{-\beta W}$ has to be replaced by
the product of exponentials as it appears in the characteristic
function of work, Eq. (\ref{eq:G[u;lambda]}). This then leads to a desired generating functional
relation
\begin{align}
\left\langle \exp\left[\int_{0}^{\tau}\mathrm{d}s u_t \mathcal
B^H_t\right] e^{-\beta \mathcal H^H(\lambda_t)}e^{\beta \mathcal
H(\lambda_0)}\right\rangle_{\lambda}&= \nonumber\\
\left\langle \exp\left[\int_{0}^{\tau}\mathrm{d}t \widetilde
u_t\varepsilon_\mathcal B \mathcal
B^H_t\right]\right\rangle_{\widetilde \lambda}&e^{-\beta \Delta
F},
\label{eq:q-J-gen-functional-identity}
\end{align}
where $\mathcal{B}$ is an observable with definite parity
$\varepsilon_\mathcal{B}$ (i.e., $\Theta \mathcal B
\Theta^\dagger=\varepsilon_{\mathcal B}\mathcal B$),
$\mathcal{B}^H_t$ denotes the observable $\mathcal B$ in the
Heisenberg representation, Eq. (\ref{eq:Heisenberg}), $u_t$ is a
real function, and $\widetilde u_t=u_{\tau-t}$. This can be proved
by using the quantum microreversibility principle, Eq.
(\ref{eq:q-microreversibility}), in a similar way as in the
classical derivation, Eq. (\ref{eq:BKderivation}).

The derivation of
Eq. (\ref{eq:q-J-gen-functional-identity}) was  provided by
\textcite{Andrieux08PRL100}, who therefrom also recovered the
formula of Kubo, Eq. (\ref{eq:Green-Kubo}), and the Onsager-Casimir
reciprocity relations.\footnote{\textcite{Andrieux08PRL100} also
allowed for a possible dependence of the Hamiltonian on a magnetic
field $\mathcal H=\mathcal H(\lambda_t,\mathbf B)$. Then it is
meant that the dynamical evolution of $\mathcal B$ in the right hand
side is governed by the Hamiltonian  $\mathcal H(\widetilde
\lambda_t,-\mathbf B)$, i.e., besides inverting the protocol, the
magnetic field needs to be inverted as well.}

These relations are obtained by means of functional derivatives
of Eq. (\ref{eq:q-J-gen-functional-identity}) with respect to the
force fields, $\lambda_t$, and test fields, $u_t$, at $\lambda_t=u_t=0$ \cite{Andrieux08PRL100}.
Relations and symmetries for higher order response functions
follow in an analogous way as in the classical case, Eq.
(\ref{eq:BK-gen-functional-identity}), by means of higher order
functional derivatives with respect to the force field $\lambda_t$.
Such relations were investigated experimentally by
\textcite{Nakamura10PRL104,Nakamura11arXiv}, see
Sec. \ref{sec:XFT}, Eq. (\ref{eq:S012}).

Within the exclusive viewpoint approach, the counterparts of Eqs.
(\ref{eq:G[u;lambda]}, \ref{eq:Z0G=ZfG},
\ref{eq:q-J-w-fluc-theo-lambda}, \ref{eq:q-J-identity},
\ref{eq:q-J-gen-functional-identity}), are obtained by replacing,
$\mathcal H^H$ with $\mathcal H_0^H$, $\mathcal H(\lambda_0)$ with
$\mathcal H_0$, $\mathcal{Z}(\lambda_t)$ with $\mathcal{Z}_0=\Tr
e^{-\beta \mathcal H_0}$, and setting accordingly $\Delta F$ to zero
\cite{Campisi10arXiv}.

\subsection{Microreversibility, conditional probabilities and entropy}
\label{subsec:microrev+condprobs}
For a Hamiltonian $\mathcal H(\lambda_t)$ with nondegenerate
instantaneous spectrum for all times $t$ and instantaneous
eigenvectors $|\psi_n^{\lambda_t}\rangle$ the conditional
probability $p_{m|n}[\lambda]$, Eq. (\ref{eq:p(m|n)}), is given by
the simple expression: $ p_{m|n}[\lambda]=|\langle
\psi_m^{\lambda_\tau}|U_{\tau,0}[\lambda]|\psi_n^{\lambda_0}\rangle|^2
\label{eq:<psi|U|psi>} $. As a consequence of the assumed invariance
of the Hamiltonian, Eq. (\ref{eq:[Theta-H]}), the eigenstates
$|\psi_n^{\lambda_t}\rangle$, are invariant under the action of the
the time reversal operator up to a phase, $ \Theta
|\psi_n^{\lambda_t}\rangle=e^{i\varphi_n^t}
|\psi_n^{\lambda_t}\rangle $. Then, expressing microreversibility as
$U_{\tau,0}[\lambda]=\Theta^\dagger  U_{0,\tau}[\widetilde \lambda] \Theta$, see Eq.
(\ref{eq:q-microreversibility}), the following symmetry relation is
obtained for the conditional probabilities:
\begin{equation}
p_{m|n}[\lambda]=p_{n|m}[\widetilde \lambda] .
\label{eq:pmn=pnm}
\end{equation}
Note the exchanged position of $m$ and $n$ in the two sides of
this equation. From Eq. (\ref{eq:pmn=pnm}) the Tasaki-Crooks
fluctuation theorem is readily obtained for a canonical initial state,
using Eq. (\ref{eq:p[w;lambda]}).

Using instead an initial microcanonical state at energy $E$,
described by the density matrix\footnote{Strictly speaking, in order
to obtain well defined expressions the $\delta$-function has to be
understood as a sharply peaked function with infinite support.}
\begin{equation}
 \rho_0(E)=\delta[E-\mathcal H(\lambda_0)]/\omega(E,\lambda_0)\, ,
\end{equation}
where $\omega(E,\lambda_t)=\Tr\, \delta(E-\mathcal H(\lambda_t))$, we obtain \cite{Talkner08PRE77}:
\begin{align}
\frac{p[E,W;\lambda]}{p[E+W,-W;\widetilde \lambda]}=e^{[S(E+W,\lambda_\tau)-S(E,\lambda_0)]/k_B}\, ,
\label{eq:mcFT}
\end{align}
where $S(E,\lambda_t)=k_B\ln\omega(E,\lambda_t)$, denotes
Boltzmann's thermodynamic equilibrium entropy. The corresponding
classical derivation was provided by \textcite{Cleuren06PRL96}.
A classical microcanonical version of the Jarzynski equality
was put forward by \textcite{Adib05PRE71} for non-Hamiltonian
iso-energetic dynamics. It was recently generalized to energy
controlled systems by \textcite{Katsuda11arXiv}.

\subsection{Weak coupling case}
\label{subsec:weak}
In the previous sections \ref{subsec:workPDF},
\ref{subsec:workCharFun}, \ref{subsec:q-GenFun}, we studied a
quantum mechanical system at canonical equilibrium at time $t=0$.
During the subsequent action of the protocol it is assumed to be
completely isolated from its surrounding apart from the influence of the
external work source and hence to undergo a unitary time evolution.
The quality of this approximation depends on the relative strength
of the interaction between the system and its environment, compared
to typical energies of the isolated system as well as on the
duration of the protocol.  In general, though, a treatment that
takes into account possible environmental interactions is necessary.
As will be shown below, the interaction  with a thermal bath does
{\it not} lead to a modification of the Jarzynski equality, Eq.
(\ref{eq:q-J-identity}), nor of the quantum work fluctuation
relation, Eq. (\ref{eq:q-J-w-fluc-theo-lambda}), both in the cases of
weak and strong coupling \cite{Talkner09JSM09, Campisi09PRL102}; a
main finding which holds true as well for classical systems
\cite{Jarzynski04JSM04}. In this section we address the weak
coupling case, while the more intricate case of strong coupling is
discussed in the next section.

\begin{figure}[t]
\includegraphics[width=7cm]{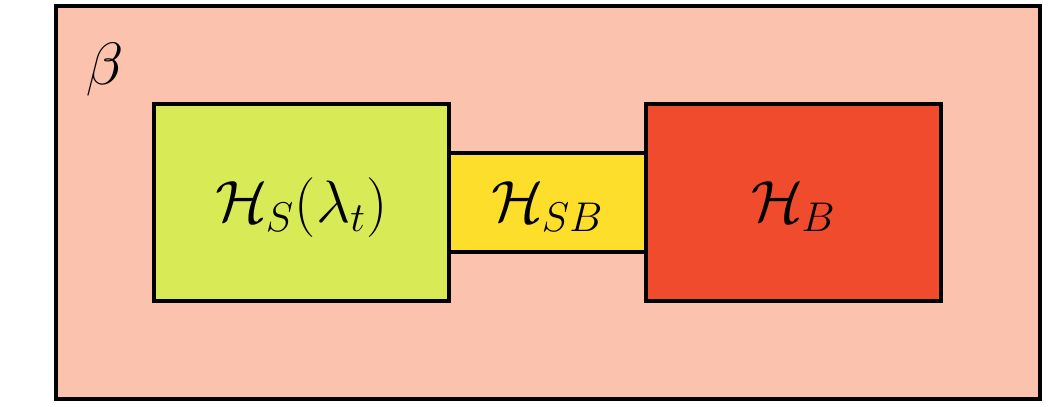}
\caption{(Color online) Driven open system. A driven system
[represented by its Hamiltonian $\mathcal H_S(\lambda_t)$] is
coupled to a bath [represented by $\mathcal H_B$] via the
interaction Hamiltonian $\mathcal H_{SB}$. The compound system
is in vanishingly weak contact with a super-bath, that provides
the initial canonical state at inverse temperature $\beta$, Eq.
(\ref{eq:varrho-coupling}).}
\label{fig:opensystem}
\end{figure}

We consider a driven system described by the time dependent
Hamiltonian $\mathcal H_S(\lambda_t)$, in contact with a thermal
bath with time independent Hamiltonian $\mathcal H_B$, see Fig.
\ref{fig:opensystem}. The Hamiltonian of the compound system is
\begin{equation}
 \mathcal H(\lambda_t)= \mathcal H_S(\lambda_t)+\mathcal H_B
+\mathcal H_{SB} \, ,
\end{equation}
where the energy contribution stemming from $\mathcal H_{SB}$ is
assumed to be much smaller than the energies of the system and bath
resulting from $\mathcal H_S(\lambda_t)$ and $\mathcal H_B$. The parameter
$\lambda$ that is manipulated according to a protocol solely enters
in the system Hamiltonian, $\mathcal H_S(\lambda_t)$.
The compound system is assumed to be initially ($t=0$) in the
canonical state
\begin{equation}
 \varrho(\lambda_0)= e^{-\beta \mathcal
H(\lambda_0)}/Y(\lambda_0) ,
 \label{eq:varrho-coupling}
\end{equation}
where $Y(\lambda_t)= \Tr e^{-\beta \mathcal H(\lambda_t)}$ is the
corresponding  partition function. This initial state may be
provided by contact with a super-bath at inverse temperature
$\beta$, see Fig. \ref{fig:opensystem}. It is then assumed that
either the contact is removed for $t\geq 0$ or that the super-bath
is so weakly coupled to the compound system that it bears no
influence on its dynamics over the time span $0$ to $\tau$.

Because the system and the environmental
Hamiltonians commute with each other, their energies can be
simultaneously measured. We denote the eigenvalues of $\mathcal
H_S(\lambda_t)$ as $E_i^{\lambda_t}$, and those of $\mathcal H_B$ as
$E_\alpha^B$. In analogy with the isolated case we assume that at
time $t=0$ a joint measurement of $\mathcal H_S(\lambda_0)$ and
$\mathcal H_B$ is performed, with outcomes $E_n^{\lambda_0}$,
$E^B_\nu$. A second joint measurement of $\mathcal
H_S(\lambda_\tau)$ and $\mathcal H_B$ at $t=\tau$ yields the outcomes
$E_m^{\lambda_\tau}$, $E^B_\mu$.

In analogy to the energy change of an isolated system, the differences of the eigenvalues of system and bath Hamiltonians yield the energy changes of system and bath, $\Delta E$ and $\Delta E^B$, respectively, in a single realization of the protocol, i.e.,
\begin{align}
\Delta E &=  E_m^{\lambda_\tau}-E_n^{\lambda_0} \, ,\\
\Delta E^B &= E^B_\mu-E^B_\nu \, .
\end{align}

In the weak coupling limit, the change of the energy content of the total system is given by the sum of the energy changes of the system and bath energies apart from a negligibly small contribution due to the interaction  Hamiltonian $\mathcal{H}_{SB}$. The work $w$ performed on the system coincides with the change of the total energy because the force is assumed to act only directly on the system. For the same reason, the change of the bath energy is solely due to an energy exchange with the system and hence, can be interpreted as negative heat, $\Delta E^B =-Q$. Accordingly we have\footnote{By use of the probability distribution in Eq. (\ref{eq:p(E,Q)}), the averaged quantity
$\langle \Delta E \rangle_\lambda= \int d(\Delta E) dQ\,
p[\Delta E, Q; \lambda] \Delta E= \Tr\,  \varrho_\tau
\mathcal{H}_S(\lambda_\tau)- \Tr\,  \varrho(\lambda_0)
\mathcal{H}_S (\lambda_0)$ cannot, in general, be interpreted as a change in thermodynamic internal energy. The reason is that the final state, $\varrho_\tau$,
reached at the end of the protocol is typically not a state of thermodynamic equilibrium; hence its themrodynamic internal energy is not defined.
}
\begin{equation}
 \Delta E=w+Q \, .
 \label{eq:FirstLaw}
\end{equation}

Following the analogy with the isolated case, we consider the
joint probability distribution function  $p[\Delta E,Q;\lambda]$
that the system energy changes by $\Delta E$ and the heat $Q$
is exchanged, under the protocol $\lambda$:
\begin{align}
p[\Delta E,Q;\lambda]=\sum_{m,n,\mu,\nu}&\delta[\Delta E-E_m^{\lambda_\tau}+E_n^{\lambda_0} ]\delta [Q +
E^B_\mu-E^B_\nu] \nonumber \\
& \times p_{m\mu|n\nu}[\lambda]p^0_{n\nu} \, ,
\label{eq:p(E,Q)}
\end{align}
where $p_{m\mu|n\nu}[\lambda]$ is the conditional probability to
obtain the outcome $E_m^{\lambda_\tau}$, $E^B_\mu$ at $\tau$,
provided that the outcome $E_n^{\lambda_0}$, $E^B_\nu$ was
obtained at time $t=0$, whereas $p^0_{n\nu}$ is the probability to
find the outcome $E_n^{\lambda_0}$, $E^B_\nu$ in the first
measurement. The conditional probability $p_{m\mu|n\nu}[\lambda]$
can be expressed in terms of the projectors on the
common eigenstates of $\mathcal H_S(\lambda_t),\mathcal H_B$, and
the unitary evolution generated by the total Hamiltonian
$\mathcal H(\lambda_t)$ \cite{Talkner09JSM09}.

By taking the Fourier transform of $p[\Delta E,Q;\lambda]$ with
respect to both $\Delta E$ and $Q$ one obtains the characteristic
function of system energy change and heat exchange, reading
\begin{align}
  G[u,v;\lambda]=\int \mathrm{d}(\Delta E) \mathrm{d}Q e^{i(u\Delta E+vQ)}
p[\Delta E,Q;\lambda] ,
\end{align}
which can be further simplified and cast, as in the isolated
case, in the form of a two-time quantum correlation function
\cite{Talkner09JSM09}:
\begin{equation}
   G[u,v;\lambda]=\langle e^{i[u\mathcal H_{S
\tau}^H(\lambda_\tau)-v\mathcal H_{B\tau}^H]}
e^{-i[u\mathcal H_S(\lambda_0)-v\mathcal H_B]}
\rangle  ,
\label{eq:G[u,v,lambda]}
\end{equation}
where the average is over the state $\bar{\varrho}(\lambda_0)$,
that is the diagonal part of $\varrho(\lambda_0)$, Eq.
(\ref{eq:varrho-coupling}), with respect to $\{\mathcal
H_S(\lambda_0),\mathcal H_B\}$. Notably, in the limit of weak
coupling this state $\bar{ \varrho}(\lambda_0)$ approximately
factorizes into the product of the equilibrium states of system
and bath with the deviations being of second order in the system-bath
interaction \cite{Talkner09JSM09}.

Using Eqs.
(\ref{eq:G[u,v,lambda]}) in combination with
microreversibility, Eq. (\ref{eq:q-microreversibility}), leads, in
analogy with Eq. (\ref{eq:Z0G=ZfG}), to
\begin{equation}
\mathcal Z_S(\lambda_0) G[u,v;\lambda]=\mathcal
Z_S(\lambda_\tau)G[-u+i\beta,-v-i\beta;\widetilde \lambda]
\label{eq:ZG[u,v]=ZG[-u+ibeta,-v-ibeta]}\, ,
\end{equation}
where
\begin{equation}
 \mathcal Z_S(\lambda_t) = \Tr_S e^{-\beta \mathcal  H_S(\lambda_t)}\, ,
\label{eq:Zweak}
\end{equation}
with $\Tr_S$ denoting the trace over the system Hilbert space.
Upon applying an inverse Fourier transform of Eq.
(\ref{eq:ZG[u,v]=ZG[-u+ibeta,-v-ibeta]}) one arrives at the
following  relation:
\begin{equation}
\frac{ p[\Delta E,Q;\lambda]}{p[-\Delta E,-Q;\widetilde
\lambda]}=e^{\beta (\Delta E-Q-\Delta F_S)} ,
\label{eq:q-J-flutheo-UQ}
\end{equation}
where
\begin{equation}
 \Delta F_S=-\beta^{-1}\ln[\mathcal
Z_S(\lambda_\tau)/\mathcal Z_S(\lambda_0)]
\end{equation}
denotes the system
free energy difference. Eq. (\ref{eq:q-J-flutheo-UQ}) generalizes
the Tasaki-Crooks fluctuation theorem, Eq.
(\ref{eq:q-J-w-fluc-theo-lambda}), to the case where the system
can exchange heat with a thermal bath.

Performing the change of the variable $\Delta E \rightarrow w$,
Eq. (\ref{eq:FirstLaw}), in
Eq. (\ref{eq:q-J-flutheo-UQ}) leads to the following fluctuation
relation for the joint probability density function of work and
heat, namely:
\begin{equation}
\frac{p[w,Q;\lambda]}{p[-w,-Q;\widetilde \lambda]}=e^{\beta
(w-\Delta F_S)} .
\label{eq:q-J-flutheo-wQ}
\end{equation}
Notably, the right hand side does not depend on the heat $Q$ but
depends on the work $w$ only. This fact implies that the marginal
probability density of work $ p[w;\lambda]=\int dQ p[w,Q;\lambda]$
obeys the Tasaki-Crooks relation:
\begin{equation}
\frac{p[w;\lambda]}{p[-w;\widetilde \lambda]}=e^{\beta (w-\Delta F_S)} .
\label{eq:q-J-fluctheo-marginal}
\end{equation}
Subsequently the Jarzynski equality, $ \langle e^{-\beta w}
\rangle=e^{-\beta \Delta F_S} $, is also satisfied. Thus the
fluctuation relation, Eq. (\ref{eq:q-J-w-fluc-theo-lambda}), and  the
Jarzynski equality, Eq. (\ref{eq:q-J-identity}), keep holding,
unaltered, also in the case of weak coupling. This result was
originally found upon assuming a Markovian quantum dynamics for the
reduced system dynamics $S$.\footnote{See \cite{Mukamel03PRL90,DeRoeck04PRE69,Esposito06PRE73,Crooks08JSM08}.}
With the above derivation we followed
\textcite{Talkner09JSM09} in which one does not rely on a Markovian
quantum evolution and consequently  the results hold true as well
for a general non-Markovian reduced quantum dynamics of the system
dynamics $S$.

\subsection{Strong coupling case}
\label{subsec:q-strong}
In the case of strong coupling, the system-bath interaction energy
is non-negligible, and therefore it is no longer possible to
identify the heat as the energy change of the bath. How to
define heat in a strongly coupled driven system and whether it is
possible to define it at all currently remain open problems. This,
however does not hinder the possibility to prove that the work
fluctuation relation, Eq. (\ref{eq:q-J-fluctheo-marginal}), remains
valid also in the case of strong coupling. For this purpose it
suffices to properly identify the work $w$ done on, and the free
energy $F_S$ of an open system, without entering the issue of what
heat means in  a strong-coupling situation. As for the classical
case, see Sec. \ref{subsec:Inc-Exc}, the system Hamiltonian
$\mathcal H_S(\lambda_t)$ is the only time dependent part of the
total Hamiltonian $\mathcal H(\lambda_t)$. Therefore the work done
on the open quantum system, coincides with the work done on the
total system, as in the weak coupling case treated in the
previous section, Sec. \ref{subsec:weak}. Consequently,
the work done on an open quantum system in a single realization is
\begin{equation}
w=\mathcal E^{\lambda_\tau}_m-\mathcal E^{\lambda_0}_n
\label{eq:q-incl-work-open}
\end{equation}
where $\mathcal E^{\lambda_t}_l$ are the eigenvalues of the total
Hamiltonian $\mathcal H(\lambda_t)$.

Regarding the proper identification of the free energy of an open
quantum system, the situation is more involved because the bare
partition function
$\mathcal Z_S^0(\lambda_t)=\Tr_S e^{-\beta \mathcal H_S(\lambda_t)}$
cannot take into account the full effect of the environment in any
case other than the limiting situation of weak coupling. For strong
coupling the equilibrium statistical mechanical description has to
be based on a partition function of the open quantum system that is
given as the ratio of the partition functions of the total system
and the isolated environment,\footnote{See \cite{Ford85PRL21,
Hoerhammer08JSP133, Hanggi05CHAOS2, Hanggi06APPB37, Ingold02LNP611,
Nieuwenhuizen02PRE66, Hanggi98Book, Grabert88PREP168,
Grabert84ZPB55, Caldeira83AP149, Feynman63AP24, Campisi09PRL102,
Campisi09JPA42, Campisi10CP375}.} i.e.:
\begin{equation}
\label{eq:ZSstrong}
\mathcal{Z}_S(\lambda_t) ={Y}(\lambda_t)/{\mathcal{Z}_B}  ,
\end{equation}
where $\mathcal Z_B = \Tr_B e^{-\beta \mathcal H_B}$ and
$Y(\lambda_t)=\Tr e^{-\beta \mathcal H(\lambda_t)}$ with $\Tr_B$,
 $\Tr$ denoting the trace over the bath Hilbert space and the
total Hilbert space, respectively.
It must be stressed that in general, the partition function
$\mathcal Z_S(\lambda_t)$ of an open quantum system differs from
its partition function in absence of a bath:
\begin{equation}
 \mathcal Z_S(\lambda_t) \neq \Tr_S e^{-\beta \mathcal H_S(\lambda_t)}\, .
\end{equation}
The equality is restored, though, in the limit of of weak coupling.

The free energy of an open quantum system follows according to
the standard rule of equilibrium statistical mechanics as
\begin{equation}
  F_S(\lambda_t)= F(\lambda_t)-F_B=-\frac{1}{\beta}\ln\frac{Y(\lambda_t)}{\mathcal Z_B} .
  \label{eq:q-FS-strong}
\end{equation}
In this way the influences of the bath on the thermodynamic
properties of the system are properly taken into account. Besides,
Eq. (\ref{eq:q-FS-strong}) complies with all the Grand Laws of
thermodynamics \cite{Campisi09PRL102}.

For a total system initially prepared in the Gibbs state, Eq.
(\ref{eq:varrho-coupling}), the Tasaki-Crooks fluctuation theorem,
Eq. (\ref{eq:q-J-w-fluc-theo-lambda}), applies yielding
\begin{equation}
\frac{p[w;\lambda]}{p[-w;\widetilde \lambda]}=
\frac{Y(\lambda_\tau)}{Y(\lambda_0)}e^{\beta w} .
\end{equation}
Since $\mathcal Z_B$ does not depend on time, the salient
relation
\begin{equation}
 Y(\lambda_\tau)/Y(\lambda_0)=\mathcal Z_S(\lambda_\tau)
/\mathcal Z_S(\lambda_0)
\end{equation}
holds, leading to:
\begin{equation}
 \frac{p[w;\lambda]}{p[-w;\widetilde \lambda]}=
\frac{\mathcal Z_S(\lambda_\tau)}{\mathcal Z_S(\lambda_0)}e^{\beta w}=
e^{\beta (w- \Delta F_S)}
\end{equation}
where $\Delta F_S=F_S(\lambda_\tau)-F_S(\lambda_0)$ is the proper
free energy difference of an open quantum system. Since $w$
coincides with the work performed on the open system the
Tasaki-Crooks theorem, Eq. (\ref{eq:q-J-fluctheo-marginal}), is
recovered  in the case of strong coupling \cite{Campisi09PRL102}.

\section{Quantum exchange fluctuation relations}
\label{sec:XFT}

The transport of energy and matter between two reservoirs
that stay at different temperatures and chemical potentials
represents an important experimental set-up, see also Sec.
\ref{sec:exp} below, as well as a central problem of non-equilibrium
thermodynamics \cite{deGroot84Book}. Here, the
two-measurement scheme described above in conjunction
with the principle of microreversibility leads to fluctuation
relations similar to the Tasaki-Crooks relation, Eq.
(\ref{eq:q-J-w-fluc-theo-lambda}),
for the probabilities of energy and matter exchanges.
The resulting fluctuation relations have been referred
to as ``exchange fluctuation theorems'' \cite{Jarzynski04PRL92},
to distinguish them from the ``work fluctuation theorems''.

\begin{figure}[t]
\includegraphics[width=50mm]{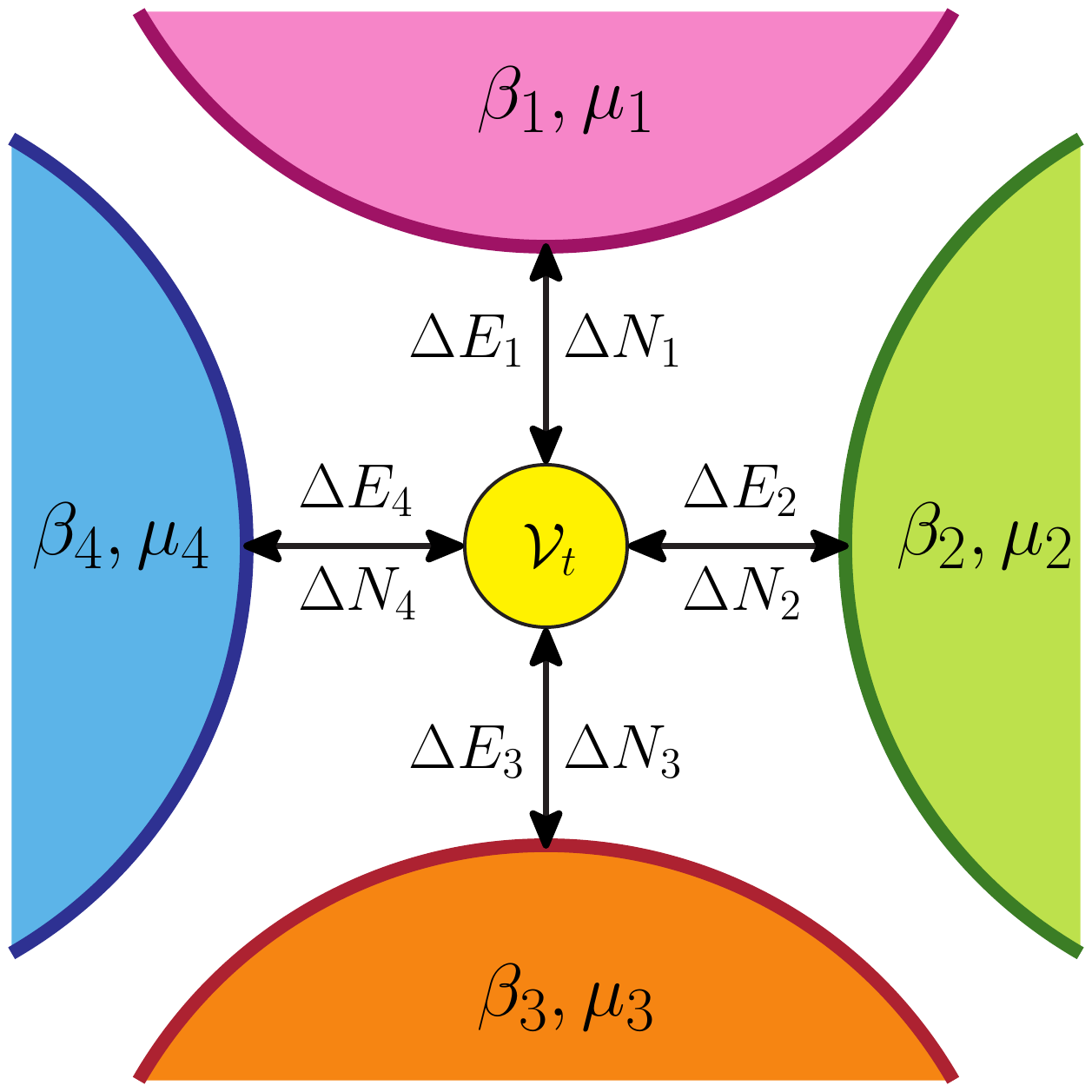}
\caption{(Color online) Exchange fluctuation relation set-up. Several
reservoirs (large semicircles) interact via the coupling $\mathcal
V_t$ (symbolized by the small circle), which is switched on at time
$t=0$ and switched off at $t=\tau$. During the on-period $(0,\tau)$
the reservoirs exchange energy and matter with each other. The
resulting net energy change of the $i$-th reservoir is $\Delta E_i$
and its particle content changes by $\Delta N_i$. Initially the
reservoirs have prescribed temperatures, $T_i=(k_B \beta_i)^{-1}$,
and chemical potentials, $\mu_i$, of the particle species that are
exchanged, respectively.} \label{fig:XFT}
\end{figure}

The first quantum exchange fluctuation theorem was put forward
by \textcite{Jarzynski04PRL92}. It applies to two systems initially
at different temperatures that are allowed to interact over the
lapse of time $(0,\tau)$, via a possibly time-dependent interaction.
This situation was later generalized by   \textcite{Saito08PRB78} and \textcite{Andrieux09NJP11},
to allow for the exchange of energy and particles between several
interacting systems initially at different temperatures and chemical
potentials; see the schematic illustration in Fig. \ref{fig:XFT}.
The total Hamiltonian $\mathcal H(\mathcal V_t)$ consisting of
$s$ subsystems is:
\begin{equation}
\mathcal H(\mathcal V_t)= \sum_{i=1}^s \mathcal H_{i}+\mathcal
V_t\, ,
\end{equation}
where $\mathcal H_i$ is the Hamiltonian of the $i$-th system, and
$\mathcal V_t$ describes the interaction between the subsystems,
which sets in at time $t=0$ and ends at time $t=\tau$. Consequently
$\mathcal V_t=0 $ for $t\notin (0,\tau)$, and in particular
$\mathcal V_0=\mathcal V_\tau =0 $.
As before, it is very important to distinguish between the values
$\mathcal{V}_t$ at a specific time and the whole protocol $\mathcal
V$.

Initially, the subsystems are supposed to be isolated from each
other and to stay in a factorized grand canonical state
\begin{equation}
\varrho_0=\prod_i \varrho_i = \prod_i{e^{-\beta_i [\mathcal H _i-\mu_i \mathcal
N_i]}}/{\Xi_i}\, ,
\label{eq:varrho_0-factorized}
\end{equation}
with $\mu_i,\beta_i,\Xi_i =\Tr_i e^{-\beta_i (\mathcal H_i-\mu_i
\mathcal N_i)}$ the chemical potential, inverse temperature, and
grand potential, respectively, of subsystem $i$. Here,
$\mathcal{N}_i$ and $\Tr_i$ denote the particle number operator and
the trace of the $i^{\text{th}}$ subsystem, respectively.

We also assume that in absence of interaction the particle numbers
in each subsystem are conserved, i.e., $[\mathcal
H_i,\mathcal{N}_i]=0$. Since operators acting on Hilbert spaces of
different subsystems commute, we find $[\mathcal
H_i,\mathcal{N}_j]=0$, $[\mathcal N_i,\mathcal N_j]=0$, and
$[\mathcal H_i,\mathcal{H}_j]=0$ for any $i,j$. Accordingly, one may
measure all the $\mathcal H_i$'s and all the $\mathcal N_i$'s
simultaneously. Adopting the two-measurement scheme discussed above
in the context the work fluctuation relation, we make a first
measurement of all the $\mathcal H_i$'s and all the $\mathcal N_i$'s
at $t=0$. Accordingly, the wave function collapses onto a common
eigenstate $|\psi_n\rangle $ of all these observable with
eigenvalues $E_n^i$, $N_n^i$. Subsequently, this wave function
evolves according to the evolution $U_{t,0}[\mathcal V]$ generated by the total
Hamiltonian, until time $\tau$ when a second measurement of all
$\mathcal H_i$'s and  $\mathcal N_i$'s is performed leading to a
wave function collapse onto an eigenstate $|\psi_m \rangle$, with
eigenvalues    $E_m^i$, $N_m^i$. As in the case studied in  Sec.
\ref{subsec:weak}, the joint probability density of energy and
particle exchanges $p[\Delta \mathbf E,\Delta \mathbf N; \mathcal
V]$ completely describes the effect of the interaction protocol
$\mathcal V$
\begin{align}
p[\Delta \mathbf E,\Delta \mathbf N; \mathcal V]=\sum_{m,n}
\prod_i
\delta(\Delta E_i-  E^i_m+ E^i_n)
\nonumber \\
\times \delta(\Delta
N_i- N^i_m+ N^i_n)
p_{m|n}[\mathcal V]p_n^0\, ,
\end{align}
where $p_{m|n}[\mathcal V]$ is the transition probability from
state $|\psi_n\rangle $ to
$|\psi_m\rangle$ and
$p_n^0=\Pi_i
{e^{-\beta_i [E_n^i-\mu_iN_n^i]}}/{\Xi_i}
$
is the initial distribution of energies and particles.
Here the symbols $\Delta\mathbf E$ and $\Delta \mathbf N$, are
short notations for the individual energy and particle number
changes of all subsystems $\Delta E_1,\Delta E_2, \dots , \Delta
E_s$ and $\Delta N_1,\Delta N_2, \dots, \Delta N_s$,
respectively.

Assuming that the total Hamiltonian commutes with the time reversal
operator at any instant of time, and using the time reversal
property of the transition probabilities, Eq. (\ref{eq:pmn=pnm}),
one obtains that
\begin{align}
\frac{p[\Delta \mathbf E,\Delta \mathbf N; \mathcal V]}{p[-\Delta
\mathbf E,-\Delta \mathbf N; \widetilde{\mathcal V}]}
=\prod_i e^{\beta_i [\Delta E_i-\mu_i \Delta N_i]}.
\label{eq:XFT-general}
\end{align}
This equation was derived by \textcite{Andrieux09NJP11}, and
expresses the exchange fluctuation relation for the case of transport
of energy and matter.

In the case of a single isolated system ($s=1$), it reduces to the
Tasaki-Crooks work fluctuation theorem, Eq.
(\ref{eq:q-J-w-fluc-theo-lambda}), upon rewriting $\Delta E_1=w$ and
assuming that the total number of particles is conserved also when
the interaction is switched on. i.e.,  $[\mathcal H(\mathcal V_t),\mathcal
N]=0$, to obtain $\Delta N=0$. The free energy difference does not
appear in Eq. (\ref{eq:XFT-general}) because we have assumed the
protocol $\mathcal V$ to be cyclic.

In the case of two weakly interacting systems ($s=2$, $\mathcal
V$ small), that do not exchange particles, it reduces to the
fluctuation theorem of \textcite{Jarzynski04PRL92} for heat
exchange:
\begin{equation}
\frac{p[Q;\mathcal V]}{p[-Q;\widetilde{\mathcal V}]}=
e^{(\beta_1-\beta_2)Q} ,
\label{eq:XFT-heat}
\end{equation}
where $Q=\Delta E_1=-\Delta E_2$, with the second equality
following from the assumed weak interaction.

In case of two weakly interacting systems ($s=2$, $\mathcal V$
small) that do exchange particles, and are initially at same
temperature, yielding $Q\simeq0$, the fluctuation relation takes on
the form:
\begin{equation}
\frac{p[q;\mathcal V]}{p[-q;\widetilde{\mathcal V}]}=
e^{\beta(\mu_1-\mu_2)q},
\label{eq:XFT-matter}
\end{equation}
where $q \equiv \Delta N_1=-\Delta N_2$.

One basic assumption leading to the exchange fluctuation relation, Eq.
(\ref{eq:XFT-general}), is that the initial state is a factorized
state, in which the various subsystems are uncorrelated from each
other. In most experimental situations, however, unavoidable
interactions between the systems, would lead to some correlations,
and a consequent deviation from the assumed factorized state, Eq.
(\ref{eq:varrho_0-factorized}). The resulting deviation from the
exchange fluctuation relation, Eq. (\ref{eq:XFT-general}), is
expected to vanish for observation times $\tau$ larger than some
characteristic time scale $\tau_c$, determined by the specific
physical properties of the experimental set-up
\cite{Esposito09RMP81, Andrieux09NJP11, Campisi10PRL105}:
\begin{equation}
\frac{p[\Delta \mathbf E,\Delta \mathbf N; \mathcal V]}{p[-\Delta
\mathbf E,-\Delta \mathbf N; \widetilde{\mathcal V}]}
\xrightarrow{\tau \gg \tau_c}
\prod_i e^{\beta_i (\Delta E_i-\mu_i \Delta N_i)}.
\label{eq:SSFT}
\end{equation}
For those large times $t \gg \tau_c$ a non equilibrium steady state
sets in under the condition that the reservoirs are chosen macroscopic.
For this reason Eq. (\ref{eq:SSFT}) is referred to as a ``steady
state fluctuation relation''. This is in contrast to the other
fluctuation relations discussed above, which instead are valid for
any observation time $\tau$, and are accordingly referred to as
transient fluctuation relations. \textcite{Saito07PRL99} provided an
explicit demonstration of Eq. (\ref{eq:SSFT}) for the quantum heat
transfer across a harmonic chain connecting two thermal reservoirs
at different temperatures. \textcite{Ren10PRL104} reported on the
breakdown of Eq. (\ref{eq:SSFT}) induced by a nonvanishing
Berry-phase heat pumping.  The latter occurs when the temperatures
of the two baths are adiabatically modulated in time.


\section{Experiments}
\label{sec:exp}

\subsection{Work fluctuation relations}
Regarding the experimental validation of the work fluctuation
relation, a fundamental difference exists between the classical and
the quantum regime. In classical experiments work is accessible by
measuring the trajectory $\mathbf x_t$ of the possibly open system,
and integrating the instantaneous power according to $W=\int
\mathrm{d}t\partial H_S/\partial t$, Eq. (\ref{eq:Wm=W}). In clear
contrast, in quantum mechanics the work is obtained as the
difference of two measurements of the energy, and an ``integrated
power'' expression does not exists for the work, see Sec. \ref{subsec:work-not-observable}.

Following closely the prescriptions of the theory one should perform
the following steps in order to experimentally verify the work
fluctuation relation, Eq. (\ref{eq:q-J-w-fluc-theo-lambda}): (i)
Prepare a quantum system in the canonical state, Eq.
(\ref{eq:varrho_0}), at time $t = 0$. (ii) Measure the energy at
$t=0$. (iii) Drive the system by means of some forcing protocol
$\lambda_t$ for times $t$ between $0$ and $\tau$, and make sure that
during this time the system is well insulated from its environment.
(iv) Measure the energy again at $\tau$ and record the work $w$,
according to Eq. (\ref{eq:q-in-work}). (v) Repeat this procedure
many times and construct the histogram of the statistics of work as
an estimate of the work pdf $p[w;\lambda]$. In order to determine
the backward probability the same type of experiments has to be
repeated with the inverted protocol, starting from an equilibrium
state at inverse temperature $\beta$ and at those parameter values
that are reached at the end of the forward protocol.

\subsubsection{Proposal for an experiment employing trapped cold
ions}
\textcite{Huber08PRL101} suggest an experiment that follows exactly
the procedure described above. They propose to implement a quantum
harmonic oscillator by optically trapping an ion in the quadratic
potential generated by a laser trap, using the set-up developed by
\textcite{Schulz08NJP10}. In principle, the set-up of
\textcite{Schulz08NJP10} allows, on one hand, to drive the system by
changing in time the stiffness of the trap, and, on the other hand,
to probe whether the ion is in a certain Fock state $|n\rangle$;
i.e., in an energy eigenstate of the harmonic oscillator. The
measurement apparatus may be understood as a single Fock state
``filter'', whose outcome is ``yes" or ``no'', depending on whether
the ion is or is not in the probed state.  Thus the experimentalist probes each possible outcome $(n,m)$, where $(n,m)$ denotes the Fock states at
time $t=0$ and $t=\tau$, respectively. Then, the relative frequency of the outcome $(n,m)$ occurrence is recored by repeating the driving protocol many times
always preparing the system in the same canonical initial state.
In this way the joint probabilities $p_{m|n}[\lambda]p_n^0$ are measured.

The work histogram is then constructed as an estimate of the work pdf,
Eq. (\ref{eq:p[w;lambda]}), thus providing experimental access
to the fluctuation relation, Eq. (\ref{eq:q-J-w-fluc-theo-lambda}).
Likewise the relative frequency of the outcomes having $n$ as the initial state gives the experimental initial population $p_n^0$.
Thus, with this experiment one can actually check the symmetry relation of the conditional probabilities, $p_{m|n}[\lambda]=p_{n|m}[\widetilde \lambda]$,
Eq. (\ref{eq:pmn=pnm}) and compare their experimental values
with the known theoretical values \cite{Husimi53PTP9,Deffner08PRE77,Talkner08PRE78,Talkner09PRE79}.

Another suitable quantum system to test quantum fluctuation relations
are  quantum versions of nanomechanical oscillator set-ups that with
present day nanotechnology are at the verge of entering the
quantum regime.\footnote{Note the exciting recent advancements
obtained with the works \cite{LaHaye04SCIENCE304,
Kippenberg08SCIENCE321, Anetsberger09NATPHYS5, Oconnel10Nature464}.}
In these systems work protocols  can be imposed by optomechanical means.

\subsubsection{Proposal for an experiment employing circuit
quantum electrodynamics}
Currently, the experiment proposed by \textcite{Huber08PRL101} has
not yet been carried out. An analogous experiment could, in
principle, be performed in a circuit quantum electrodynamics (QED)
set-up as the one described in (\citealt{Hofheinz08NAT454},
\citeyear{Hofheinz09NAT459}). The set-up consists of a Cooper pair
box qubit (a two states quantum system) that can be coupled to and
de-coupled from a superconducting 1D transmission line, where the
latter mimics a quantum harmonic oscillator. With this architecture
it is possible to implement various functions with  very high degree of
accuracy. Among them the following tasks are of special interest in
the present context: (i) Creation of pure Fock states $|n\rangle$,
i.e., the energy eigenstates of the quantum harmonic oscillator in
the resonator. (ii) Measurement of photon statistics $p_m$, i.e.,
measurements of the population of each quantum state $|m\rangle$ of
the oscillator. (iii) Driving the resonator by means of an external
field.

\textcite{Hofheinz08NAT454} report, for example, on the
creation of the ground Fock state $|0\rangle$, followed by a driving
protocol $\lambda$ (a properly engineered microwave pulse applied to
the resonator) that ``displaces'' the oscillator and creates a
coherent state $|\alpha\rangle$, whose photon statistics
$p_{m|0}[\lambda]$ was  measured with good accuracy up to $n_{max}
\sim 10$. In more recent experiments \cite{Hofheinz09NAT459} the
accuracy was improved and $n_{max}$ has raised to $\sim 15$. The
quantity $p_{m|0}[\lambda]$ is actually the conditional probability
to find the state $|m\rangle$ at time $t=\tau$, given that the
system was in the state $|0\rangle $ at time $t=0$. Thus, by
preparing the oscillator in the Fock state $|n\rangle$ instead of
the ground state $|0\rangle$, and repeating the same driving and
readout as before, the matrix $p_{m|n}[\lambda]$ can be determined
experimentally. Accordingly one can test the validity of the
symmetry relation $p_{m|n}[\lambda]=p_{n|m}[\widetilde \lambda]$,
Eq. (\ref{eq:pmn=pnm}), which in turn implies the work fluctuation
relation, see Sec. \ref{subsec:microrev+condprobs}. 
At variance with the proposal of \textcite{Huber08PRL101}, in this
case the initial state would not be randomly sampled from a
canonical state, but would be rather created deterministically by
the experimenter.

The theoretical values of transition probabilities
for this case corresponding to a displacement of the oscillator were
first reported by \cite{Husimi53PTP9}, see also
\cite{Campisi08PRE78b}. \textcite{Talkner08PRE78} provided an
analytical expression for the characteristic function of work and
investigated in detail the work probability distribution function
and its dependence on the initial state, such as for example
canonical, microcanonical, and coherent state.

So far we have addressed possible experimental tests of the
Tasaki-Crooks work fluctuation theorem, Eq.
(\ref{eq:q-J-w-fluc-theo-lambda}), for isolated systems. The case of
open systems, interacting with a thermal bath, poses extra
difficulties related to the fact that in order to measure the work
in this case one should make two measurements of the energy of the
total macroscopic system, made up of the system of interest and its
environment. This presents an extra obstacle that at the moment
seems difficult to surmount except for a situation at (i)
weak coupling and (ii) $Q \sim 0$, then yielding, together
with Eq. (\ref{eq:FirstLaw}) $ w\sim \Delta E $.

\subsection{Exchange fluctuation relations}
Like the quantum work fluctuation relations, the quantum exchange
fluctuation relations are understood in terms of two point quantum
measurements. In an experimental test, the net amount of energy
and/or number of particles (depending which of the three exchange
fluctuation relations, Eqs. (\ref{eq:XFT-general}, \ref{eq:XFT-heat},
\ref{eq:XFT-matter}), is studied) has to be measured in each
subsystems twice, at the beginning and at the end of the protocol.
However typically these are macroscopic reservoirs, whose energy and
particle number measurement is practically impossible.\footnote{See
\cite{Andrieux09NJP11,Campisi10PRL105,Esposito09RMP81}, and also
\cite{Jarzynski00JSP98} regarding the classical case.} Thus,
seemingly, the verification of the exchange fluctuation relations
would be even more problematic than the validation of the quantum
work fluctuation relations. Indeed, while experimental tests of the
work fluctuation relations have not been reported yet, experiments
concerning quantum exchange fluctuation relations have been already
performed. In the following we shall discuss two of them, one by  \textcite{Utsumi10PRB81} and the other by
\textcite{Nakamura10PRL104}. In doing so we demonstrate how the
obstacle of energy/particle content measurement of macroscopic
reservoirs was circumvented.

\subsubsection{An electron counting statistics experiment}
\begin{figure}[t]
\includegraphics[width=5cm]{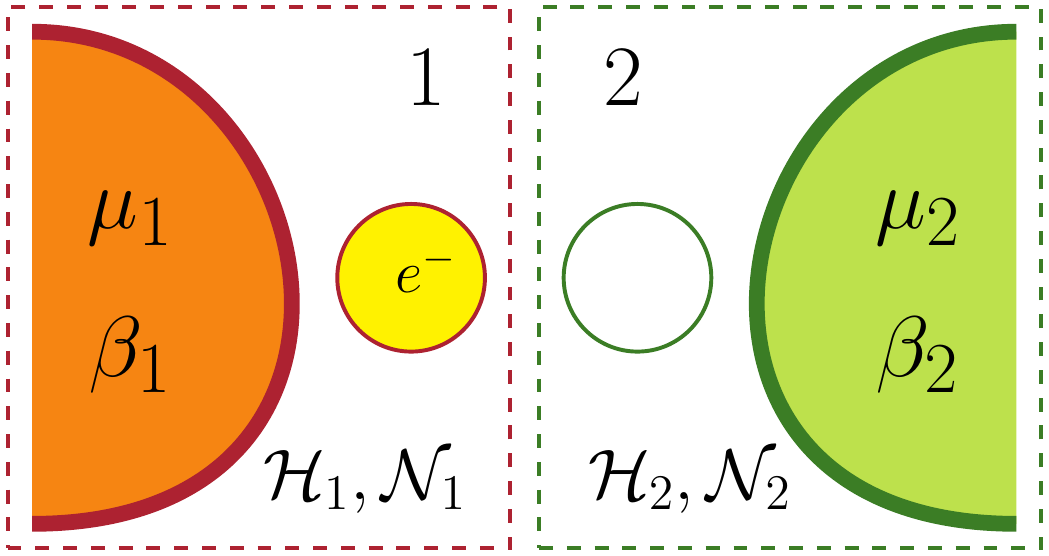}
\caption{(Color online) Scheme of a bidirectional counting
statistics experiment. Two leads (large
semicircles) with different electronic chemical potentials
($\mu_1\neq\mu_2$) and same temperature ($\beta_1=\beta_2$) are
connected through a double quantum dot (small circles), whose
quantum state is continuously monitored. The state (1,0), i.e.,
``one electron in the left dot, no electrons in the right dot''
is depicted. The transition from this state to the state (0,1)
signals the exchange of one electron from subsystem 1 to
subsystem 2. $\mathcal H_{1,2}$ and $\mathcal{N}_{1,2}$ denote the
Hamiltonian and electron number operators of the subsystems,
respectively.}
\label{fig:bidirectional-counting}
\end{figure}

\textcite{Utsumi10PRB81} have recently performed an experimental
verification of the particle exchange fluctuation relation, Eq.
(\ref{eq:XFT-matter}), using bi-directional electron counting
statistics \cite{Fujisawa06SCIENCE312}. The experimental set-up
consists of two electron reservoirs (leads) at the same
temperature. The two leads are connected via a double quantum dot,
see Fig. \ref{fig:bidirectional-counting}.
When an electric potential difference $V=\mu_1-\mu_2$ is applied to
the leads, a net flow of electrons starts transporting charges from
one lead to the other, via lead-dot and dot-dot quantum tunnelings.
The measurement apparatus consists of a secondary circuit in which a
current flows due to an applied voltage. Thanks to a properly
engineered coupling between secondary circuit and the double quantum
dot, the current in the circuit depends on the quantum state of the
double dot. The latter has four relevant states, which we shall
denote as $|00\rangle,|01\rangle,|10\rangle,|11\rangle$
corresponding respectively to: no electrons in the left dot and no
electrons in the right dot, no electrons in the left dot and one
electron in the right dot, etc.,. Each of these states leads to a
different value of the current in the secondary circuit. In the
experiment an electric potential difference is applied to the two
leads for a time $\tau$. During this time the state of the double
quantum dot is monitored by registering the current in the secondary
circuit. This current was found to switch between the four values
corresponding to the four quantum states mentioned above. The
outcome of the experiment is a sequence $I_k$ of current values,
with $I_k$ taking only four possible values. In other terms, the
outcome of the experiment consists of a sequence $\{(l,r)\}_k$ (with
$l,r=0,1$) of joint eigenvalues  of two commuting observables
$\mathcal L, \mathcal R$ specifying the occupation of the left ($l$)
and right ($r$) dots by single electrons at the time of the
$k^{\text{th}}$ measurement. The presence of an exchange of entries
within one time step of the form $(1,0)_n,(0,1)_{n+1}$ signals the
transfer of one electron from left to right, and vice versa
$(0,1)_n,(1,0)_{n+1}$ the transfer from right to left. Thus, given a
sequence $\{(l,r)\}_k$, the total number $q[\{(l,r)\}_k]$ of
electrons transferred from left to right is obtained by subtracting
the total number of right-to-left transfers from the total number of
left-to-right transfers. It was found that, for observation times
larger than a characteristic time $\tau_c$, the fluctuation relation,
$p[q]=p[-q]e^{\beta V q}$, Eq. (\ref{eq:XFT-matter}), was satisfied
with the actual temperature of the leads replaced by an effective
temperature, see Fig \ref{fig:histogram}. The renormalization of
temperature was explained as an effect due to an exchange of
electrons occurring between the dots and the secondary circuit
\cite{Utsumi10PRB81}.

The question however remains of how to connect
this experiment in which the flux of electrons through an interface
is monitored and the theory, leading to Eq. (\ref{eq:XFT-matter}),
which instead prescribes only two measurements of total particle
numbers in the reservoirs. The answer was given in
\textcite{Campisi10PRL105}, who showed that the exchange fluctuation
relation, Eq. (\ref{eq:XFT-general}), remains valid, if in addition
to the two measurements of total energy and particle numbers
occurring at $0$ and $\tau$, the evolution of a quantum system is
interrupted by means of projective quantum measurements of any
observable $\mathcal A$ that commutes with the quantum time reversal
operator $\Theta$. In other words, while the forward and backward
probabilities are affected by the occurrence of intermediate
measurement processes, their ratio remains unaltered.

In the experiment of \textcite{Utsumi10PRB81} one does not need
to measure the initial and final content of particles in the
reservoirs because the number of exchanged particles is inferred
from the sequence of intermediate measurements outcomes
$\{(l,r)\}_k$. Thus, thanks to the fact that quantum measurements
do not alter the fluctuation relation, one may overcome the
problem of measuring the energy and number of particles of the
macroscopic reservoirs, by monitoring instead the flux through a
microscopic junction.

\subsubsection{Nonlinear response relations in a quantum
coherent conductor}
\label{subsec:nonlinear}
As discussed in the introduction, the original motivation for the
study of fluctuation relations was to overcome the limitations of
linear response theory and to obtain relations connecting
higher order response functions  to fluctuation properties of the
unperturbed system.
As an indirect and partial confirmation of the fluctuation relations higher order static fluctuation-response relations can be tested experimentally.

Such a  validation was recently accomplished in coherent
quantum transport
experiments by  \citeauthor{Nakamura10PRL104} (\citeyear{Nakamura10PRL104}, \citeyear{Nakamura11arXiv}),
where the average current $I$ and the zero-frequency current
noise power $S$ generated
in an Aharonov-Bohm ring were investigated as a function of an
applied dc voltage $V$, and magnetic field $\mathbf B$. In the
nonlinear response regime, the current and noise power may be
expressed as power series of the applied voltage:
\begin{align}
I(V,\mathbf{B}) &= G_1(\mathbf{B}) V+\frac{G_2(\mathbf{B})}{2}
V^2+\frac{G_3(\mathbf{B})}{3!}  V^3 + \dots \\
S(V,\mathbf{B})&= S_0(\mathbf{B})+ S_1(\mathbf{B})
V+\frac{S_2(\mathbf{B})}{2} V^2 + \dots
\end{align}
where the coefficients depend on the applied magnetic field
$\mathbf{B}$.
The steady state fluctuation theorem, Eq. (\ref{eq:SSFT}), then
predicts the following fluctuation relations \cite{Saito08PRB78}
\begin{eqnarray}
 S_0 = 4k_B TG_1, \quad
 S_1^S = 2k_B T G_2^S, \quad
 S_1^A = 6 k_B T G_2^A \quad
\label{eq:S012}
\end{eqnarray}
where $S^S_i=S_i(\mathbf{B})+S_i(-\mathbf{B})$,
$S^{A}_i=S_i(\mathbf{B})-S_i(-\mathbf{B})$, and analogous
definitions for $G^{S}_i$ and $G^{A}_i$. The first equation in
(\ref{eq:S012}) is the Johnson-Nyquist relation
\cite{Johnson28PR32,Nyquist28PR32}. In the experiment by
\textcite{Nakamura10PRL104} good quantitative agreement with the
first and the third expressions in (\ref{eq:S012}) was established,
whereas, for the time being,  only qualitative agreement was found
with the second relation.

The higher order static fluctuation dissipation relations (\ref{eq:S012}) were obtained from a steady state fluctuation theorem for particle exchange under the simplifying assumption that no heat exchange occurs \cite{Nakamura10PRL104}. Then the probability of transferring $q$ particles is related to the probability of the reverse transfer by $p(q) = p(-q) e^{A q}$ where $A = \beta V = \beta (\mu_1- \mu_2)$ is the so-called affinity. If both sides are multiplied by $q$ and integrated over $q$ a comparison of equal powers of applied voltage $V$ yields Eq. (\ref{eq:S012}) \cite{Nakamura11arXiv}.
An alternative approach, that also allows  to include the effect of heat conduction, is offered by the fluctuation theorems for currents in open quantum systems. This objective has been put forward by  \textcite{Saito08PRB78} and also by \textcite{Andrieux09NJP11}, based on a generating function  approach in the spirit of Eq. (\ref{eq:q-J-gen-functional-identity}).

\section{Outlook}
\label{sec:outlook}
In closing this colloquium we stress that the known fluctuation
relations are based on two facts: (a) microreversibility for
non-autonomous Hamiltonian systems Eq.
(\ref{eq:q-microreversibility}), and (b) the special nature of the
initial equilibrium states which is expressible in either
micro-canonical, canonical or grand-canonical form, or products
thereof. The final state reached at the end of a protocol though is
in no way restricted. It evolves from the initial state according to
the governing dynamical laws under a prescribed protocol. In general
this final state may markedly differ from any kind of equilibrium
state.

For quantum mechanical systems it also is of utmost importance to
correctly identify the work performed on a system as the difference
between the energy of the system at the end and the beginning of the
protocol. In case of open systems the difference of the energies of
the total system at the end and beginning of the protocol coincides
with the work done on the open system as long as the forces
exclusively act on this open system. With the free energy of an open
system determined as the difference of free energies of the total
system and that of the isolated environment the quantum and
classical Jarzynski equality and the Tasaki-Crooks theorem continue
to hold true even for systems strongly interacting with their
environment. Deviations from the fluctuation relations however must
be expected if protocol forces  not only act on the system alone but
as well directly on the environmental degrees of freedom; for
example, if a time-dependent system-bath interaction protocol is
applied.

The most general and compact formulation of quantum work fluctuation
relations also containing the Onsager-Casimir reciprocity relations
and nonlinear response to all orders, is the Andrieux-Gaspard
relation, Eq. (\ref{eq:q-J-gen-functional-identity}) which
represents the proper quantum version of the classical
Bochkov-Kuzovlev formula \cite{Bochkov77SPJETP45}, Eq.
(\ref{eq:BK-gen-functional-identity}). These relations provide a
complete theoretical understanding of those nonequilibrium
situations that emerge from arbitrary time-dependent perturbations
of equilibrium initial states.

Less understood are exchange fluctuation relations with their
important applications to counting statistics
\cite{Esposito09RMP81}. The theory there so far is restricted to
situations where the initial state factorizes into grand-canonical
states of reservoirs at different temperatures or chemical
potentials. The interaction between these reservoirs is turned on
and it is assumed that it will lead to a steady state within the
duration of the protocol. Experimentally, it is in general difficult
to exactly follow this prescription and therefore a comparison of
theory and experiment is only meaningful for the steady state.
Alternative derivations of exchange relations for more realistic,
non-factorizing initial states would certainly be of much interest.
In this context, the issue of deriving quantum fluctuation relations
for open  systems that initially are in nonequilibrium steady
quantum transport states constitutes a most interesting challenge.
Likewise, from the theoretical point of view little is known thus
far about quantum effects for transport in presence of time
dependent reservoirs, for example using a varying temperature and/or
chemical potentials \cite{Ren10PRL104}.

The experimental applications and validation schemes involving
nonlinear quantum fluctuation relations still are in a state of
infancy, as detailed with Sec. \ref{sec:exp}, so that there is plenty of room
for advancements. The major obstacle for the experimental
verification of the work fluctuation relation is posed by the
necessity of performing quantum projective measurements of energy.
Besides the proposal of \textcite{Huber08PRL101} employing trapped
ions, we suggested here the scheme of a possible experiment
employing circuit-QED architectures. In regard to exchange
fluctuation relations instead, the main problem is related to the
difficulty of measuring microscopic changes of macroscopic
quantities pertaining to heat and matter reservoirs. Continuous
measurements of fluxes  seemingly provide a practical and efficient
loophole for this dilemma \cite{Campisi10PRL105}.

The idea that useful work may be obtained by using information
\cite{Maruyama09RMP81} has established a connection between the
topical fields of quantum information theory \cite{Vedral02RMP74}
and quantum fluctuation relations. \citet{Piechocinska00PRA61} and
\textcite{Kawai07PRL98} used fluctuation relations and information
theoretic measures to derive Landauer's principle.
A generalization of the Jarzynski equality to the case of feedback
controlled systems was provided in the classical case
by \textcite{Sagawa10PRL104}, and in the quantum case
by \textcite{Morikuni11arXiv}. Recently
\textcite{Deffner10PRL105} gave bounds on the entropy production in
terms of quantum information concepts. In similar spirit,
\textcite{Hide10PRA81} presented a method by relating relative
quantum entropy to the  quantum Jarzynski fluctuation identity in
order to quantify  multi-partite entanglement within different
thermal quantum  states.
A practical application of the Jarzynski equality in quantum
computation was showed by \textcite{Ohzeki10PRL105}.

In conclusion, the authors are confident in their belief that this
topic of quantum fluctuation relations will exhibit an ever growing
activity within nanosciences and further may invigorate  readers to
pursue still own research and experiments as this theme certainly
offers many more surprises and unforeseen applications.

\begin{acknowledgments}
The authors thank \textcite{Utsumi10PRB81} for providing the data
for Fig. \ref{fig:histogram}.
This work was supported by the cluster of excellence
Nanosystems Initiative Munich (NIM) and the Volkswagen Foundation
(project I/83902).
\end{acknowledgments}

\appendix

\section{Derivation of the Bochkov-Kuzvlev relation}
\label{app:BK}

We report below the steps leading to Eq.
(\ref{eq:BK-gen-functional-identity})
\begin{align}
&\left \langle \exp\left[{\int_{0}^{\tau} \! \mathrm{d}s \, u_s
B_s}\right] e^{-\beta W_0}\right\rangle_{\lambda}=\nonumber\\
=&\int \!\mathrm{d}\mathbf{z}_0
\frac{e^{-\beta[H_0(\mathbf{z}_0)+W_0]}}{Z(t_0)}\exp
\left[\int_{0}^{\tau}\! \mathrm{d} s\,  u_s
B(\varphi_{s,0}[\mathbf{z}_0;\lambda])\right]
\nonumber\\
=& \int \!\mathrm{d}\mathbf{z}_\tau \rho_0(\mathbf{z}_\tau) \exp
\left[\int_{0}^{\tau}\!\!\!\! \mathrm{d}s \, u_s
 B(\varepsilon\varphi_{\tau-s,0}[\varepsilon
\mathbf{z}_\tau;\varepsilon_Q\widetilde \lambda])\right]
\nonumber\\
=&\int \! \mathrm{d}\mathbf{z}_\tau' \rho_0(\mathbf{z}'_\tau)
\exp \left[\int_{0}^{\tau}\!\!\!\!\mathrm{d}r\, u_{\tau-r}
 \varepsilon_B
B(\varphi_{r,0}[\mathbf{z}'_\tau;\varepsilon_Q\widetilde
\lambda])\right] \, ,
\label{eq:BKderivation}
\end{align}
where the first equality provides an explicit expression for the
l.h.s. of Eq.  (\ref{eq:BK-gen-functional-identity}). In going from
the second to the third line we employed the expression of work in
Eq. (\ref{eq:ex-work}), the microreversibility principle
(\ref{eq:microreversibility}) and made the change of variable
$\mathbf{z}_0\rightarrow \mathbf{z}_\tau$. The Jacobian of this
transformation is unity, because the time evolution in classical
mechanics is a canonical transformation. A further change of
variables $\mathbf{z}_\tau\rightarrow\mathbf{z}'_\tau=\varepsilon
\mathbf{z}_\tau$, whose Jacobian is unity as well, and the change
$s\rightarrow r=\tau-s$, yields the expression in the last line,
that coincides with the right hand side of Eq.
(\ref{eq:BK-gen-functional-identity}). In the last line we used the
property $\rho_0(\mathbf{z})=\rho_0(\varepsilon \mathbf z)$,
inherited by $\rho_0=e^{-\beta H_0}/Z_0$ from the assumed time
reversal invariance of the Hamiltonian, $H(\mathbf{z})=H(\varepsilon
\mathbf z)$.

\section{Quantum microreversibility}
\label{app:q-microreversibility}

In order to prove the quantum principle of microreversibility, we
first discretize time and express the time evolution operator
$U_{t,0}[\widetilde \lambda]$ as a time ordered product
\cite{Schleich01Book}
\begin{equation}
U_{\tau-t,0}[\widetilde \lambda]= \lim_{N\rightarrow \infty}
 e^{-\frac{i}{\hbar} \mathcal{H} (\widetilde \lambda_{\tau-N\varepsilon})\varepsilon}
    \dots e^{-\frac{i}{\hbar} \mathcal{H} (\widetilde \lambda_{\varepsilon})\varepsilon} e^{-\frac{i}{\hbar}
\mathcal{H} (\widetilde \lambda_0)\varepsilon}\, ,
\end{equation}
where $\varepsilon = t/N$ denotes the time step.
Using Eq. (\ref{eq:lambda-tilde}), we obtain
\begin{equation}
U_{\tau-t,0}[\widetilde \lambda]= \lim_{N\rightarrow \infty}
 e^{-\frac{i}{\hbar} \mathcal{H} (\lambda_{N\varepsilon})\varepsilon}
    \dots e^{-\frac{i}{\hbar} \mathcal{H} (\lambda_{\tau-\varepsilon})\varepsilon} e^{-\frac{i}{\hbar}
\mathcal{H} (\lambda_\tau)\varepsilon}\, .
\end{equation}
Thus:
\begin{align}
&\Theta^{\dagger}U_{\tau-t,0}[\widetilde \lambda] \Theta= \\
&\lim_{N\rightarrow \infty}
  \Theta^{\dagger} e^{-\frac{i}{\hbar} \mathcal{H} ( \lambda_{N\varepsilon})\varepsilon} \Theta
\Theta^{\dagger}  e^{-\frac{i}{\hbar} \mathcal{H}(\lambda_{N\varepsilon+\varepsilon})\varepsilon}
\Theta \dots \Theta^{\dagger} e^{-\frac{i}{\hbar} \mathcal{H}
(\lambda_{\tau})\varepsilon} \Theta\, ,\nonumber
\end{align}
where we inserted $\Theta  \Theta^{\dagger}={\mathbb 1}$, $N-1$
times. Assuming that $\mathcal \mathcal{H}(\lambda_t)$ commutes at
all times with the time reversal operator $\Theta$, Eq.
(\ref{eq:[Theta-H]}), we find
\begin{equation}
  \Theta^{\dagger} e^{-\frac{i}{\hbar} \mathcal{H}(\lambda_t)u}\Theta =e^{\frac{i}{\hbar}
\mathcal{H}(\lambda_t)u^*} \, ,
\label{eq:Theta-e^iH-Theta}
\end{equation}
for any complex number $u$. Using this equation and the fact that
$\varepsilon$ is real-valued, $\varepsilon^*=\varepsilon$, we obtain Eq.
(\ref{eq:q-microreversibility}):
\begin{align}
 \Theta^{\dagger}U_{\tau-t,0}[\widetilde \lambda] \Theta  &=
\lim_{N\rightarrow
  \infty}  e^{\frac{i}{\hbar} \mathcal{H} (\lambda_{N\varepsilon})\varepsilon} e^{\frac{i}{\hbar}
\mathcal{H}(\lambda_{N\varepsilon+\varepsilon})\varepsilon}  \dots e^{\frac{i}{\hbar} \mathcal{H}
  (\lambda_{\tau})\varepsilon} \nonumber\\
=  \lim_{N\rightarrow \infty}&\left[ e^{-\frac{i}{\hbar} \mathcal{H}
(\lambda_{\tau})\varepsilon} \dots e^{-\frac{i}{\hbar}
    \mathcal{H}(\lambda_{N\varepsilon+\varepsilon})\varepsilon} e^{-\frac{i}{\hbar} \mathcal{H}
(\lambda_{N\varepsilon})\varepsilon}\right]^\dagger   \nonumber \\
&= U_{\tau,t}^\dagger[\lambda]  = U_{t,\tau}[\lambda] \, .
\end{align}

\section{Tasaki-Crooks relation for the characteristic function}
\label{app:Z0G=ZfG}
From Eq. (\ref{eq:G[u;lambda]}) we have
\begin{equation}
 \mathcal Z(\lambda_0) G[u;\lambda]=\Tr \,U^{\dagger}_{\tau,0}[\lambda]
e^{i u \mathcal H(\lambda_\tau)}
  U_{\tau,0}[\lambda] e^{-i u \mathcal H(\lambda_0)}
e^{-\beta \mathcal H(\lambda_0)}.
\end{equation}
For $t=0$, the microreversibility principle, Eq.
(\ref{eq:q-microreversibility}), becomes
$U_{0,\tau}[\lambda]=U_{\tau,0}^\dagger[\lambda]=\Theta^\dagger
U_{\tau,0}[\widetilde \lambda]\Theta $. Therefore,
\begin{align}
& \mathcal Z(\lambda_0)G[u;\lambda]= \nonumber \\
& \Tr \,\Theta^\dagger U_{\tau,0}[\widetilde \lambda]
  \Theta e^{i u \mathcal H(\lambda_\tau)} \Theta^\dagger
U^{\dagger}_{\tau,0}[\widetilde\lambda] \Theta e^{-i u \mathcal H(\lambda_0)}
e^{-\beta \mathcal H(\lambda_0)} \Theta ^\dagger\Theta\;,
\end{align}
where we inserted
$\Theta^\dagger\Theta=\mathbb{1}$ under the trace.
Using Eq. (\ref{eq:Theta-e^iH-Theta}), we obtain
\begin{align}
 \mathcal Z(\lambda_0) &G[u;\lambda]=\\
&\Tr \,\Theta^\dagger U_{\tau,0}[\widetilde\lambda]
  e^{-i u^* \mathcal H(\lambda_\tau)}
U_{\tau,0}^{\dagger}[\widetilde\lambda] e^{i u^* \mathcal H(\lambda_0)}
e^{-\beta
\mathcal H(\lambda_0)}\Theta.\nonumber
\end{align}
The anti-linearity of $\Theta$ implies, for any trace class operator
$\mathcal A$: $ \Tr \,\Theta^\dagger \mathcal A \Theta = \Tr \,
\mathcal A^{\dagger} \, . $ Using this we can write
\begin{equation}
\mathcal Z(\lambda_0) G[u;\lambda]=\Tr \,e^{-\beta \mathcal
H(\lambda_0)}  e^{-i u \mathcal H(\lambda_0)}
  U_{\tau,0}[\widetilde\lambda]  e^{i u \mathcal
H(\lambda_\tau)}   U^{\dagger}_{\tau,0}[\widetilde\lambda] \; .
\end{equation}
Using the cyclic property of the trace one then obtains the
important result
\begin{align}
\mathcal Z(\lambda_0) G[u;\lambda]& \nonumber \\
=\Tr \, U^{\dagger}_{\tau,0}[\widetilde\lambda]&e^{i(-u+i\beta)
    \mathcal{H}(\lambda_0)}
U_{\tau,0}[\widetilde\lambda] e^{-i(-u+i\beta)
\mathcal{H}(\lambda_0)}  e^{-\beta \mathcal{H}(\lambda_0)}
\nonumber\\
=&\mathcal Z(\lambda_\tau) G[-u+i\beta;\widetilde \lambda]\;
.
\end{align}

\bibliographystyle{apsrmplong}

\begin{thebibliography}{131}
\expandafter\ifx\csname natexlab\endcsname\relax\def\natexlab#1{#1}\fi
\expandafter\ifx\csname bibnamefont\endcsname\relax
  \def\bibnamefont#1{#1}\fi
\expandafter\ifx\csname bibfnamefont\endcsname\relax
  \def\bibfnamefont#1{#1}\fi
\expandafter\ifx\csname citenamefont\endcsname\relax
  \def\citenamefont#1{#1}\fi
\expandafter\ifx\csname url\endcsname\relax
  \def\url#1{\texttt{#1}}\fi
\expandafter\ifx\csname urlprefix\endcsname\relax\def\urlprefix{URL }\fi
\providecommand{\bibinfo}[2]{#2}
\providecommand{\eprint}[2][]{\url{#2}}

\bibitem[{\citenamefont{Adib}(2005)}]{Adib05PRE71}
\bibinfo{author}{\bibnamefont{Adib}, \bibfnamefont{A.~B.}},
  \bibinfo{year}{2005}, {``}\bibinfo{title}{Entropy and density of states from
  isoenergetic nonequilibrium processes},{''} \bibinfo{journal}{Phys. Rev. E}
  \textbf{\bibinfo{volume}{71}},  \bibinfo{pages}{056128}.

\bibitem[{\citenamefont{Adib}(2009)}]{Adib09JCP130}
\bibinfo{author}{\bibnamefont{Adib}, \bibfnamefont{A.~B.}},
  \bibinfo{year}{2009}, {``}\bibinfo{title}{Comment on ``On the Crooks
  fluctuation theorem and the Jarzynski equality'' [J. Chem. Phys.
  \textbf{129}, 091101 (2008)]},{''} \bibinfo{journal}{J. Chem. Phys.}
  \textbf{\bibinfo{volume}{130}},  \bibinfo{pages}{247101}.

\bibitem[{\citenamefont{Allahverdyan and
  Nieuwenhuizen}(2005)}]{Allahverdyan05PRE71}
\bibinfo{author}{\bibnamefont{Allahverdyan}, \bibfnamefont{A.~E.}}, and
  \bibinfo{author}{\bibfnamefont{T.~M.} \bibnamefont{Nieuwenhuizen}},
  \bibinfo{year}{2005}, {``}\bibinfo{title}{Fluctuations of work from quantum
  subensembles: The case against quantum work-fluctuation theorems},{''}
  \bibinfo{journal}{Phys. Rev. E} \textbf{\bibinfo{volume}{71}},
  \bibinfo{pages}{066102}.

\bibitem[{\citenamefont{Andrieux and Gaspard}(2008)}]{Andrieux08PRL100}
\bibinfo{author}{\bibnamefont{Andrieux}, \bibfnamefont{D.}}, and
  \bibinfo{author}{\bibfnamefont{P.}~\bibnamefont{Gaspard}},
  \bibinfo{year}{2008}, {``}\bibinfo{title}{Quantum Work Relations and Response
  Theory},{''} \bibinfo{journal}{Phys. Rev. Lett.}
  \textbf{\bibinfo{volume}{100}},  \bibinfo{pages}{230404}.

\bibitem[{\citenamefont{Andrieux} \emph{et~al.}(2009)\citenamefont{Andrieux,
  Gaspard, Monnai, and Tasaki}}]{Andrieux09NJP11}
\bibinfo{author}{\bibnamefont{Andrieux}, \bibfnamefont{D.}},
  \bibinfo{author}{\bibfnamefont{P.}~\bibnamefont{Gaspard}},
  \bibinfo{author}{\bibfnamefont{T.}~\bibnamefont{Monnai}}, and
  \bibinfo{author}{\bibfnamefont{S.}~\bibnamefont{Tasaki}},
  \bibinfo{year}{2009}, {``}\bibinfo{title}{The fluctuation theorem for
  currents in open quantum systems},{''} \bibinfo{journal}{New J. Phys.}
  \textbf{\bibinfo{volume}{11}},  \bibinfo{pages}{043014}.

\bibitem[{\citenamefont{{Anetsberger}}
  \emph{et~al.}(2009)\citenamefont{{Anetsberger}, {Arcizet}, {Unterreithmeier},
  {Rivi{\`e}re}, {Schliesser}, {Weig}, {Kotthaus}, and
  {Kippenberg}}}]{Anetsberger09NATPHYS5}
\bibinfo{author}{\bibnamefont{{Anetsberger}}, \bibfnamefont{G.}},
  \bibinfo{author}{\bibfnamefont{O.}~\bibnamefont{{Arcizet}}},
  \bibinfo{author}{\bibfnamefont{Q.~P.} \bibnamefont{{Unterreithmeier}}},
  \bibinfo{author}{\bibfnamefont{R.}~\bibnamefont{{Rivi{\`e}re}}},
  \bibinfo{author}{\bibfnamefont{A.}~\bibnamefont{{Schliesser}}},
  \bibinfo{author}{\bibfnamefont{E.~M.} \bibnamefont{{Weig}}},
  \bibinfo{author}{\bibfnamefont{J.~P.} \bibnamefont{{Kotthaus}}}, and
  \bibinfo{author}{\bibfnamefont{T.~J.} \bibnamefont{{Kippenberg}}},
  \bibinfo{year}{2009}, {``}\bibinfo{title}{Near-field cavity optomechanics
  with nanomechanical oscillators},{''} \bibinfo{journal}{Nature Physics}
  \textbf{\bibinfo{volume}{5}},  \bibinfo{pages}{909--914}.

\bibitem[{\citenamefont{Bernard and Callen}(1959)}]{Bernard59RMP31}
\bibinfo{author}{\bibnamefont{Bernard}, \bibfnamefont{W.}}, and
  \bibinfo{author}{\bibfnamefont{H.~B.} \bibnamefont{Callen}},
  \bibinfo{year}{1959}, {``}\bibinfo{title}{Irreversible Thermodynamics of
  Nonlinear Processes and Noise in Driven Systems},{''} \bibinfo{journal}{Rev.
  Mod. Phys.} \textbf{\bibinfo{volume}{31}},  \bibinfo{pages}{1017--1044}.

\bibitem[{\citenamefont{{Bochkov} and {Kuzovlev}}(1977)}]{Bochkov77SPJETP45}
\bibinfo{author}{\bibnamefont{{Bochkov}}, \bibfnamefont{G.~N.}}, and
  \bibinfo{author}{\bibfnamefont{Y.~E.} \bibnamefont{{Kuzovlev}}},
  \bibinfo{year}{1977}, {``}\bibinfo{title}{General theory of thermal
  fluctuations in nonlinear systems},{''} \bibinfo{journal}{Zh. Eksp. Teor.
  Fiz.} \textbf{\bibinfo{volume}{72}},  \bibinfo{pages}{238--247}
  \bibinfo{note}{[Sov. Phys. JETP \textbf{45}, 125--130 (1977)]}.

\bibitem[{\citenamefont{Bochkov and Kuzovlev}(1978)}]{Bochkov78RQE21}
\bibinfo{author}{\bibnamefont{Bochkov}, \bibfnamefont{G.~N.}}, and
  \bibinfo{author}{\bibfnamefont{Y.~E.} \bibnamefont{Kuzovlev}},
  \bibinfo{year}{1978}, {``}\bibinfo{title}{Nonlinear stochastic models of
  oscillator systems},{''} \bibinfo{journal}{Radiophysics and Quantum
  Electronics} \textbf{\bibinfo{volume}{21}},  \bibinfo{pages}{1019--1032}.

\bibitem[{\citenamefont{{Bochkov} and {Kuzovlev}}(1979)}]{Bochkov79SPJETP49}
\bibinfo{author}{\bibnamefont{{Bochkov}}, \bibfnamefont{G.~N.}}, and
  \bibinfo{author}{\bibfnamefont{Y.~E.} \bibnamefont{{Kuzovlev}}},
  \bibinfo{year}{1979}, {``}\bibinfo{title}{Fluctuation-dissipation relations
  for nonequilibrium processes in open systems},{''} \bibinfo{journal}{Zh.
  Eksp. Teor. Fiz.} \textbf{\bibinfo{volume}{76}},  \bibinfo{pages}{1071--1088}
  \bibinfo{note}{[Sov. Phys. JETP \textbf{49}, 543--551 (1979)]}.

\bibitem[{\citenamefont{{Bochkov} and
  {Kuzovlev}}(1981{\natexlab{a}})}]{Bochkov81aPHYSA106}
\bibinfo{author}{\bibnamefont{{Bochkov}}, \bibfnamefont{G.~N.}}, and
  \bibinfo{author}{\bibfnamefont{Y.~E.} \bibnamefont{{Kuzovlev}}},
  \bibinfo{year}{1981}{\natexlab{a}}, {``}\bibinfo{title}{Nonlinear
  fluctuation-dissipation relations and stochastic models in nonequilibrium
  thermodynamics I. Generalized fluctuation-dissipation theorem},{''}
  \bibinfo{journal}{Physica A} \textbf{\bibinfo{volume}{106}},
  \bibinfo{pages}{443--479}.

\bibitem[{\citenamefont{{Bochkov} and
  {Kuzovlev}}(1981{\natexlab{b}})}]{Bochkov81bPHYSA106}
\bibinfo{author}{\bibnamefont{{Bochkov}}, \bibfnamefont{G.~N.}}, and
  \bibinfo{author}{\bibfnamefont{Y.~E.} \bibnamefont{{Kuzovlev}}},
  \bibinfo{year}{1981}{\natexlab{b}}, {``}\bibinfo{title}{Nonlinear
  fluctuation-dissipation relations and stochastic models in nonequilibrium
  thermodynamics II. Kinetic potential and variational principles for nonlinear
  irreversible processes},{''} \bibinfo{journal}{Physica A}
  \textbf{\bibinfo{volume}{106}},  \bibinfo{pages}{480--520}.

\bibitem[{\citenamefont{{Caldeira} and {Leggett}}(1983)}]{Caldeira83AP149}
\bibinfo{author}{\bibnamefont{{Caldeira}}, \bibfnamefont{A.~O.}}, and
  \bibinfo{author}{\bibfnamefont{A.~J.} \bibnamefont{{Leggett}}},
  \bibinfo{year}{1983}, {``}\bibinfo{title}{{Quantum tunnelling in a
  dissipative system}},{''} \bibinfo{journal}{Ann. Phys. (N.Y.)}
  \textbf{\bibinfo{volume}{149}},  \bibinfo{pages}{374--456}.

\bibitem[{\citenamefont{Callen and Welton}(1951)}]{Callen51PR83}
\bibinfo{author}{\bibnamefont{Callen}, \bibfnamefont{H.~B.}}, and
  \bibinfo{author}{\bibfnamefont{T.~A.} \bibnamefont{Welton}},
  \bibinfo{year}{1951}, {``}\bibinfo{title}{Irreversibility and Generalized
  Noise},{''} \bibinfo{journal}{Phys. Rev.} \textbf{\bibinfo{volume}{83}},
  \bibinfo{pages}{34--40}.

\bibitem[{\citenamefont{Campisi}(2008)}]{Campisi08PRE78b}
\bibinfo{author}{\bibnamefont{Campisi}, \bibfnamefont{M.}},
  \bibinfo{year}{2008}, {``}\bibinfo{title}{Increase of Boltzmann entropy in a
  quantum forced harmonic oscillator},{''} \bibinfo{journal}{Phys. Rev. E}
  \textbf{\bibinfo{volume}{78}},  \bibinfo{pages}{051123}.

\bibitem[{\citenamefont{Campisi}
  \emph{et~al.}(2009{\natexlab{a}})\citenamefont{Campisi, Talkner, and
  H{\"a}nggi}}]{Campisi09PRL102}
\bibinfo{author}{\bibnamefont{Campisi}, \bibfnamefont{M.}},
  \bibinfo{author}{\bibfnamefont{P.}~\bibnamefont{Talkner}}, and
  \bibinfo{author}{\bibfnamefont{P.}~\bibnamefont{H{\"a}nggi}},
  \bibinfo{year}{2009}{\natexlab{a}}, {``}\bibinfo{title}{Fluctuation Theorem
  for Arbitrary Open Quantum Systems},{''} \bibinfo{journal}{Phys. Rev. Lett.}
  \textbf{\bibinfo{volume}{102}},  \bibinfo{pages}{210401}.

\bibitem[{\citenamefont{Campisi}
  \emph{et~al.}(2009{\natexlab{b}})\citenamefont{Campisi, Talkner, and
  H{\"a}nggi}}]{Campisi09JPA42}
\bibinfo{author}{\bibnamefont{Campisi}, \bibfnamefont{M.}},
  \bibinfo{author}{\bibfnamefont{P.}~\bibnamefont{Talkner}}, and
  \bibinfo{author}{\bibfnamefont{P.}~\bibnamefont{H{\"a}nggi}},
  \bibinfo{year}{2009}{\natexlab{b}}, {``}\bibinfo{title}{Thermodynamics and
  fluctuation theorems for a strongly coupled open quantum system: an exactly
  solvable case},{''} \bibinfo{journal}{J. Phys. A: Math. Theo.}
  \textbf{\bibinfo{volume}{42}},  \bibinfo{pages}{392002}.

\bibitem[{\citenamefont{{Campisi}}
  \emph{et~al.}(2010{\natexlab{a}})\citenamefont{{Campisi}, {Talkner}, and
  {H{\"a}nggi}}}]{Campisi10PRL105}
\bibinfo{author}{\bibnamefont{{Campisi}}, \bibfnamefont{M.}},
  \bibinfo{author}{\bibfnamefont{P.}~\bibnamefont{{Talkner}}}, and
  \bibinfo{author}{\bibfnamefont{P.}~\bibnamefont{{H{\"a}nggi}}},
  \bibinfo{year}{2010}{\natexlab{a}}, {``}\bibinfo{title}{{Fluctuation theorems
  for continuously monitored quantum fluxes}},{''} \bibinfo{journal}{Phys. Rev.
  Lett.} \textbf{\bibinfo{volume}{105}},  \bibinfo{pages}{140601}.

\bibitem[{\citenamefont{{Campisi}}
  \emph{et~al.}(2011{\natexlab{a}})\citenamefont{{Campisi}, {Talkner}, and
  {H{\"a}nggi}}}]{Campisi10arXiv}
\bibinfo{author}{\bibnamefont{{Campisi}}, \bibfnamefont{M.}},
  \bibinfo{author}{\bibfnamefont{P.}~\bibnamefont{{Talkner}}}, and
  \bibinfo{author}{\bibfnamefont{P.}~\bibnamefont{{H{\"a}nggi}}},
  \bibinfo{year}{2011}{\natexlab{a}}, {``}\bibinfo{title}{{Quantum
  Bochkov-Kuzovlev Work Fluctuation Theorems}},{''} \bibinfo{journal}{Phil. Trans. R. Soc, A} \textbf{\bibinfo{volume}{369}},  \bibinfo{pages}{291-306}.

\bibitem[{\citenamefont{{Campisi}} \emph{et~al.}(2011{\natexlab{b}})\citenamefont{{Campisi},
  {Talkner}, and {Hanggi}}}]{Campisi11arXiv}
\bibinfo{author}{\bibnamefont{{Campisi}}, \bibfnamefont{M.}},
  \bibinfo{author}{\bibfnamefont{P.}~\bibnamefont{{Talkner}}}, and
  \bibinfo{author}{\bibfnamefont{P.}~\bibnamefont{{Hanggi}}},
  \bibinfo{year}{2011}{\natexlab{b}}, {``}\bibinfo{title}{Influence of measurements on the
  statistics of work performed on a quantum system},{''}
\bibinfo{journal}{Phys. Rev. E} \textbf{\bibinfo{volume}{83}},
  \bibinfo{pages}{041114}.

\bibitem[{\citenamefont{Campisi} \emph{et~al.}(2010)\citenamefont{Campisi,
  Zueco, and Talkner}}]{Campisi10CP375}
\bibinfo{author}{\bibnamefont{Campisi}, \bibfnamefont{M.}},
  \bibinfo{author}{\bibfnamefont{D.}~\bibnamefont{Zueco}}, and
  \bibinfo{author}{\bibfnamefont{P.}~\bibnamefont{Talkner}},
  \bibinfo{year}{2010}, {``}\bibinfo{title}{Thermodynamic anomalies in open
  quantum systems: Strong coupling effects in the isotropic XY model},{''}
  \bibinfo{journal}{Chem. Phys.} \textbf{\bibinfo{volume}{375}},
  \bibinfo{pages}{187--194}.

\bibitem[{\citenamefont{Casimir}(1945)}]{Casimir45RMP17}
\bibinfo{author}{\bibnamefont{Casimir}, \bibfnamefont{H.~B.~G.}},
  \bibinfo{year}{1945}, {``}\bibinfo{title}{On Onsager's Principle of
  Microscopic Reversibility},{''} \bibinfo{journal}{Rev. Mod. Phys.}
  \textbf{\bibinfo{volume}{17}},  \bibinfo{pages}{343--350}.

\bibitem[{\citenamefont{Chen}(2008{\natexlab{a}})}]{Chen08bJCP129}
\bibinfo{author}{\bibnamefont{Chen}, \bibfnamefont{L.~Y.}},
  \bibinfo{year}{2008}{\natexlab{a}}, {``}\bibinfo{title}{Nonequilibrium
  fluctuation-dissipation theorem of Brownian dynamics},{''}
  \bibinfo{journal}{J. Chem. Phys.} \textbf{\bibinfo{volume}{129}},
  \bibinfo{pages}{144113}.

\bibitem[{\citenamefont{Chen}(2008{\natexlab{b}})}]{Chen08aJCP129}
\bibinfo{author}{\bibnamefont{Chen}, \bibfnamefont{L.~Y.}},
  \bibinfo{year}{2008}{\natexlab{b}}, {``}\bibinfo{title}{On the Crooks
  fluctuation theorem and the Jarzynski equality},{''} \bibinfo{journal}{J.
  Chem. Phys.} \textbf{\bibinfo{volume}{129}},  \bibinfo{pages}{091101}.

\bibitem[{\citenamefont{Chen}(2009)}]{Chen09JCP130}
\bibinfo{author}{\bibnamefont{Chen}, \bibfnamefont{L.~Y.}},
  \bibinfo{year}{2009}, {``}\bibinfo{title}{Response to ``Comment on `On the
  Crooks fluctuation theorem and the Jaraynski equality' and `Nonequilibrium
  fluctuation dissipation theorem of Brownian dynamics' [J. Chem. Phys.
  \textbf{130}, 107101 (2009)]''},{''} \bibinfo{journal}{J. Chem. Phys.}
  \textbf{\bibinfo{volume}{130}},  \bibinfo{pages}{107102}.

\bibitem[{\citenamefont{Cleuren} \emph{et~al.}(2006)\citenamefont{Cleuren,
  Van~den Broeck, and Kawai}}]{Cleuren06PRL96}
\bibinfo{author}{\bibnamefont{Cleuren}, \bibfnamefont{B.}},
  \bibinfo{author}{\bibfnamefont{C.}~\bibnamefont{Van~den Broeck}}, and
  \bibinfo{author}{\bibfnamefont{R.}~\bibnamefont{Kawai}},
  \bibinfo{year}{2006}, {``}\bibinfo{title}{Fluctuation and Dissipation of Work
  in a Joule Experiment},{''} \bibinfo{journal}{Phys. Rev. Lett.}
  \textbf{\bibinfo{volume}{96}},  \bibinfo{pages}{050601}.

\bibitem[{\citenamefont{Cohen-Tannoudji}
  \emph{et~al.}(1977)\citenamefont{Cohen-Tannoudji, Diu, and
  Lalo{\"e}}}]{CohenTannoudji77Book}
\bibinfo{author}{\bibnamefont{Cohen-Tannoudji}, \bibfnamefont{C.}},
  \bibinfo{author}{\bibfnamefont{B.}~\bibnamefont{Diu}}, and
  \bibinfo{author}{\bibfnamefont{F.}~\bibnamefont{Lalo{\"e}}},
  \bibinfo{year}{1977}, \emph{\bibinfo{title}{Quantum Mechanics}}
  (\bibinfo{publisher}{Wiley}, \bibinfo{address}{New York}),
  \bibinfo{note}{vol. 1, p. 318}.

\bibitem[{\citenamefont{{Collin}} \emph{et~al.}(2005)\citenamefont{{Collin},
  {Ritort}, {Jarzynski}, {Smith}, {Tinoco}, and {Bustamante}}}]{Collin05NAT437}
\bibinfo{author}{\bibnamefont{{Collin}}, \bibfnamefont{D.}},
  \bibinfo{author}{\bibfnamefont{F.}~\bibnamefont{{Ritort}}},
  \bibinfo{author}{\bibfnamefont{C.}~\bibnamefont{{Jarzynski}}},
  \bibinfo{author}{\bibfnamefont{S.~B.} \bibnamefont{{Smith}}},
  \bibinfo{author}{\bibfnamefont{I.}~\bibnamefont{{Tinoco}}}, and
  \bibinfo{author}{\bibfnamefont{C.}~\bibnamefont{{Bustamante}}},
  \bibinfo{year}{2005}, {``}\bibinfo{title}{{Verification of the Crooks
  fluctuation theorem and recovery of RNA folding free energies}},{''}
  \bibinfo{journal}{Nature} \textbf{\bibinfo{volume}{437}},
  \bibinfo{pages}{231--234}.

\bibitem[{\citenamefont{Crooks}(1999)}]{Crooks99PRE60}
\bibinfo{author}{\bibnamefont{Crooks}, \bibfnamefont{G.~E.}},
  \bibinfo{year}{1999}, {``}\bibinfo{title}{Entropy production fluctuation
  theorem and the nonequilibrium work relation for free energy
  differences},{''} \bibinfo{journal}{Phys. Rev. E}
  \textbf{\bibinfo{volume}{60}},  \bibinfo{pages}{2721--2726}.

\bibitem[{\citenamefont{Crooks}(2008)}]{Crooks08JSM08}
\bibinfo{author}{\bibnamefont{Crooks}, \bibfnamefont{G.~E.}},
  \bibinfo{year}{2008}, {``}\bibinfo{title}{On the Jarzynski relation for
  dissipative quantum dynamics},{''} \bibinfo{journal}{J. Stat. Mech.: Theory
  Exp.}  \bibinfo{pages}{P10023}.

\bibitem[{\citenamefont{Crooks}(2009)}]{Crooks09JCP130}
\bibinfo{author}{\bibnamefont{Crooks}, \bibfnamefont{G.~E.}},
  \bibinfo{year}{2009}, {``}\bibinfo{title}{Comment regarding ``On the Crooks
  fluctuation theorem and the Jarzynski equality'' [J. Chem. Phys.
  \textbf{129}, 091101 (2008)] and ``Nonequilibrium fluctuation-dissipation
  theorem of Brownian dynamics'' [J. Chem. Phys. \textbf{129}, 144113
  (2008)]},{''} \bibinfo{journal}{J. Chem. Phys.}
  \textbf{\bibinfo{volume}{130}},  \bibinfo{pages}{107101}.

\bibitem[{\citenamefont{{de Groot} and Mazur}(1984)}]{deGroot84Book}
\bibinfo{author}{\bibnamefont{{de Groot}}, \bibfnamefont{S.~R.}}, and
  \bibinfo{author}{\bibfnamefont{P.}~\bibnamefont{Mazur}},
  \bibinfo{year}{1984}, \emph{\bibinfo{title}{Non-equilibrium Thermodynamics}}
  (\bibinfo{publisher}{Dover}, \bibinfo{address}{New York}).

\bibitem[{\citenamefont{{de Roeck} and Maes}(2004)}]{DeRoeck04PRE69}
\bibinfo{author}{\bibnamefont{{de Roeck}}, \bibfnamefont{W.}}, and
  \bibinfo{author}{\bibfnamefont{C.}~\bibnamefont{Maes}}, \bibinfo{year}{2004},
  {``}\bibinfo{title}{Quantum version of free-energy--irreversible-work
  relations},{''} \bibinfo{journal}{Phys. Rev. E}
  \textbf{\bibinfo{volume}{69}},  \bibinfo{pages}{026115}.

\bibitem[{\citenamefont{Deffner} \emph{et~al.}(2010)\citenamefont{Deffner,
  Abah, and Lutz}}]{Deffner10CP375}
\bibinfo{author}{\bibnamefont{Deffner}, \bibfnamefont{S.}},
  \bibinfo{author}{\bibfnamefont{O.}~\bibnamefont{Abah}}, and
  \bibinfo{author}{\bibfnamefont{E.}~\bibnamefont{Lutz}}, \bibinfo{year}{2010},
  {``}\bibinfo{title}{Quantum work statistics of linear and nonlinear
  parametric oscillators},{''} \bibinfo{journal}{Chem. Phys.}
  \textbf{\bibinfo{volume}{375}},  \bibinfo{pages}{200 -- 208}.

\bibitem[{\citenamefont{Deffner and Lutz}(2008)}]{Deffner08PRE77}
\bibinfo{author}{\bibnamefont{Deffner}, \bibfnamefont{S.}}, and
  \bibinfo{author}{\bibfnamefont{E.}~\bibnamefont{Lutz}}, \bibinfo{year}{2008},
  {``}\bibinfo{title}{Nonequilibrium work distribution of a quantum harmonic
  oscillator},{''} \bibinfo{journal}{Phys. Rev. E}
  \textbf{\bibinfo{volume}{77}},  \bibinfo{pages}{021128}.

\bibitem[{\citenamefont{Deffner and Lutz}(2010)}]{Deffner10PRL105}
\bibinfo{author}{\bibnamefont{Deffner}, \bibfnamefont{S.}}, and
  \bibinfo{author}{\bibfnamefont{E.}~\bibnamefont{Lutz}}, \bibinfo{year}{2010},
  {``}\bibinfo{title}{Generalized Clausius Inequality for Nonequilibrium
  Quantum Processes},{''} \bibinfo{journal}{Phys. Rev. Lett.}
  \textbf{\bibinfo{volume}{105}},  \bibinfo{pages}{170402}.

\bibitem[{\citenamefont{Dittrich} \emph{et~al.}(1998)\citenamefont{Dittrich,
  H{\"a}nggi, Ingold, Kramer, Sch{\"o}n, and Zwerger}}]{Hanggi98Book}
\bibinfo{author}{\bibnamefont{Dittrich}, \bibfnamefont{T.}},
  \bibinfo{author}{\bibfnamefont{P.}~\bibnamefont{H{\"a}nggi}},
  \bibinfo{author}{\bibfnamefont{G.-L.} \bibnamefont{Ingold}},
  \bibinfo{author}{\bibfnamefont{B.}~\bibnamefont{Kramer}},
  \bibinfo{author}{\bibfnamefont{G.}~\bibnamefont{Sch{\"o}n}}, and
  \bibinfo{author}{\bibfnamefont{W.}~\bibnamefont{Zwerger}},
  \bibinfo{year}{1998}, \emph{\bibinfo{title}{Quantum Transport and
  Dissipation}} (\bibinfo{publisher}{Wiley-VCH}, \bibinfo{address}{Weinheim}).

\bibitem[{\citenamefont{Douarche} \emph{et~al.}(2005)\citenamefont{Douarche,
  Ciliberto, Petrosyan, and Rabbiosi}}]{Douarche05EPL70}
\bibinfo{author}{\bibnamefont{Douarche}, \bibfnamefont{F.}},
  \bibinfo{author}{\bibfnamefont{S.}~\bibnamefont{Ciliberto}},
  \bibinfo{author}{\bibfnamefont{A.}~\bibnamefont{Petrosyan}}, and
  \bibinfo{author}{\bibfnamefont{I.}~\bibnamefont{Rabbiosi}},
  \bibinfo{year}{2005}, {``}\bibinfo{title}{An experimental test of the
  Jarzynski equality in a mechanical experiment},{''}
  \bibinfo{journal}{Europhys. Lett.} \textbf{\bibinfo{volume}{70}},
  \bibinfo{pages}{593--599}.

\bibitem[{\citenamefont{Efremov}(1969)}]{Efremov69ZETF55}
\bibinfo{author}{\bibnamefont{Efremov}, \bibfnamefont{G.~F.}},
  \bibinfo{year}{1969}, {``}\bibinfo{title}{A fluctuation dissipation theorem
  for nonlinear media},{''} \bibinfo{journal}{Zh. Eksp. Teor. Fiz.}
  \textbf{\bibinfo{volume}{55}},  \bibinfo{pages}{2322} \bibinfo{note}{[Sov.
  Phys. JETP, \textbf{28}, 1232 (1969)]}.

\bibitem[{\citenamefont{Einstein}(1905)}]{Einstein05AP17}
\bibinfo{author}{\bibnamefont{Einstein}, \bibfnamefont{A.}},
  \bibinfo{year}{1905}, {``}\bibinfo{title}{{\"U}ber die von der
  molekularkinetischen Theorie der W{\"a}rme geforderte Bewegung von in
  ruhenden Fl{\"u}ssigkeiten suspendierten Teilchen},{''}
  \bibinfo{journal}{Ann. Phys.} \textbf{\bibinfo{volume}{17}},
  \bibinfo{pages}{549--560} \bibinfo{note}{, translated into English in
  \cite{Einstein26Book}}.

\bibitem[{\citenamefont{Einstein}(1906{\natexlab{a}})}]{Einstein06AP19a}
\bibinfo{author}{\bibnamefont{Einstein}, \bibfnamefont{A.}},
  \bibinfo{year}{1906}{\natexlab{a}}, {``}\bibinfo{title}{Eine neue Bestimmung
  der Molek{\"u}ldimensionen},{''} \bibinfo{journal}{Ann. Phys.}
  \textbf{\bibinfo{volume}{19}},  \bibinfo{pages}{289--306} \bibinfo{note}{,
  translated into English in \cite{Einstein26Book}}.

\bibitem[{\citenamefont{Einstein}(1906{\natexlab{b}})}]{Einstein06AP19b}
\bibinfo{author}{\bibnamefont{Einstein}, \bibfnamefont{A.}},
  \bibinfo{year}{1906}{\natexlab{b}}, {``}\bibinfo{title}{Zur Theorie der
  Brownschen Bewegung},{''} \bibinfo{journal}{Ann. Phys.}
  \textbf{\bibinfo{volume}{19}},  \bibinfo{pages}{371--381} \bibinfo{note}{,
  translated into English in \cite{Einstein26Book}}.

\bibitem[{\citenamefont{Einstein}(1926)}]{Einstein26Book}
\bibinfo{author}{\bibnamefont{Einstein}, \bibfnamefont{A.}},
  \bibinfo{year}{1926}, \emph{\bibinfo{title}{Investigations on the Theory of
  Brownian Movement}} (\bibinfo{publisher}{Methuen},
  \bibinfo{address}{London}), \bibinfo{note}{reprinted (Dover, New York),
  1956}.

\bibitem[{\citenamefont{Esposito} \emph{et~al.}(2009)\citenamefont{Esposito,
  Harbola, and Mukamel}}]{Esposito09RMP81}
\bibinfo{author}{\bibnamefont{Esposito}, \bibfnamefont{M.}},
  \bibinfo{author}{\bibfnamefont{U.}~\bibnamefont{Harbola}}, and
  \bibinfo{author}{\bibfnamefont{S.}~\bibnamefont{Mukamel}},
  \bibinfo{year}{2009}, {``}\bibinfo{title}{Nonequilibrium fluctuations,
  fluctuation theorems, and counting statistics in quantum systems},{''}
  \bibinfo{journal}{Rev. Mod. Phys.} \textbf{\bibinfo{volume}{81}},
  \bibinfo{pages}{1665--1702}.

\bibitem[{\citenamefont{Esposito and Mukamel}(2006)}]{Esposito06PRE73}
\bibinfo{author}{\bibnamefont{Esposito}, \bibfnamefont{M.}}, and
  \bibinfo{author}{\bibfnamefont{S.}~\bibnamefont{Mukamel}},
  \bibinfo{year}{2006}, {``}\bibinfo{title}{Fluctuation theorems for quantum
  master equations},{''} \bibinfo{journal}{Phys. Rev. E}
  \textbf{\bibinfo{volume}{73}},  \bibinfo{pages}{046129}.

\bibitem[{\citenamefont{Evans} \emph{et~al.}(1993)\citenamefont{Evans, Cohen,
  and Morriss}}]{Evans93PRL71}
\bibinfo{author}{\bibnamefont{Evans}, \bibfnamefont{D.~J.}},
  \bibinfo{author}{\bibfnamefont{E.~G.~D.} \bibnamefont{Cohen}}, and
  \bibinfo{author}{\bibfnamefont{G.~P.} \bibnamefont{Morriss}},
  \bibinfo{year}{1993}, {``}\bibinfo{title}{Probability of second law
  violations in shearing steady states},{''} \bibinfo{journal}{Phys. Rev.
  Lett.} \textbf{\bibinfo{volume}{71}},  \bibinfo{pages}{2401--2404}.

\bibitem[{\citenamefont{{Feynman} and {Vernon}}(1963)}]{Feynman63AP24}
\bibinfo{author}{\bibnamefont{{Feynman}}, \bibfnamefont{R.~P.}}, and
  \bibinfo{author}{\bibfnamefont{J.~F.~L.} \bibnamefont{{Vernon}}},
  \bibinfo{year}{1963}, {``}\bibinfo{title}{{The theory of a general quantum
  system interacting with a linear dissipative system}},{''}
  \bibinfo{journal}{Ann. Phys. (N.Y.)} \textbf{\bibinfo{volume}{24}},
  \bibinfo{pages}{118--173}.

\bibitem[{\citenamefont{Ford} \emph{et~al.}(1985)\citenamefont{Ford, Lewis, and
  O'Connell}}]{Ford85PRL21}
\bibinfo{author}{\bibnamefont{Ford}, \bibfnamefont{G.~W.}},
  \bibinfo{author}{\bibfnamefont{J.~T.} \bibnamefont{Lewis}}, and
  \bibinfo{author}{\bibfnamefont{R.~F.} \bibnamefont{O'Connell}},
  \bibinfo{year}{1985}, {``}\bibinfo{title}{Quantum oscillator in a blackbody
  radiation field},{''} \bibinfo{journal}{Phys. Rev. Lett.}
  \textbf{\bibinfo{volume}{55}},  \bibinfo{pages}{2273--2276}.

\bibitem[{\citenamefont{Fujisawa} \emph{et~al.}(2006)\citenamefont{Fujisawa,
  Hayashi, Tomita, and Hirayama}}]{Fujisawa06SCIENCE312}
\bibinfo{author}{\bibnamefont{Fujisawa}, \bibfnamefont{T.}},
  \bibinfo{author}{\bibfnamefont{T.}~\bibnamefont{Hayashi}},
  \bibinfo{author}{\bibfnamefont{R.}~\bibnamefont{Tomita}}, and
  \bibinfo{author}{\bibfnamefont{Y.}~\bibnamefont{Hirayama}},
  \bibinfo{year}{2006}, {``}\bibinfo{title}{{Bidirectional counting of single
  electrons}},{''} \bibinfo{journal}{Science} \textbf{\bibinfo{volume}{312}},
  \bibinfo{pages}{1634--1636}.

\bibitem[{\citenamefont{Gallavotti and Cohen}(1995)}]{Gallavotti95PRL74}
\bibinfo{author}{\bibnamefont{Gallavotti}, \bibfnamefont{G.}}, and
  \bibinfo{author}{\bibfnamefont{E.~G.~D.} \bibnamefont{Cohen}},
  \bibinfo{year}{1995}, {``}\bibinfo{title}{Dynamical Ensembles in
  Nonequilibrium Statistical Mechanics},{''} \bibinfo{journal}{Phys. Rev.
  Lett.} \textbf{\bibinfo{volume}{74}},
  \bibinfo{pages}{2694--2697}.

\bibitem[{\citenamefont{{Grabert}} \emph{et~al.}(1988)\citenamefont{{Grabert},
  {Schramm}, and {Ingold}}}]{Grabert88PREP168}
\bibinfo{author}{\bibnamefont{{Grabert}}, \bibfnamefont{H.}},
  \bibinfo{author}{\bibfnamefont{P.}~\bibnamefont{{Schramm}}}, and
  \bibinfo{author}{\bibfnamefont{G.-L.} \bibnamefont{{Ingold}}},
  \bibinfo{year}{1988}, {``}\bibinfo{title}{{Quantum Brownian motion: The
  functional integral approach}},{''} \bibinfo{journal}{Phys. Rep.}
  \textbf{\bibinfo{volume}{168}},  \bibinfo{pages}{115--207}.

\bibitem[{\citenamefont{{Grabert}} \emph{et~al.}(1984)\citenamefont{{Grabert},
  {Weiss}, and {Talkner}}}]{Grabert84ZPB55}
\bibinfo{author}{\bibnamefont{{Grabert}}, \bibfnamefont{H.}},
  \bibinfo{author}{\bibfnamefont{U.}~\bibnamefont{{Weiss}}}, and
  \bibinfo{author}{\bibfnamefont{P.}~\bibnamefont{{Talkner}}},
  \bibinfo{year}{1984}, {``}\bibinfo{title}{{Quantum theory of the damped
  harmonic oscillator}},{''} \bibinfo{journal}{Z. Phys. B}
  \textbf{\bibinfo{volume}{55}},  \bibinfo{pages}{87--94}.

\bibitem[{\citenamefont{Green}(1952)}]{Green52JCP20}
\bibinfo{author}{\bibnamefont{Green}, \bibfnamefont{M.~S.}},
  \bibinfo{year}{1952}, {``}\bibinfo{title}{Markoff Random Processes and the
  Statistical Mechanics of Time-Dependent Phenomena},{''} \bibinfo{journal}{J.
  Chem. Phys.} \textbf{\bibinfo{volume}{20}},  \bibinfo{pages}{1281--1295}.

\bibitem[{\citenamefont{Green}(1954)}]{Green54JCP22}
\bibinfo{author}{\bibnamefont{Green}, \bibfnamefont{M.~S.}},
  \bibinfo{year}{1954}, {``}\bibinfo{title}{Markov Random Processes and the
  Statistical Mechanics of Time-Dependent Phenomena.II. Irreversible Processes
  in Fluids},{''} \bibinfo{journal}{J. Chem. Phys.}
  \textbf{\bibinfo{volume}{22}},  \bibinfo{pages}{398--413}.

\bibitem[{\citenamefont{Hahn and Then}(2009)}]{Hahn09PRE79}
\bibinfo{author}{\bibnamefont{Hahn}, \bibfnamefont{A.~M.}}, and
  \bibinfo{author}{\bibfnamefont{H.}~\bibnamefont{Then}}, \bibinfo{year}{2009},
  {``}\bibinfo{title}{Using bijective maps to improve free-energy
  estimates},{''} \bibinfo{journal}{Phys. Rev. E}
  \textbf{\bibinfo{volume}{79}},  \bibinfo{pages}{011113}.

\bibitem[{\citenamefont{Hahn and Then}(2010)}]{Hahn10PRE81}
\bibinfo{author}{\bibnamefont{Hahn}, \bibfnamefont{A.~M.}}, and
  \bibinfo{author}{\bibfnamefont{H.}~\bibnamefont{Then}}, \bibinfo{year}{2010},
  {``}\bibinfo{title}{Measuring the convergence of Monte Carlo free-energy
  calculations},{''} \bibinfo{journal}{Phys. Rev. E}
  \textbf{\bibinfo{volume}{81}},  \bibinfo{pages}{041117}.

\bibitem[{\citenamefont{H\"anggi}(1978)}]{Hanggi78HPA51}
\bibinfo{author}{\bibnamefont{H\"anggi}, \bibfnamefont{P.}},
  \bibinfo{year}{1978}, {``}\bibinfo{title}{Stochastic Processes II: Response
  theory and fluctuation theorems},{''} \bibinfo{journal}{Helv. Phys. Acta}
  \textbf{\bibinfo{volume}{51}},  \bibinfo{pages}{202--219}.

\bibitem[{\citenamefont{H\"anggi}(1982)}]{Hanggi82PRA25}
\bibinfo{author}{\bibnamefont{H\"anggi}, \bibfnamefont{P.}},
  \bibinfo{year}{1982}, {``}\bibinfo{title}{Nonlinear fluctuations: The problem
  of deterministic limit and reconstruction of stochastic dynamics},{''}
  \bibinfo{journal}{Phys. Rev. A} \textbf{\bibinfo{volume}{25}},
  \bibinfo{pages}{1130--1136}.

\bibitem[{\citenamefont{H{\"a}nggi and Ingold}({2005})}]{Hanggi05CHAOS2}
\bibinfo{author}{\bibnamefont{H{\"a}nggi}, \bibfnamefont{P.}}, and
  \bibinfo{author}{\bibfnamefont{G.}~\bibnamefont{Ingold}},
  \bibinfo{year}{{2005}}, {``}\bibinfo{title}{{Fundamental aspects of quantum
  Brownian motion}},{''} \bibinfo{journal}{Chaos}
  \textbf{\bibinfo{volume}{{15}}},  \bibinfo{pages}{026105}.

\bibitem[{\citenamefont{H{\"a}nggi and Ingold}({2006})}]{Hanggi06APPB37}
\bibinfo{author}{\bibnamefont{H{\"a}nggi}, \bibfnamefont{P.}}, and
  \bibinfo{author}{\bibfnamefont{G.-L.} \bibnamefont{Ingold}},
  \bibinfo{year}{{2006}}, {``}\bibinfo{title}{{Quantum Brownian motion and the
  third law of thermodynamics}},{''} \bibinfo{journal}{Acta Phys. Pol. B}
  \textbf{\bibinfo{volume}{{37}}},  \bibinfo{pages}{1537}.

\bibitem[{\citenamefont{H{\"a}nggi and Thomas}(1982)}]{Hanggi82PREP88}
\bibinfo{author}{\bibnamefont{H{\"a}nggi}, \bibfnamefont{P.}}, and
  \bibinfo{author}{\bibfnamefont{H.}~\bibnamefont{Thomas}},
  \bibinfo{year}{1982}, {``}\bibinfo{title}{Stochastic processes: Time
  evolution, symmetries and linear response},{''} \bibinfo{journal}{Phys. Rep.}
  \textbf{\bibinfo{volume}{88}},  \bibinfo{pages}{207--319}.

\bibitem[{\citenamefont{Hide and Vedral}(2010)}]{Hide10PRA81}
\bibinfo{author}{\bibnamefont{Hide}, \bibfnamefont{J.}}, and
  \bibinfo{author}{\bibfnamefont{V.}~\bibnamefont{Vedral}},
  \bibinfo{year}{2010}, {``}\bibinfo{title}{Detecting entanglement with
  Jarzynski's equality},{''} \bibinfo{journal}{Phys. Rev. A}
  \textbf{\bibinfo{volume}{81}},  \bibinfo{pages}{062303}.

\bibitem[{\citenamefont{{Hofheinz}}
  \emph{et~al.}(2009)\citenamefont{{Hofheinz}, {Wang}, {Ansmann}, {Bialczak},
  {Lucero}, {Neeley}, {O'Connell}, {Sank}, {Wenner}, {Martinis}, and
  {Cleland}}}]{Hofheinz09NAT459}
\bibinfo{author}{\bibnamefont{{Hofheinz}}, \bibfnamefont{M.}},
  \bibinfo{author}{\bibfnamefont{H.}~\bibnamefont{{Wang}}},
  \bibinfo{author}{\bibfnamefont{M.}~\bibnamefont{{Ansmann}}},
  \bibinfo{author}{\bibfnamefont{R.~C.} \bibnamefont{{Bialczak}}},
  \bibinfo{author}{\bibfnamefont{E.}~\bibnamefont{{Lucero}}},
  \bibinfo{author}{\bibfnamefont{M.}~\bibnamefont{{Neeley}}},
  \bibinfo{author}{\bibfnamefont{A.~D.} \bibnamefont{{O'Connell}}},
  \bibinfo{author}{\bibfnamefont{D.}~\bibnamefont{{Sank}}},
  \bibinfo{author}{\bibfnamefont{J.}~\bibnamefont{{Wenner}}},
  \bibinfo{author}{\bibfnamefont{J.~M.} \bibnamefont{{Martinis}}}, and
  \bibinfo{author}{\bibfnamefont{A.~N.} \bibnamefont{{Cleland}}},
  \bibinfo{year}{2009}, {``}\bibinfo{title}{{Synthesizing arbitrary quantum
  states in a superconducting resonator}},{''} \bibinfo{journal}{Nature}
  \textbf{\bibinfo{volume}{459}},  \bibinfo{pages}{546--549}.

\bibitem[{\citenamefont{{Hofheinz}}
  \emph{et~al.}(2008)\citenamefont{{Hofheinz}, {Weig}, {Ansmann}, {Bialczak},
  {Lucero}, {Neeley}, {O'Connell}, {Wang}, {Martinis}, and
  {Cleland}}}]{Hofheinz08NAT454}
\bibinfo{author}{\bibnamefont{{Hofheinz}}, \bibfnamefont{M.}},
  \bibinfo{author}{\bibfnamefont{E.~M.} \bibnamefont{{Weig}}},
  \bibinfo{author}{\bibfnamefont{M.}~\bibnamefont{{Ansmann}}},
  \bibinfo{author}{\bibfnamefont{R.~C.} \bibnamefont{{Bialczak}}},
  \bibinfo{author}{\bibfnamefont{E.}~\bibnamefont{{Lucero}}},
  \bibinfo{author}{\bibfnamefont{M.}~\bibnamefont{{Neeley}}},
  \bibinfo{author}{\bibfnamefont{A.~D.} \bibnamefont{{O'Connell}}},
  \bibinfo{author}{\bibfnamefont{H.}~\bibnamefont{{Wang}}},
  \bibinfo{author}{\bibfnamefont{J.~M.} \bibnamefont{{Martinis}}}, and
  \bibinfo{author}{\bibfnamefont{A.~N.} \bibnamefont{{Cleland}}},
  \bibinfo{year}{2008}, {``}\bibinfo{title}{Generation of Fock states in a
  superconducting quantum circuit},{''} \bibinfo{journal}{Nature}
  \textbf{\bibinfo{volume}{454}},  \bibinfo{pages}{310--314}.

\bibitem[{\citenamefont{H{\"o}rhammer and
  B{\"u}ttner}(2008)}]{Hoerhammer08JSP133}
\bibinfo{author}{\bibnamefont{H{\"o}rhammer}, \bibfnamefont{C.}}, and
  \bibinfo{author}{\bibfnamefont{H.}~\bibnamefont{B{\"u}ttner}},
  \bibinfo{year}{2008}, {``}\bibinfo{title}{{Information and Entropy in Quantum
  Brownian Motion Thermodynamic Entropy versus von Neumann Entropy}},{''}
  \bibinfo{journal}{J. Stat. Phys.} \textbf{\bibinfo{volume}{{133}}},
  \bibinfo{pages}{1161--1174}.

\bibitem[{\citenamefont{Horowitz and Jarzynski}(2007)}]{Horowitz07JSM07}
\bibinfo{author}{\bibnamefont{Horowitz}, \bibfnamefont{J.}}, and
  \bibinfo{author}{\bibfnamefont{C.}~\bibnamefont{Jarzynski}},
  \bibinfo{year}{2007}, {``}\bibinfo{title}{Comparison of work fluctuation
  relations},{''} \bibinfo{journal}{J. Stat. Mech.: Theory Exp.}
  \bibinfo{pages}{P11002}.

\bibitem[{\citenamefont{Horowitz and Jarzynski}(2008)}]{Horowitz08PRL101}
\bibinfo{author}{\bibnamefont{Horowitz}, \bibfnamefont{J.}}, and
  \bibinfo{author}{\bibfnamefont{C.}~\bibnamefont{Jarzynski}},
  \bibinfo{year}{2008}, {``}\bibinfo{title}{Comment on ``Failure of the
  work-Hamiltonian connection for free-energy calculations''},{''}
  \bibinfo{journal}{Phys. Rev. Lett.} \textbf{\bibinfo{volume}{101}},
  \bibinfo{pages}{098901}.

\bibitem[{\citenamefont{Huber} \emph{et~al.}(2008)\citenamefont{Huber,
  Schmidt-Kaler, Deffner, and Lutz}}]{Huber08PRL101}
\bibinfo{author}{\bibnamefont{Huber}, \bibfnamefont{G.}},
  \bibinfo{author}{\bibfnamefont{F.}~\bibnamefont{Schmidt-Kaler}},
  \bibinfo{author}{\bibfnamefont{S.}~\bibnamefont{Deffner}}, and
  \bibinfo{author}{\bibfnamefont{E.}~\bibnamefont{Lutz}}, \bibinfo{year}{2008},
  {``}\bibinfo{title}{Employing trapped cold ions to verify the quantum
  Jarzynski equality},{''} \bibinfo{journal}{Phys. Rev. Lett.}
  \textbf{\bibinfo{volume}{101}},  \bibinfo{pages}{070403}.

\bibitem[{\citenamefont{Husimi}(1953)}]{Husimi53PTP9}
\bibinfo{author}{\bibnamefont{Husimi}, \bibfnamefont{K.}},
  \bibinfo{year}{1953}, {``}\bibinfo{title}{Miscellanea in Elementary Quantum
  Mechanics, I},{''} \bibinfo{journal}{Progress of Theoretical Physics}
  \textbf{\bibinfo{volume}{9}},  \bibinfo{pages}{238--244}.

\bibitem[{\citenamefont{Ingold}({2002})}]{Ingold02LNP611}
\bibinfo{author}{\bibnamefont{Ingold}, \bibfnamefont{G.-L.}},
  \bibinfo{year}{{2002}} \bibinfo{journal}{Lect. Notes Phys.}
  \textbf{\bibinfo{volume}{611}},  \bibinfo{pages}{1--53}.

\bibitem[{\citenamefont{Jarzynski}(1997)}]{Jarzynski97PRL78}
\bibinfo{author}{\bibnamefont{Jarzynski}, \bibfnamefont{C.}},
  \bibinfo{year}{1997}, {``}\bibinfo{title}{Nonequilibrium equality for free
  energy differences},{''} \bibinfo{journal}{Phys. Rev. Lett.}
  \textbf{\bibinfo{volume}{78}},  \bibinfo{pages}{2690--2693}.

\bibitem[{\citenamefont{Jarzynski}(2000)}]{Jarzynski00JSP98}
\bibinfo{author}{\bibnamefont{Jarzynski}, \bibfnamefont{C.}},
  \bibinfo{year}{2000}, {``}\bibinfo{title}{Hamiltonian derivation of a
  detailed fluctuation theorem},{''} \bibinfo{journal}{J. Stat. Phys.}
  \textbf{\bibinfo{volume}{98}},  \bibinfo{pages}{77--102}.

\bibitem[{\citenamefont{Jarzynski}(2002)}]{Jarzynski02PRE65}
\bibinfo{author}{\bibnamefont{Jarzynski}, \bibfnamefont{C.}},
  \bibinfo{year}{2002}, {``}\bibinfo{title}{Targeted free energy
  perturbation},{''} \bibinfo{journal}{Phys. Rev. E}
  \textbf{\bibinfo{volume}{65}},  \bibinfo{pages}{046122}.

\bibitem[{\citenamefont{Jarzynski}(2004)}]{Jarzynski04JSM04}
\bibinfo{author}{\bibnamefont{Jarzynski}, \bibfnamefont{C.}},
  \bibinfo{year}{2004}, {``}\bibinfo{title}{Nonequilibrium work theorem for a
  system strongly coupled to a thermal environment},{''} \bibinfo{journal}{J.
  Stat. Mech.: Theory Exp.}  \bibinfo{pages}{P09005}.

\bibitem[{\citenamefont{{Jarzynski}}(2007)}]{Jarzynski07CRPHYS8}
\bibinfo{author}{\bibnamefont{{Jarzynski}}, \bibfnamefont{C.}},
  \bibinfo{year}{2007}, {``}\bibinfo{title}{Comparison of far-from-equilibrium
  work relations},{''} \bibinfo{journal}{C. R. Phys.}
  \textbf{\bibinfo{volume}{8}},  \bibinfo{pages}{495--506}.

\bibitem[{\citenamefont{Jarzynski}(2008)}]{Jarzynski08EPJB64}
\bibinfo{author}{\bibnamefont{Jarzynski}, \bibfnamefont{C.}},
  \bibinfo{year}{2008}, {``}\bibinfo{title}{Nonequilibrium work relations:
  foundations and applications},{''} \bibinfo{journal}{Eur. Phys. J. B}
  \textbf{\bibinfo{volume}{64}},  \bibinfo{pages}{331--340}.

\bibitem[{\citenamefont{Jarzynski}(2011)}]{Jarzynski11ARCMP2}
\bibinfo{author}{\bibnamefont{Jarzynski}, \bibfnamefont{C.}},
  \bibinfo{year}{2011}, {``}\bibinfo{title}{Equalities and inequalities: Irreversibility and the second law of thermodynamics at the nanoscale},{''} \bibinfo{journal}{Annu. Rev. Condens. Matter Phys.}
  \textbf{\bibinfo{volume}{2}},  \bibinfo{pages}{329--351}.

\bibitem[{\citenamefont{Jarzynski and W{\'o}jcik}(2004)}]{Jarzynski04PRL92}
\bibinfo{author}{\bibnamefont{Jarzynski}, \bibfnamefont{C.}}, and
  \bibinfo{author}{\bibfnamefont{D.~K.} \bibnamefont{W{\'o}jcik}},
  \bibinfo{year}{2004}, {``}\bibinfo{title}{Classical and quantum fluctuation
  theorems for heat exchange},{''} \bibinfo{journal}{Phys. Rev. Lett.}
  \textbf{\bibinfo{volume}{92}},  \bibinfo{pages}{230602}.

\bibitem[{\citenamefont{Johnson}(1928)}]{Johnson28PR32}
\bibinfo{author}{\bibnamefont{Johnson}, \bibfnamefont{J.~B.}},
  \bibinfo{year}{1928}, {``}\bibinfo{title}{Thermal agitation of electricity in
  conductors},{''} \bibinfo{journal}{Phys. Rev.} \textbf{\bibinfo{volume}{32}},
   \bibinfo{pages}{97--109}.

\bibitem[{\citenamefont{{Katsuda} and {Ohzeki}}(2011)}]{Katsuda11arXiv}
\bibinfo{author}{\bibnamefont{{Katsuda}}, \bibfnamefont{H.}}, and
  \bibinfo{author}{\bibfnamefont{M.}~\bibnamefont{{Ohzeki}}},
  \bibinfo{year}{2011}, {``}\bibinfo{title}{{Jarzynski Equality for an
  Energy-Controlled System}},{''}
\bibinfo{journal}{J. Phys. Soc. Jpn.} \textbf{\bibinfo{volume}{80}},
   \bibinfo{pages}{045003}.


\bibitem[{\citenamefont{Kawai} \emph{et~al.}(2007)\citenamefont{Kawai,
  Parrondo, and den Broeck}}]{Kawai07PRL98}
\bibinfo{author}{\bibnamefont{Kawai}, \bibfnamefont{R.}},
  \bibinfo{author}{\bibfnamefont{J.~M.~R.} \bibnamefont{Parrondo}}, and
  \bibinfo{author}{\bibfnamefont{C.~V.} \bibnamefont{den Broeck}},
  \bibinfo{year}{2007}, {``}\bibinfo{title}{Dissipation: The Phase-Space
  Perspective},{''} \bibinfo{journal}{Phys. Rev. Lett.}
  \textbf{\bibinfo{volume}{98}},  \bibinfo{pages}{080602}.

\bibitem[{\citenamefont{Khinchin}(1949)}]{Khinchin49Book}
\bibinfo{author}{\bibnamefont{Khinchin}, \bibfnamefont{A.}},
  \bibinfo{year}{1949}, \emph{\bibinfo{title}{Mathematical foundations of
  statistical mechanics}} (\bibinfo{publisher}{Dover}, \bibinfo{address}{New
  York}).

\bibitem[{\citenamefont{Kippenberg and Vahala}(2008)}]{Kippenberg08SCIENCE321}
\bibinfo{author}{\bibnamefont{Kippenberg}, \bibfnamefont{T.~J.}}, and
  \bibinfo{author}{\bibfnamefont{K.~J.} \bibnamefont{Vahala}},
  \bibinfo{year}{2008}, {``}\bibinfo{title}{Cavity optomechanics: Back-action
  at the mesoscale},{''} \bibinfo{journal}{Science}
  \textbf{\bibinfo{volume}{321}},  \bibinfo{pages}{1172--1176}.

\bibitem[{\citenamefont{Kobe}(1981)}]{Kobe81AJP49}
\bibinfo{author}{\bibnamefont{Kobe}, \bibfnamefont{D.~H.}},
  \bibinfo{year}{1981}, {``}\bibinfo{title}{Gauge-invariant classical
  Hamiltonian formulation of the electrodynamics of nonrelativistic
  particles},{''} \bibinfo{journal}{Am. J. Phys.}
  \textbf{\bibinfo{volume}{49}},  \bibinfo{pages}{581--588}.

\bibitem[{\citenamefont{{Kubo}}(1957)}]{Kubo57aJPSJ12}
\bibinfo{author}{\bibnamefont{{Kubo}}, \bibfnamefont{R.}},
  \bibinfo{year}{1957}, {``}\bibinfo{title}{Statistical-mechanical theory of
  irreversible processes. I},{''} \bibinfo{journal}{J. Phys. Soc. Jpn.}
  \textbf{\bibinfo{volume}{12}},  \bibinfo{pages}{570--586}.

\bibitem[{\citenamefont{Kurchan}(2000)}]{Kurchan00arXiv}
\bibinfo{author}{\bibnamefont{Kurchan}, \bibfnamefont{J.}},
  \bibinfo{year}{2000}, {``}\bibinfo{title}{A quantum fluctuation theorem},{''}
  \eprint{arXiv:cond-mat/0007360}.

\bibitem[{\citenamefont{LaHaye} \emph{et~al.}(2004)\citenamefont{LaHaye, Buu,
  Camarota, and Schwab}}]{LaHaye04SCIENCE304}
\bibinfo{author}{\bibnamefont{LaHaye}, \bibfnamefont{M.~D.}},
  \bibinfo{author}{\bibfnamefont{O.}~\bibnamefont{Buu}},
  \bibinfo{author}{\bibfnamefont{B.}~\bibnamefont{Camarota}}, and
  \bibinfo{author}{\bibfnamefont{K.~C.} \bibnamefont{Schwab}},
  \bibinfo{year}{2004}, {``}\bibinfo{title}{Approaching the Quantum Limit of a
  Nanomechanical Resonator},{''} \bibinfo{journal}{Science}
  \textbf{\bibinfo{volume}{304}},  \bibinfo{pages}{74--77}.

\bibitem[{\citenamefont{{Liphardt}}
  \emph{et~al.}(2002)\citenamefont{{Liphardt}, {Dumont}, {Smith}, {Tinoco}, and
  {Bustamante}}}]{Liphardt02SCIENCE296}
\bibinfo{author}{\bibnamefont{{Liphardt}}, \bibfnamefont{J.}},
  \bibinfo{author}{\bibfnamefont{S.}~\bibnamefont{{Dumont}}},
  \bibinfo{author}{\bibfnamefont{S.~B.} \bibnamefont{{Smith}}},
  \bibinfo{author}{\bibfnamefont{I.}~\bibnamefont{{Tinoco}}}, and
  \bibinfo{author}{\bibfnamefont{C.}~\bibnamefont{{Bustamante}}},
  \bibinfo{year}{2002}, {``}\bibinfo{title}{{Equilibrium information from
  nonequilibrium measurements in an experimental test of Jarzynski's
  equality}},{''} \bibinfo{journal}{Science} \textbf{\bibinfo{volume}{296}},
  \bibinfo{pages}{1832--1836}.

\bibitem[{\citenamefont{Marconi} \emph{et~al.}(2008)\citenamefont{Marconi,
  Puglisi, Rondoni, and Vulpiani}}]{Marconi08PREP111}
\bibinfo{author}{\bibnamefont{Marconi}, \bibfnamefont{U.~M.~B.}},
  \bibinfo{author}{\bibfnamefont{A.}~\bibnamefont{Puglisi}},
  \bibinfo{author}{\bibfnamefont{L.}~\bibnamefont{Rondoni}}, and
  \bibinfo{author}{\bibfnamefont{A.}~\bibnamefont{Vulpiani}},
  \bibinfo{year}{2008}, {``}\bibinfo{title}{Fluctuation-dissipation: Response
  theory in statistical physics},{''} \bibinfo{journal}{Phys. Rep.}
  \textbf{\bibinfo{volume}{461}},  \bibinfo{pages}{111--195}.

\bibitem[{\citenamefont{Maruyama} \emph{et~al.}(2009)\citenamefont{Maruyama,
  Nori, and Vedral}}]{Maruyama09RMP81}
\bibinfo{author}{\bibnamefont{Maruyama}, \bibfnamefont{K.}},
  \bibinfo{author}{\bibfnamefont{F.}~\bibnamefont{Nori}}, and
  \bibinfo{author}{\bibfnamefont{V.}~\bibnamefont{Vedral}},
  \bibinfo{year}{2009}, {``}\bibinfo{title}{Colloquium: The physics of
  Maxwell's demon and information},{''} \bibinfo{journal}{Rev. Mod. Phys.}
  \textbf{\bibinfo{volume}{81}},  \bibinfo{pages}{1--23}.

\bibitem[{\citenamefont{Messiah}(1962)}]{Messiah62Book}
\bibinfo{author}{\bibnamefont{Messiah}, \bibfnamefont{A.}},
  \bibinfo{year}{1962}, \emph{\bibinfo{title}{Quantum Mechanics}}
  (\bibinfo{publisher}{North Holland}, \bibinfo{address}{Amsterdam}).

\bibitem[{\citenamefont{Minh and Adib}(2008)}]{Minh08PRL100}
\bibinfo{author}{\bibnamefont{Minh}, \bibfnamefont{D.~D.~L.}}, and
  \bibinfo{author}{\bibfnamefont{A.~B.} \bibnamefont{Adib}},
  \bibinfo{year}{2008}, {``}\bibinfo{title}{Optimized free energies from
  bidirectional single-molecule force spectroscopy},{''}
  \bibinfo{journal}{Phys. Rev. Lett.} \textbf{\bibinfo{volume}{100}},
  \bibinfo{pages}{180602}.

\bibitem[{\citenamefont{{Morikuni} and {Tasaki}}(2011)}]{Morikuni11arXiv}
\bibinfo{author}{\bibnamefont{{Morikuni}}, \bibfnamefont{Y.}}, and
  \bibinfo{author}{\bibfnamefont{H.}~\bibnamefont{{Tasaki}}},
  \bibinfo{year}{2011}, {``}\bibinfo{title}{{Quantum Jarzynski-Sagawa-Ueda
  relations}},{''} \bibinfo{journal}{J. Stat. Phys.} \textbf{\bibinfo{volume}{143}},  \bibinfo{pages}{1}.

\bibitem[{\citenamefont{Mukamel}(2003)}]{Mukamel03PRL90}
\bibinfo{author}{\bibnamefont{Mukamel}, \bibfnamefont{S.}},
  \bibinfo{year}{2003}, {``}\bibinfo{title}{Quantum extension of the Jarzynski
  relation: Analogy with stochastic dephasing},{''} \bibinfo{journal}{Phys.
  Rev. Lett.} \textbf{\bibinfo{volume}{90}},  \bibinfo{pages}{170604}.

\bibitem[{\citenamefont{Nakamura} \emph{et~al.}(2010)\citenamefont{Nakamura,
  Yamauchi, Hashisaka, Chida, Kobayashi, Ono, Leturcq, Ensslin, Saito, Utsumi,
  and Gossard}}]{Nakamura10PRL104}
\bibinfo{author}{\bibnamefont{Nakamura}, \bibfnamefont{S.}},
  \bibinfo{author}{\bibfnamefont{Y.}~\bibnamefont{Yamauchi}},
  \bibinfo{author}{\bibfnamefont{M.}~\bibnamefont{Hashisaka}},
  \bibinfo{author}{\bibfnamefont{K.}~\bibnamefont{Chida}},
  \bibinfo{author}{\bibfnamefont{K.}~\bibnamefont{Kobayashi}},
  \bibinfo{author}{\bibfnamefont{T.}~\bibnamefont{Ono}},
  \bibinfo{author}{\bibfnamefont{R.}~\bibnamefont{Leturcq}},
  \bibinfo{author}{\bibfnamefont{K.}~\bibnamefont{Ensslin}},
  \bibinfo{author}{\bibfnamefont{K.}~\bibnamefont{Saito}},
  \bibinfo{author}{\bibfnamefont{Y.}~\bibnamefont{Utsumi}}, and
  \bibinfo{author}{\bibfnamefont{A.~C.} \bibnamefont{Gossard}},
  \bibinfo{year}{2010}, {``}\bibinfo{title}{Nonequilibrium fluctuation
  relations in a quantum coherent conductor},{''} \bibinfo{journal}{Phys. Rev.
  Lett.} \textbf{\bibinfo{volume}{104}},  \bibinfo{pages}{080602}.

\bibitem[{\citenamefont{{Nakamura}}
  \emph{et~al.}(2011)\citenamefont{{Nakamura}, {Yamauchi}, {Hashisaka},
  {Chida}, {Kobayashi}, {Ono}, {Leturcq}, {Ensslin}, {Saito}, {Utsumi}, and
  {Gossard}}}]{Nakamura11arXiv}
\bibinfo{author}{\bibnamefont{{Nakamura}}, \bibfnamefont{S.}},
  \bibinfo{author}{\bibfnamefont{Y.}~\bibnamefont{{Yamauchi}}},
  \bibinfo{author}{\bibfnamefont{M.}~\bibnamefont{{Hashisaka}}},
  \bibinfo{author}{\bibfnamefont{K.}~\bibnamefont{{Chida}}},
  \bibinfo{author}{\bibfnamefont{K.}~\bibnamefont{{Kobayashi}}},
  \bibinfo{author}{\bibfnamefont{T.}~\bibnamefont{{Ono}}},
  \bibinfo{author}{\bibfnamefont{R.}~\bibnamefont{{Leturcq}}},
  \bibinfo{author}{\bibfnamefont{K.}~\bibnamefont{{Ensslin}}},
  \bibinfo{author}{\bibfnamefont{K.}~\bibnamefont{{Saito}}},
  \bibinfo{author}{\bibfnamefont{Y.}~\bibnamefont{{Utsumi}}}, and
  \bibinfo{author}{\bibfnamefont{A.~C.} \bibnamefont{{Gossard}}},
  \bibinfo{year}{2011}, {``}\bibinfo{title}{{Fluctuation Theorem and
  Microreversibility in a Quantum Coherent Conductor}},{''}
  \bibinfo{journal}{Phys. Rev. B} \textbf{\bibinfo{volume}{83}},  \bibinfo{pages}{155431}.

\bibitem[{\citenamefont{Nieuwenhuizen and
  Allahverdyan}(2002)}]{Nieuwenhuizen02PRE66}
\bibinfo{author}{\bibnamefont{Nieuwenhuizen}, \bibfnamefont{T.~M.}}, and
  \bibinfo{author}{\bibfnamefont{A.~E.} \bibnamefont{Allahverdyan}},
  \bibinfo{year}{2002}, {``}\bibinfo{title}{Statistical thermodynamics of
  quantum Brownian motion: Construction of perpetuum mobile of the second
  kind},{''} \bibinfo{journal}{Phys. Rev. E} \textbf{\bibinfo{volume}{66}},
  \bibinfo{pages}{036102}.

\bibitem[{\citenamefont{Nyquist}(1928)}]{Nyquist28PR32}
\bibinfo{author}{\bibnamefont{Nyquist}, \bibfnamefont{H.}},
  \bibinfo{year}{1928}, {``}\bibinfo{title}{Thermal agitation of electric
  charge in conductors},{''} \bibinfo{journal}{Phys. Rev.}
  \textbf{\bibinfo{volume}{32}},  \bibinfo{pages}{110--113}.

\bibitem[{\citenamefont{O'Connell} \emph{et~al.}(2010)\citenamefont{O'Connell,
  Hofheinz, Ansmann, Bialczak, Lenander, E., Neeley, D., Wang, Weides, Wenner,
  Martinis} \emph{et~al.}}]{Oconnel10Nature464}
\bibinfo{author}{\bibnamefont{O'Connell}, \bibfnamefont{A.~D.}},
  \bibinfo{author}{\bibfnamefont{M.}~\bibnamefont{Hofheinz}},
  \bibinfo{author}{\bibfnamefont{M.}~\bibnamefont{Ansmann}},
  \bibinfo{author}{\bibfnamefont{R.~C.} \bibnamefont{Bialczak}},
  \bibinfo{author}{\bibfnamefont{M.}~\bibnamefont{Lenander}},
  \bibinfo{author}{\bibfnamefont{E.~L.} \bibnamefont{E.}},
  \bibinfo{author}{\bibfnamefont{M.}~\bibnamefont{Neeley}},
  \bibinfo{author}{\bibfnamefont{D.~S.} \bibnamefont{D.}},
  \bibinfo{author}{\bibfnamefont{H.}~\bibnamefont{Wang}},
  \bibinfo{author}{\bibfnamefont{M.}~\bibnamefont{Weides}},
  \bibinfo{author}{\bibfnamefont{J.}~\bibnamefont{Wenner}},
  \bibinfo{author}{\bibfnamefont{J.~M.} \bibnamefont{Martinis}}, \emph{et~al.},
  \bibinfo{year}{2010}, {``}\bibinfo{title}{Quantum ground state and
  single-phonon control of a mechanical resonator},{''}
  \bibinfo{journal}{Nature} \textbf{\bibinfo{volume}{464}},
  \bibinfo{pages}{697--703}.

\bibitem[{\citenamefont{Ohzeki}(2010)}]{Ohzeki10PRL105}
\bibinfo{author}{\bibnamefont{Ohzeki}, \bibfnamefont{M.}},
  \bibinfo{year}{2010}, {``}\bibinfo{title}{Quantum Annealing with the
  Jarzynski Equality},{''} \bibinfo{journal}{Phys. Rev. Lett.}
  \textbf{\bibinfo{volume}{105}},  \bibinfo{pages}{050401}.

\bibitem[{\citenamefont{Onsager}(1931{\natexlab{a}})}]{Onsager31PR37}
\bibinfo{author}{\bibnamefont{Onsager}, \bibfnamefont{L.}},
  \bibinfo{year}{1931}{\natexlab{a}}, {``}\bibinfo{title}{Reciprocal relations
  in irreversible processes. I.},{''} \bibinfo{journal}{Phys. Rev.}
  \textbf{\bibinfo{volume}{37}},  \bibinfo{pages}{405--426}.

\bibitem[{\citenamefont{Onsager}(1931{\natexlab{b}})}]{Onsager31PR38}
\bibinfo{author}{\bibnamefont{Onsager}, \bibfnamefont{L.}},
  \bibinfo{year}{1931}{\natexlab{b}}, {``}\bibinfo{title}{Reciprocal relations
  in irreversible processes. II.},{''} \bibinfo{journal}{Phys. Rev.}
  \textbf{\bibinfo{volume}{38}},  \bibinfo{pages}{2265--2279}.

\bibitem[{\citenamefont{Peliti}(2008{\natexlab{a}})}]{Peliti08PRL101}
\bibinfo{author}{\bibnamefont{Peliti}, \bibfnamefont{L.}},
  \bibinfo{year}{2008}{\natexlab{a}}, {``}\bibinfo{title}{Comment on ``Failure
  of the Work-Hamiltonian Connection for Free-Energy Calculations''},{''}
  \bibinfo{journal}{Phys. Rev. Lett.} \textbf{\bibinfo{volume}{101}},
  \bibinfo{pages}{098903}.

\bibitem[{\citenamefont{Peliti}(2008{\natexlab{b}})}]{Peliti08JSM08}
\bibinfo{author}{\bibnamefont{Peliti}, \bibfnamefont{L.}},
  \bibinfo{year}{2008}{\natexlab{b}}, {``}\bibinfo{title}{On the
  work-Hamiltonian connection in manipulated systems},{''} \bibinfo{journal}{J.
  Stat. Mech.: Theory. Exp.}  \bibinfo{pages}{P05002}.

\bibitem[{\citenamefont{Piechocinska}(2000)}]{Piechocinska00PRA61}
\bibinfo{author}{\bibnamefont{Piechocinska}, \bibfnamefont{B.}},
  \bibinfo{year}{2000}, {``}\bibinfo{title}{Information erasure},{''}
  \bibinfo{journal}{Phys. Rev. A} \textbf{\bibinfo{volume}{61}},
  \bibinfo{pages}{062314}.

\bibitem[{\citenamefont{Prost} \emph{et~al.}(2009)\citenamefont{Prost, Joanny,
  and Parrondo}}]{Prost09PRL103}
\bibinfo{author}{\bibnamefont{Prost}, \bibfnamefont{J.}},
  \bibinfo{author}{\bibfnamefont{J.-F.} \bibnamefont{Joanny}}, and
  \bibinfo{author}{\bibfnamefont{J.~M.~R.} \bibnamefont{Parrondo}},
  \bibinfo{year}{2009}, {``}\bibinfo{title}{Generalized Fluctuation-Dissipation
  Theorem for Steady-State Systems},{''} \bibinfo{journal}{Phys. Rev. Lett.}
  \textbf{\bibinfo{volume}{103}},
  \bibinfo{pages}{090601}.

\bibitem[{\citenamefont{Ren} \emph{et~al.}(2010)\citenamefont{Ren, H\"anggi,
  and Li}}]{Ren10PRL104}
\bibinfo{author}{\bibnamefont{Ren}, \bibfnamefont{J.}},
  \bibinfo{author}{\bibfnamefont{P.}~\bibnamefont{H\"anggi}}, and
  \bibinfo{author}{\bibfnamefont{B.}~\bibnamefont{Li}}, \bibinfo{year}{2010},
  {``}\bibinfo{title}{Berry-Phase-Induced Heat Pumping and Its Impact on the
  Fluctuation Theorem},{''} \bibinfo{journal}{Phys. Rev. Lett.}
  \textbf{\bibinfo{volume}{104}},  \bibinfo{pages}{170601}.

\bibitem[{\citenamefont{Rondoni and
  Mej{\'i}a-Monasterio}(2007)}]{Rondoni07NL20}
\bibinfo{author}{\bibnamefont{Rondoni}, \bibfnamefont{L.}}, and
  \bibinfo{author}{\bibfnamefont{C.}~\bibnamefont{Mej{\'i}a-Monasterio}},
  \bibinfo{year}{2007}, {``}\bibinfo{title}{Fluctuations in nonequilibrium
  statistical mechanics: models, mathematical theory, physical mechanisms},{''}
  \bibinfo{journal}{Nonlinearity} \textbf{\bibinfo{volume}{20}},
  \bibinfo{pages}{R1--R37}.

\bibitem[{\citenamefont{Sagawa and Ueda}(2010)}]{Sagawa10PRL104}
\bibinfo{author}{\bibnamefont{Sagawa}, \bibfnamefont{T.}}, and
  \bibinfo{author}{\bibfnamefont{M.}~\bibnamefont{Ueda}}, \bibinfo{year}{2010},
  {``}\bibinfo{title}{Generalized Jarzynski Equality under Nonequilibrium
  Feedback Control},{''} \bibinfo{journal}{Phys. Rev. Lett.}
  \textbf{\bibinfo{volume}{104}},  \bibinfo{pages}{090602}.

\bibitem[{\citenamefont{Saito and Dhar}(2007)}]{Saito07PRL99}
\bibinfo{author}{\bibnamefont{Saito}, \bibfnamefont{K.}}, and
  \bibinfo{author}{\bibfnamefont{A.}~\bibnamefont{Dhar}}, \bibinfo{year}{2007},
  {``}\bibinfo{title}{Fluctuation Theorem in Quantum Heat Conduction},{''}
  \bibinfo{journal}{Phys. Rev. Lett.} \textbf{\bibinfo{volume}{99}},
  \bibinfo{pages}{180601}.

\bibitem[{\citenamefont{Saito and Utsumi}(2008)}]{Saito08PRB78}
\bibinfo{author}{\bibnamefont{Saito}, \bibfnamefont{K.}}, and
  \bibinfo{author}{\bibfnamefont{Y.}~\bibnamefont{Utsumi}},
  \bibinfo{year}{2008}, {``}\bibinfo{title}{Symmetry in full counting
  statistics, fluctuation theorem, and relations among nonlinear transport
  coefficients in the presence of a magnetic field},{''}
  \bibinfo{journal}{Phys. Rev. B} \textbf{\bibinfo{volume}{78}},
  \bibinfo{pages}{115429}.

\bibitem[{\citenamefont{Schleich}(2001)}]{Schleich01Book}
\bibinfo{author}{\bibnamefont{Schleich}, \bibfnamefont{W.~P.}},
  \bibinfo{year}{2001}, \emph{\bibinfo{title}{Quantum Optics in Phase Space}}
  (\bibinfo{publisher}{Wiley-VCH}, \bibinfo{address}{Berlin}).

\bibitem[{\citenamefont{Schulz} \emph{et~al.}(2008)\citenamefont{Schulz,
  Poschinger, Ziesel, and Schmidt-Kaler}}]{Schulz08NJP10}
\bibinfo{author}{\bibnamefont{Schulz}, \bibfnamefont{S.~A.}},
  \bibinfo{author}{\bibfnamefont{U.}~\bibnamefont{Poschinger}},
  \bibinfo{author}{\bibfnamefont{F.}~\bibnamefont{Ziesel}}, and
  \bibinfo{author}{\bibfnamefont{F.}~\bibnamefont{Schmidt-Kaler}},
  \bibinfo{year}{2008}, {``}\bibinfo{title}{Sideband cooling and coherent
  dynamics in a microchip multi-segmented ion trap},{''} \bibinfo{journal}{New
  J. Phys.} \textbf{\bibinfo{volume}{10}},  \bibinfo{pages}{045007}.



\bibitem[{\citenamefont{Seifert}(2008)}]{Seifert08EPJB64}
\bibinfo{author}{\bibnamefont{Seifert}, \bibfnamefont{U.}},
  \bibinfo{year}{2008}, {``}\bibinfo{title}{Stochastic thermodynamics: Principles and perspectives},{''} \bibinfo{journal}{Eur. Phys. J. B} \textbf{\bibinfo{volume}{64}},  \bibinfo{pages}{423-431}.



\bibitem[{\citenamefont{Stratonovich}(1992)}]{Stratonovich92Book}
\bibinfo{author}{\bibnamefont{Stratonovich}, \bibfnamefont{R.~L.}},
  \bibinfo{year}{1992}, \emph{\bibinfo{title}{Nonlinear nonequilibrium
  thermodynamics I : linear and nonlinear fluctuation-dissipation theorems}},
  volume~\bibinfo{volume}{57} of \emph{\bibinfo{series}{Springer series in
  synergetics}} (\bibinfo{publisher}{Springer-Verlag}, \bibinfo{address}{Berlin
  ; Heidelberg; New York}).

\bibitem[{\citenamefont{Stratonovich}(1994)}]{Stratonovich94Book}
\bibinfo{author}{\bibnamefont{Stratonovich}, \bibfnamefont{R.~L.}},
  \bibinfo{year}{1994}, \emph{\bibinfo{title}{Nonlinear nonequilibrium
  thermodynamics II : advanced theory}}, volume~\bibinfo{volume}{59} of
  \emph{\bibinfo{series}{Springer series in synergetics}}
  (\bibinfo{publisher}{Springer-Verlag}, \bibinfo{address}{Berlin ; Heidelberg;
  New York}).

\bibitem[{\citenamefont{Sutherland}(1902)}]{Sutherland02PHILMAG3}
\bibinfo{author}{\bibnamefont{Sutherland}, \bibfnamefont{W.}},
  \bibinfo{year}{1902}, {``}\bibinfo{title}{Ionization, ionic velocities, and
  atomic sizes},{''} \bibinfo{journal}{Phil. Mag.}
  \textbf{\bibinfo{volume}{3}},  \bibinfo{pages}{161--177}.

\bibitem[{\citenamefont{Sutherland}(1905)}]{Sutherland05PHILMAG9}
\bibinfo{author}{\bibnamefont{Sutherland}, \bibfnamefont{W.}},
  \bibinfo{year}{1905}, {``}\bibinfo{title}{Dynamical theory of diffusion for
  non-electrolytes and the molecular mass of albumin},{''}
  \bibinfo{journal}{Phil. Mag.} \textbf{\bibinfo{volume}{9}},
  \bibinfo{pages}{781--785}.

\bibitem[{\citenamefont{Talkner}
  \emph{et~al.}(2008{\natexlab{a}})\citenamefont{Talkner, Burada, and
  {H{\"a}nggi}}}]{Talkner08PRE78}
\bibinfo{author}{\bibnamefont{Talkner}, \bibfnamefont{P.}},
  \bibinfo{author}{\bibfnamefont{P.~S.} \bibnamefont{Burada}}, and
  \bibinfo{author}{\bibfnamefont{P.}~\bibnamefont{{H{\"a}nggi}}},
  \bibinfo{year}{2008}{\natexlab{a}}, {``}\bibinfo{title}{Statistics of work
  performed on a forced quantum oscillator},{''} \bibinfo{journal}{Phys. Rev.
  E} \textbf{\bibinfo{volume}{78}},  \bibinfo{pages}{011115}.

\bibitem[{\citenamefont{Talkner}
  \emph{et~al.}(2009{\natexlab{a}})\citenamefont{Talkner, Burada, and
  H\"anggi}}]{Talkner09PRE79}
\bibinfo{author}{\bibnamefont{Talkner}, \bibfnamefont{P.}},
  \bibinfo{author}{\bibfnamefont{P.~S.} \bibnamefont{Burada}}, and
  \bibinfo{author}{\bibfnamefont{P.}~\bibnamefont{H\"anggi}},
  \bibinfo{year}{2009}{\natexlab{a}}, {``}\bibinfo{title}{Erratum: Statistics
  of work performed on a forced quantum oscillator [Phys. Rev. E 78, 011115
  (2008)]},{''} \bibinfo{journal}{Phys. Rev. E} \textbf{\bibinfo{volume}{79}},
  \bibinfo{pages}{039902}.

\bibitem[{\citenamefont{Talkner}
  \emph{et~al.}(2009{\natexlab{b}})\citenamefont{Talkner, Campisi, and
  H{\"a}nggi}}]{Talkner09JSM09}
\bibinfo{author}{\bibnamefont{Talkner}, \bibfnamefont{P.}},
  \bibinfo{author}{\bibfnamefont{M.}~\bibnamefont{Campisi}}, and
  \bibinfo{author}{\bibfnamefont{P.}~\bibnamefont{H{\"a}nggi}},
  \bibinfo{year}{2009}{\natexlab{b}}, {``}\bibinfo{title}{Fluctuation theorems
  in driven open quantum systems},{''} \bibinfo{journal}{J. Stat. Mech.: Theory
  Exp.}  \bibinfo{pages}{P02025}.

\bibitem[{\citenamefont{Talkner and H{\"a}nggi}(2007)}]{Talkner07JPA40}
\bibinfo{author}{\bibnamefont{Talkner}, \bibfnamefont{P.}}, and
  \bibinfo{author}{\bibfnamefont{P.}~\bibnamefont{H{\"a}nggi}},
  \bibinfo{year}{2007}, {``}\bibinfo{title}{The Tasaki-Crooks quantum
  fluctuation theorem},{''} \bibinfo{journal}{J. Phys. A}
  \textbf{\bibinfo{volume}{40}},  \bibinfo{pages}{F569--F571}.

\bibitem[{\citenamefont{Talkner}
  \emph{et~al.}(2008{\natexlab{b}})\citenamefont{Talkner, H{\"a}nggi, and
  Morillo}}]{Talkner08PRE77}
\bibinfo{author}{\bibnamefont{Talkner}, \bibfnamefont{P.}},
  \bibinfo{author}{\bibfnamefont{P.}~\bibnamefont{H{\"a}nggi}}, and
  \bibinfo{author}{\bibfnamefont{M.}~\bibnamefont{Morillo}},
  \bibinfo{year}{2008}{\natexlab{b}}, {``}\bibinfo{title}{Microcanonical
  quantum fluctuation theorems},{''} \bibinfo{journal}{Phys. Rev. E}
  \textbf{\bibinfo{volume}{77}},  \bibinfo{pages}{051131}.

\bibitem[{\citenamefont{Talkner} \emph{et~al.}(2007)\citenamefont{Talkner,
  Lutz, and H{\"a}nggi}}]{Talkner07PRE75}
\bibinfo{author}{\bibnamefont{Talkner}, \bibfnamefont{P.}},
  \bibinfo{author}{\bibfnamefont{E.}~\bibnamefont{Lutz}}, and
  \bibinfo{author}{\bibfnamefont{P.}~\bibnamefont{H{\"a}nggi}},
  \bibinfo{year}{2007}, {``}\bibinfo{title}{Fluctuation theorems: Work is not
  an observable},{''} \bibinfo{journal}{Phys. Rev. E}
  \textbf{\bibinfo{volume}{75}},  \bibinfo{pages}{050102}.

\bibitem[{\citenamefont{Tasaki}(2000)}]{Tasaki00arXiv}
\bibinfo{author}{\bibnamefont{Tasaki}, \bibfnamefont{H.}},
  \bibinfo{year}{2000}, {``}\bibinfo{title}{Jarzynski relations for quantum
  systems and some applications},{''} \eprint{arXiv:cond-mat/0009244}.

\bibitem[{\citenamefont{Utsumi} \emph{et~al.}(2010)\citenamefont{Utsumi,
  Golubev, Marthaler, Saito, Fujisawa, and Sch{\"o}n}}]{Utsumi10PRB81}
\bibinfo{author}{\bibnamefont{Utsumi}, \bibfnamefont{Y.}},
  \bibinfo{author}{\bibfnamefont{D.~S.} \bibnamefont{Golubev}},
  \bibinfo{author}{\bibfnamefont{M.}~\bibnamefont{Marthaler}},
  \bibinfo{author}{\bibfnamefont{K.}~\bibnamefont{Saito}},
  \bibinfo{author}{\bibfnamefont{T.}~\bibnamefont{Fujisawa}}, and
  \bibinfo{author}{\bibfnamefont{G.}~\bibnamefont{Sch{\"o}n}},
  \bibinfo{year}{2010}, {``}\bibinfo{title}{Bidirectional single-electron
  counting and the fluctuation theorem},{''} \bibinfo{journal}{Phys. Rev. B}
  \textbf{\bibinfo{volume}{81}},  \bibinfo{pages}{125331}.

\bibitem[{\citenamefont{Vaikuntanathan and
  Jarzynski}(2008)}]{Vaikuntanathan08PRL100}
\bibinfo{author}{\bibnamefont{Vaikuntanathan}, \bibfnamefont{S.}}, and
  \bibinfo{author}{\bibfnamefont{C.}~\bibnamefont{Jarzynski}},
  \bibinfo{year}{2008}, {``}\bibinfo{title}{Escorted free energy simulations:
  Improving convergence by reducing dissipation},{''} \bibinfo{journal}{Phys.
  Rev. Lett.} \textbf{\bibinfo{volume}{100}},  \bibinfo{pages}{190601}.

\bibitem[{\citenamefont{Vedral}(2002)}]{Vedral02RMP74}
\bibinfo{author}{\bibnamefont{Vedral}, \bibfnamefont{V.}},
  \bibinfo{year}{2002}, {``}\bibinfo{title}{The role of relative entropy in
  quantum information theory},{''} \bibinfo{journal}{Rev. Mod. Phys.}
  \textbf{\bibinfo{volume}{74}},  \bibinfo{pages}{197--234}.

\bibitem[{\citenamefont{Vilar and Rubi}(2008{\natexlab{a}})}]{Vilar08aPRL101}
\bibinfo{author}{\bibnamefont{Vilar}, \bibfnamefont{J.~M.~G.}}, and
  \bibinfo{author}{\bibfnamefont{J.~M.} \bibnamefont{Rubi}},
  \bibinfo{year}{2008}{\natexlab{a}}, {``}\bibinfo{title}{Comment on ``Failure
  of the work-Hamiltonian connection for free-energy calculations'' --
  Reply},{''} \bibinfo{journal}{Phys. Rev. Lett.}
  \textbf{\bibinfo{volume}{101}},  \bibinfo{pages}{098902}.

\bibitem[{\citenamefont{Vilar and Rubi}(2008{\natexlab{b}})}]{Vilar08bPRL101}
\bibinfo{author}{\bibnamefont{Vilar}, \bibfnamefont{J.~M.~G.}}, and
  \bibinfo{author}{\bibfnamefont{J.~M.} \bibnamefont{Rubi}},
  \bibinfo{year}{2008}{\natexlab{b}}, {``}\bibinfo{title}{Comment on ``Failure
  of the work-Hamiltonian connection for free-energy calculations'' --
  Reply},{''} \bibinfo{journal}{Phys. Rev. Lett.}
  \textbf{\bibinfo{volume}{101}},  \bibinfo{pages}{098904}.

\bibitem[{\citenamefont{Vilar and Rubi}(2008{\natexlab{c}})}]{Vilar08PRL100}
\bibinfo{author}{\bibnamefont{Vilar}, \bibfnamefont{J.~M.~G.}}, and
  \bibinfo{author}{\bibfnamefont{J.~M.} \bibnamefont{Rubi}},
  \bibinfo{year}{2008}{\natexlab{c}}, {``}\bibinfo{title}{Failure of the
  work-Hamiltonian connection for free-energy calculations},{''}
  \bibinfo{journal}{Phys. Rev. Lett.} \textbf{\bibinfo{volume}{100}},
  \bibinfo{pages}{020601}.

\bibitem[{\citenamefont{Yukawa}(2000)}]{Yukawa00JPSJ69}
\bibinfo{author}{\bibnamefont{Yukawa}, \bibfnamefont{S.}},
  \bibinfo{year}{2000}, {``}\bibinfo{title}{A quantum analogue of the Jarzynski
  equality},{''} \bibinfo{journal}{J. Phys. Soc. Jap.}
  \textbf{\bibinfo{volume}{69}}(\bibinfo{number}{8}),
  \bibinfo{pages}{2367--2370}.

\bibitem[{\citenamefont{Zimanyi and Silbey}(2009)}]{Zimanyi09JCP130}
\bibinfo{author}{\bibnamefont{Zimanyi}, \bibfnamefont{E.~N.}}, and
  \bibinfo{author}{\bibfnamefont{R.~J.} \bibnamefont{Silbey}},
  \bibinfo{year}{2009}, {``}\bibinfo{title}{The work-Hamiltonian connection and
  the usefulness of the Jarzynski equality for free energy calculations},{''}
  \bibinfo{journal}{J. Chem. Phys.} \textbf{\bibinfo{volume}{130}},
  \bibinfo{pages}{171102}.

\end{thebibliography}
%

\onecolumngrid
\newpage
\begin{flushleft}
\includegraphics[width=\linewidth]{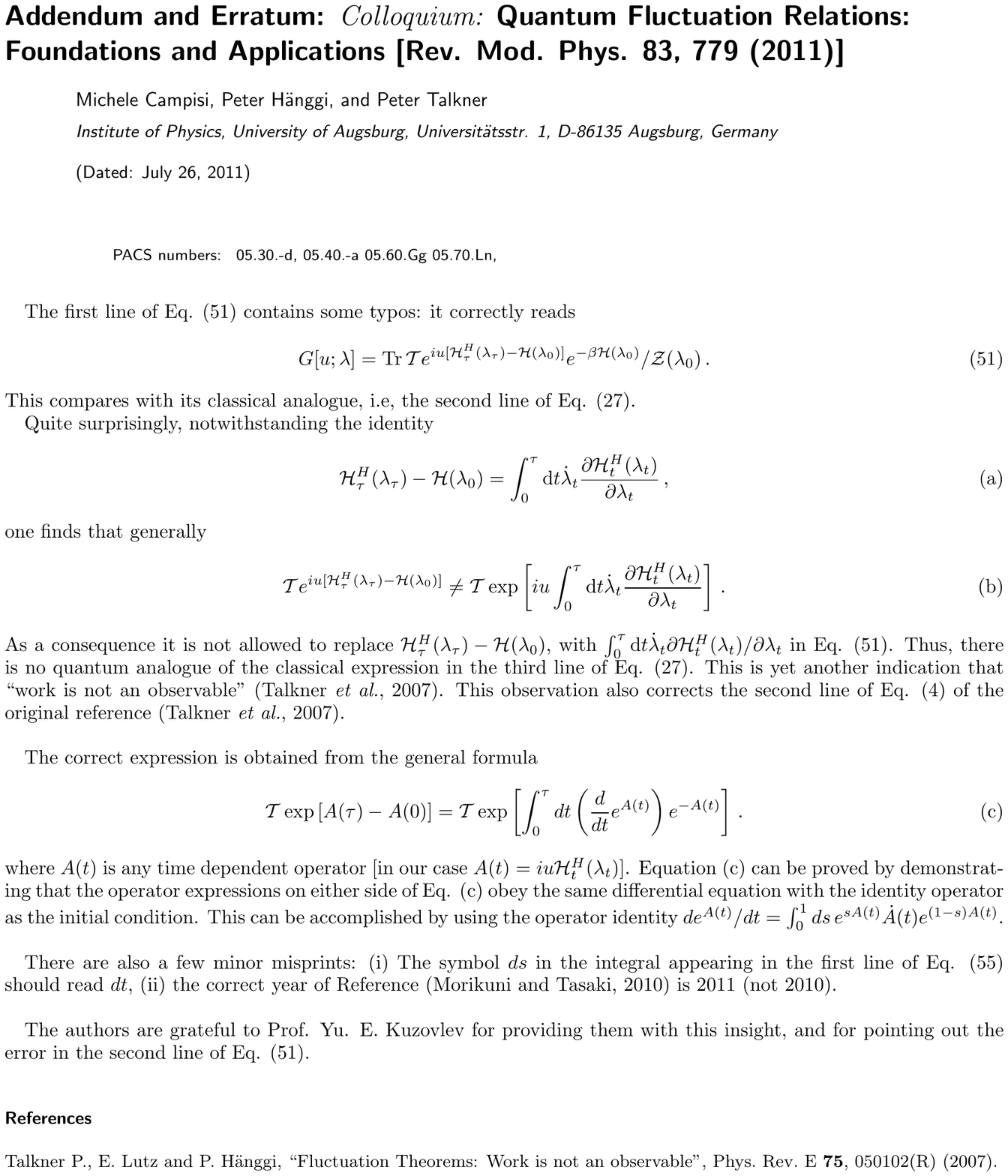}
\end{flushleft}

\end{document}